\documentclass{amsart}
\usepackage{comment}
\usepackage{geometry}
\usepackage{hyperref}
\usepackage{amsmath,amssymb,amsthm}
\usepackage{tikz}
\usetikzlibrary{calc}
\usetikzlibrary{shapes}
\usetikzlibrary{positioning}

\tikzset{every loop/.style={},
           every node/.style={minimum size=8pt,inner sep=0,outer
sep=0,circle, draw,thick},
           bv/.style={rectangle,fill=teal}, rv/.style={fill=magenta}}

\usepackage{xcolor}
\usepackage{mathrsfs}
\usepackage{enumerate}
\newcommand\blue[1]{#1}
\newcommand\newblue[1]{#1}
\newcommand\new[1]{#1}
\newcommand\red[1]{#1}
\newcommand\nomenclature[3][Default]{}

\newcommand{\Nesetril}{Ne\v{s}et\v{r}il}
\newcommand{\Fresse}{Fra\"{i}ss\'{e}}

\DeclareMathOperator{\Age}{Age}
\DeclareMathOperator{\HN}{\bB_{{\mathcal F}}^{\text{HN}}}
\DeclareMathOperator{\id}{id}

\DeclareMathOperator{\Ho}{He}

\DeclareMathOperator{\PP}{Ppp}

\DeclareMathOperator{\mmod}{mod}
\DeclareMathOperator{\ind}{ind}

\newcommand{\bA}{{\mathfrak A}}
\newcommand{\bB}{{\mathfrak B}}
\newcommand{\bC}{{\mathfrak C}}
\newcommand{\bD}{{\mathfrak D}}

\newcommand{\bF}{\mathfrak{F}}
\newcommand{\bT}{\mathfrak{T}}
\newcommand{\bS}{\mathfrak{S}}

\newcommand{\mN}{{\mathbb N}}
\renewcommand{\Pr}{\mathscr{P}}
\newcommand{\Projs}{\mathscr{P}}
\newcommand\typ{\mathrm{typ}}
\newcommand{\scrC}{\mathscr C}

\newcommand\tuple[1]{\bar #1}
\newcommand\cl[2]{#1/\kern-4pt\sim_{#2}}

\DeclareMathOperator{\Aut}{Aut}
\DeclareMathOperator{\End}{End}
\DeclareMathOperator{\Pol}{Pol}

\newcommand\ignore[1]{}
\newtheorem{theorem}{Theorem}[section]
\newtheorem{proposition}{Proposition}[section]
\newtheorem{conjecture}{Conjecture}[section]
\newtheorem{lemma}{Lemma}[section]
\newtheorem{corollary}{Corollary}[section]
\theoremstyle{definition}
\newtheorem{definition}{Definition}[section]
\theoremstyle{remark}
\newtheorem{example}{Example}[section]

\DeclareMathOperator{\Csp}{CSP}
\DeclareMathOperator{\Forb}{Forb}

\title{A universal-algebraic proof of the complexity dichotomy for Monotone Monadic SNP}
\author{Manuel Bodirsky}
\address{Institut f\"ur Algebra, TU Dresden.}
\email{manuel.bodirsky@tu-dresden.de}
\urladdr{https://www.math.tu-dresden.de/~bodirsky/}

\author{Florent Madelaine}
\address{Univ Paris Est Creteil, LACL, F-94010}
\urladdr{https://www.lacl.fr/fmadelaine/}

\author{Antoine Mottet}
\address{Department of Algebra, Faculty of Mathematics and Physics, Charles University.}
\urladdr{http://www.karlin.mff.cuni.cz/~mottet/}

\thanks{An extended abstract of this paper appeared in the proceedings of LICS 2018. Manuel Bodirsky has received funding from the European Research Council (Grant Agreement no. 681988, CSP-Infinity), and the German Research Foundation (DFG, grant number 622397).
  Antoine Mottet received funding from DFG Graduiertenkolleg 1763 (QuantLA) and the European Research Council (Grant Agreement no 771005, CoCoSym).}
\date{}

\begin{document}
\maketitle

\begin{abstract}
The logic MMSNP is a restricted fragment
of existential second-order logic which can express many interesting queries in graph theory 
and finite model theory. The logic was introduced by Feder and Vardi who showed that every MMSNP sentence is computationally equivalent to a finite-domain constraint satisfaction problem (CSP); the involved probabilistic reductions were derandomized by Kun
using explicit constructions of expander structures. 
We present a new proof of
the reduction to finite-domain CSPs
which does not rely on the results of Kun.  
The new universal-algebraic proof allows us to obtain a stronger statement and to verify the more general Bodirsky-Pinsker dichotomy conjecture for CSPs in MMSNP. 
Our approach uses the fact that every MMSNP sentence 
describes
a finite union of CSPs 
for countably infinite $\omega$-categorical structures;
moreover, by a recent result of Hubi\v{c}ka and Ne\v{s}et\v{r}il, these structures can be expanded to homogeneous structures with finite relational signature and the Ramsey property. 
\end{abstract}

\tableofcontents

\section{Introduction}
Monotone Monadic SNP (MMSNP) is a fragment of monadic existential second-order logic whose sentences describe problems of the form
``given a structure $\bA$, is there a colouring of the elements of $\bA$ that avoids some fixed family of forbidden patterns?'' Examples of such problems are the classical $k$-colorability problem for graphs
(where the forbidden patterns are edges whose endpoints have the same colour), or the problem of colouring the vertices of a graph so as to avoid monochromatic triangles (Figure~\ref{fig:ex-precol}).

MMSNP has been introduced by Feder and Vardi~\cite{FederVardi}, whose motivation was to find fragments of existential second-order logic that exhibit a complexity dichotomy between P and NP-complete.
They proved that every problem described by an MMSNP sentence is equivalent under polynomial-time randomised reductions to a constraint satisfaction problem (CSP) over a finite domain,
and conjectured that every finite-domain CSP is in P or NP-complete.
Kun~\cite{Kun} later improved the result by derandomising the equivalence, thus showing that MMSNP exhibits a complexity dichotomy if and only if the Feder-Vardi dichotomy conjecture holds.
Recently, Bulatov~\cite{BulatovFVConjecture} and Zhuk~\cite{ZhukFVConjecture} independently proved that the dichotomy conjecture indeed holds.
Both authors establish a stronger form of the dichotomy, the so-called \emph{tractability conjecture}, which gives a characterisation of the finite-domain CSPs that are solvable in polynomial time (assuming P is not NP).
This characterisation is phrased in the language of universal algebra and is moreover decidable.

\begin{figure}
    \begin{center}
     \begin{tikzpicture}
        \node[bv] (tb1) at (2,0) {};
        \node[bv] (tb2) at (3,0) {};
        \node[bv] (tb3) at (2.5,-1) {};

        \node[rv] (tr1) at (4,0) {};
        \node[rv] (tr2) at (5,0) {};
        \node[rv] (tr3) at (4.5,-1) {};

        \draw(tb1) -- (tb2) -- (tb3) -- (tb1);
        \draw(tr1) -- (tr2) -- (tr3) -- (tr1);

    \end{tikzpicture}
    \caption{The No-monochromatic-triangle problem: the input is a finite graph $G$, and the question is whether there exists a colouring
    of the vertices of $G$ with two colours that avoids monochromatic triangles.}
    \label{fig:ex-precol}
    \end{center}
    \end{figure}

More recently, MMSNP was studied in connection with \emph{ontology-mediated data access}~\cite{FeierKuusistoLutz,LutzWolter} in database theory.
This connection was used to characterise the power of ontology-mediated queries in query languages used in practice, as well as to study
the rewritability of such queries into some efficient query languages such as first-order logic, Datalog, or fragments of Datalog~\cite{FeierKuusistoLutz,LutzSabellek}.
In~\cite{FeierKuusistoLutzRewritability}, it is proved that the first-order rewritability question for MMSNP is decidable.
We obtain a different proof that follows directly from our algebraic perspective on MMSNP.
An important question that remains open  in this area (see~\cite{FeierKuusistoLutz}) is whether Datalog-rewritability of MMSNP sentences is decidable.
Like the complexity dichotomy for finite-domain CSPs, this question was answered for finite-domain CSPs using the universal-algebraic approach.

\subsection*{Infinite-Domain Constraint Satisfaction and MMSNP}
\newblue{The universal-algebraic approach can also be used to study constraint satisfaction problems over
\emph{infinite} templates $\bB$, at least if the structure $\bB$ is $\omega$-categorical. 
If $\bB$ can even be expanded to a finitely bounded homogeneous structure,
then there exists a generalisation of the tractability conjecture for
finite-domain CSPs; see
e.g.~\cite{BartoPinskerDichotomy,wonderland,BKOPP,Bodirsky-Mottet}.
The motivation for the study of CSPs within the scope of the conjecture is that they form a large natural class of decision problems (capturing, e.g., problems in artificial intelligence and database theory), and that the complexity of these problems is described by algebraic objects, allowing for a systematic study of the class.
From the theoretical perspective, this means that there is a chance of being able to understand the differences between polynomial-time tractability and NP-completeness on a deeper level.}

\newblue{The main challenges that stem from the transition from finite- to infinite-domain CSPs essentially boil down to the following:
\begin{itemize}
	\item In the finite, the universal-algebraic approach works best when the algebras under consideration are \emph{idempotent}.
	This property can be assumed without loss of generality when the templates are finite, but not when they are infinite.
	\item The algebras that arise from $\omega$-categorical templates are naturally endowed with a topology,
	which essentially encodes the difference between local behaviours and global behaviours of functions.
	This topology can be ignored for finite algebras (as it is discrete),
	but in the infinite the exact role of this topology is an important topic, and we refer the reader to~\cite{TopoIrrelevant,TopologyIsRelevant,TopoRelevant-journal}  for more details.
\end{itemize}}

\newblue{Some tools have been developed to try to overcome these challenges, and in particular Ramsey theory has been applied in this setting via the notion of \emph{canonical functions}, which we heavily use in this article.}

Dalmau and Bodirsky~\cite{BodDalJournal} showed that every problem in MMSNP is a finite union of constraint satisfaction problems for $\omega$-categorical structures. 
These structures can be expanded to finitely bounded homogeneous structures so that they fall into the scope
of the mentioned infinite-domain tractability conjecture. 
It can be seen that in order to prove the MMSNP dichotomy,
it suffices to prove the dichotomy for those MMSNP problems that are CSPs (see Section~\ref{sect:connected}).  
This poses the question whether 
the complexity of MMSNP can be studied directly using the universal-algebraic approach, rather than the reduction of Kun which
involves a complicated construction of expander structures. In particular, even though 
we now have a complexity dichotomy for MMSNP, it was hitherto unknown 
whether the CSPs in MMSNP satisfy
the infinite-domain tractability conjecture. 
This status is particularly unpleasant 
for the infinite-domain tractability conjecture, for the following reason. 
One of the origins of the conjecture is 
the observation that most of the hardness results for 
CSPs of reducts of finitely bounded homogeneous structures can be shown (as in the finite-domain classification) via reductions 
using primitive positive interpretability and homomorphic equivalence (so called \emph{pp-constructions}~\cite{wonderland}). 
The conjecture says precisely that all CSPs in this class 
whose NP-hardness cannot be shown via pp-constructions are in P~\cite{BKOPP}. 
We are only aware of one exception to this observation: MMSNP. The hardness
proofs for MMSNP involving high-girth constructions cannot be formulated using pp-constructions. This makes MMSNP an attractive topic for the universal-algebraic approach for $\omega$-categorical structures.

\subsection*{Results}

We briefly discuss the results contained in this article.
The main result of this paper is the confirmation of the infinite-domain tractability conjecture for CSPs in MMSNP.

\begin{theorem}\label{thm:main-intro}
Let $\bB$ be an $\omega$-categorical structure such that $\Csp(\bB)$ is described by an MMSNP sentence.
Then exactly one of the following holds:
\begin{itemize}
	\item There is a uniformly continuous minor-preserving map from $\Pol(\bB)$ to the clone of projections on a 2-element set, and $\Csp(\bB)$ is NP-complete, or
	\item there is no such map and $\Csp(\bB)$ is in P.
\end{itemize}
\end{theorem}
As a by-product, we obtain a new proof of the complexity dichotomy for MMSNP that does not rely on the results of Kun.
We also use our algebraic approach to MMSNP to solve an open problem posed by Lutz and Wolter~\cite{LutzWolter}.
Consider colouring problems as in Figure~\ref{fig:ex-precol} where the input structure comes with a partial colouring and the question is whether this partial colouring can be extended to a full colouring (i.e., in the case of the problem in Figure~\ref{fig:ex-precol}, the input graphs comes with some vertices labelled red or blue).
We show that each such problem is polynomial-time equivalent to the original problem with no partial colouring.
\begin{theorem}\label{thm:precolouring-intro}
Let $\Phi$ be an MMSNP sentence in strong normal form\footnote{See Section~\ref{sect:nf} for a definition. The assumption of strong normal form can be thought of as the analog of being a core for a finite structure.}. The problem described by $\Phi$ is polynomial-time equivalent to the problem with partially coloured input structures.
\end{theorem}

We also consider some meta-problems concerning MMSNP. The containment problem is the problem of given two MMSNP sentences $\Phi$ and $\Psi$, decide whether $\Phi\Rightarrow\Psi$ holds.
The first-order rewritability problem is the problem of deciding whether a given MMSNP sentence is equivalent to a first-order sentence.
The tractability problem is the problem of deciding whether a given MMSNP sentence describes a problem in P.
\begin{theorem}\label{thm:containment-intro}
The containment problem, the tractability problem, and the first-order rewritability problem for MMSNP are decidable.
\end{theorem}
The proof of Theorem~\ref{thm:containment-intro} follows from our model-theoretic and algebraic analysis of MMSNP.
We show that the containment problem reduces to a question about the existence of a special homomorphism from $\bC_\Phi$ to $\bC_\Psi$,
where $\bC_\Phi$ and $\bC_\Psi$ are two countable structures and where such a special homomorphism has a finite description.
Similarly, the rewritability problem is shown to reduce to the existence of particular homomorphisms $(\bC_\Phi)^n\to\bC_\Phi$
that have a finite description.
We note that these problems were already known to be decidable (see~\cite{FederVardi,FeierKuusistoLutzRewritability}).
However, our proof is streamlined and shows that our universal-algebraic treatment of MMSNP has the potential to solve
some of the open problems that are related to MMSNP and its extensions.

\subsection*{Overview}
Sections~\ref{sect:csps} and~\ref{sect:mmsnp} 
introduces MMSNP, CSPs, and how they relate. 
The choice of a template for a CSP in MMSNP is not unique, and the right choice of the infinite structures to work with in our analysis is one of the central topics in this article. 
In fact, there are differences between the infinite structures we work with on three  levels:
\begin{enumerate}
\item In certain proofs it is necessary to work with an expansion of the structure having a larger relational signature. We might expand the structure with unary relations that correspond to the monadic predicates 
of the MMSNP sentence. But we also need larger (first-order) expansions that make the structure homogeneous (see Section~\ref{sect:cat}), or Ramsey (Section~\ref{sect:ramsey}; this expansion is by a linear order which is not first-order definable).
We finally also work with templates for MMSNP sentences where each monadic predicate extends a unary input predicate, called \emph{precoloured} (Section~\ref{sect:precoloured}), solving an open problem from~\cite{LutzWolter}. 
\item Even when we stick with the signature of our MMSNP sentence, the template is of course not unique.
There always exists the up to isomorphism unique model-complete core template,
but this is in many situations not the most appropriate template to work with; one of the reasons is, roughly speaking, that we sometimes need to work with injective polymorphisms with certain properties 
and that the model-complete core template might not have such polymorphisms.  
\item There is a third level of difficulty: not only do we care about the templates, but we also care about the \emph{description} of the template. 
Different MMSNP $\tau$-sentences might describe the same CSP. 
Which $\omega$-categorical template we construct for an MMSNP sentence $\Phi$ might not only depend on the CSP that is described by $\Phi$, but also on the sentence $\Phi$.
Here we solve a problem that the first two authors have been discussing since 2005: we show that if $\Phi$ is even in \emph{strong normal form}
(a concept from~\cite{MadelaineUniversal} that strengthens the MMSNP normal form introduced by Feder and Vardi~\cite{FederVardi,MadelaineStewartSicomp}),
then the $\omega$-categorical $\tau$-structure that we obtain for $\Phi$ is already the model-complete core template (Theorem~\ref{thm:mc-core}). 
\end{enumerate} 
One outcome of these investigations 
is the reduction of the classification 
to the precoloured situation,
where the template also contains the monadic predicates of the MMSNP sentence in the input signature. The real classification work is then done 
in Section~\ref{sect:alg}, 
and uses the following strategy: 
\begin{enumerate}
\item We show that a CSP in MMSNP is in P if the corresponding template
has a \emph{canonical} polymorphism that behaves on the orbits of the template as a Siggers operation.
\item In order to prove that this is the only way to obtain polynomial-time tractability, we want to show that the absence of such a canonical polymorphism
is equivalent to the existence of a uniformly continuous minor-preserving map to the clone of projections, which is known to entail NP-hardness~\cite{wonderland}.
We construct this map by first defining a minor-preserving map from the clone of canonical polymorphisms of the template to the clone of projections,
followed by extending this map to the whole polymorphism clone (similarly as in~\cite{Bodirsky-Mottet}). For this, two ingredients are necessary.
\item The first one is the fact that every polymorphism of the template locally interpolates a canonical operation.
This requires proving that the template under consideration has an $\omega$-categorical Ramsey expansion, which follows from recent results of Hubi\v{c}ka and 
Ne\v{s}et\v{r}il~\cite{Hubicka-Nesetril-All-Those}.
\item The second ingredient is the fact that every polymorphism of our template canonises in essentially one way. We obtain this through an analysis of the binary symmetric relations
that are preserved by the polymorphisms of the template.
\end{enumerate}
This presentation of the strategy oversimplifies certain aspects, and
we have to defer a more precise discussion to Section~\ref{sect:alg}.
As pointed out
  above, the article associates several infinite structures with an
  MMSNP sentence: the reader may wish to refer to
  Figure~\ref{fig:recap-structures} that recaps them.

\section{Constraint Satisfaction Problems}
\label{sect:csps}
Our main result is not only the new proof
of the dichotomy for MMSNP, but also the proof that the Bodirsky-Pinsker dichotomy conjecture holds for all CSPs in MMSNP;
the dichotomy for all of MMSNP follows from this result. So we have to introduce CSPs, too, which will be done in this section. %

\nomenclature[001]{$\bA,\bB,\bC$}{structures}
Let $\bA$ and $\bB$ be two structures
with the same relational signature $\tau$.
\nomenclature[000]{$\tau, \sigma, \rho$}{signatures}
A \emph{homomorphism} from $\bA$ to $\bB$ is a map from $A$ (the
domain of $\bA)$ to $B$ (the domain of $\bB$) that preserves all
relations; \red{if a homomorphism from $\bA$ to $\bB$ exists, we write $\bA\to\bB$}.
\nomenclature[002]{$\bA\to\bB$}{existence of a homomorphism from $\bA$ to $\bB$}
An \emph{embedding} is a homomorphism which is additionally
injective and also preserves the complements of all relations; \red{if an embedding from $\bA$ to $\bB$ exists, we also say that $\bA$ \emph{embeds} into $\bB$, and write $\bA \hookrightarrow \bB$.}
\nomenclature[003]{$\bA \hookrightarrow \bB$}{existence of an embedding from $\bA$ to $\bB$}
 For a relational $\tau$-structure $\bB$ we write
 \begin{itemize}
\item $\Age(\bB)$ for the class of finite 
 $\tau$-structures that embed into $\bB$;
 \nomenclature[004]{$\Age(\bB)$}{the age of a structure}
 \item $\Csp(\bB)$ for the class of finite $\tau$-structures that map
   homomorphically to $\bB$\nomenclature[005]{$\Csp(\bB)$}{the constraint satisfaction problem of $\bB$}.
 \end{itemize}
 For example, $\Csp(K_3)$ is the 3-colouring problem: the signature $\tau := \{E\}$ is the signature of graphs, and $K_3 := (\{0,1,2\};E)$ denotes the clique with three vertices, i.e., $E^{\bB} := \{0,1,2\}^3 \setminus \{(0,0),(1,1),(2,2)\}$.

 \nomenclature[006]{$\mathcal F$}{a class of finite (forbidden) structures}
Let $\mathcal F$ be a class of finite 
relational $\tau$-structures. We write 
\begin{itemize}
\item 
$\Forb^{\ind}({\mathcal F})$
for the class of all finite $\tau$-structures that do not embed any
structure from $\mathcal F$;
 \nomenclature[007]{$\Forb^{\ind}({\mathcal F})$}{all finite structures that
   do not embed any structure from $\mathcal F$}
\item $\Forb^{\hom}({\mathcal F})$ stands for
 the class of all finite $\tau$-structures
 $\bA$ such that no structure in $\mathcal F$ 
 \emph{homomorphically} maps to $\bA$.
  \nomenclature[008]{$\Forb^{\hom}({\mathcal F})$}{all finite structures that
   do not admit a homomorphism from any structure from $\mathcal F$}
 \end{itemize}
A relational structure $\bB$ is called 
\emph{finitely bounded} if it has a finite signature $\tau$ and there exists a finite
set ${\mathcal F}$ of finite $\tau$-structures 
 (the \emph{bounds}) such that
$\Age(\bB) = \Forb^{\ind}({\mathcal F})$.

\subsection{Logic perspective}
We present the classical terminology
to pass from structures to formulas and vice versa. 
Let $\bA$ be a $\tau$-structure.
Then the \emph{canonical query}
of $\bA$ is the formula whose variables
are the elements of $\bA$, 
and which is a conjunction that contains for every $R \in \tau$ a conjunct $R(a_1,\dots,a_n)$ if and only if $(a_1,\dots,a_n) \in R^{\bA}$. 

A \emph{primitive positive $\tau$-formula}
(also known as \emph{conjunctive query} in database theory)
is a formula that can be constructed from atomic formulas using conjunction $\wedge$ and existential quantification $\exists$; in other words, it is a first-order
formula without using disjunction $\vee$, negation $\neg$, or universal quantification $\forall$. 
By renaming the existentially quantified variables and
pulling out the existential quantifiers, it is straightforward
to rewrite primitive positive formulas 
into \emph{unnested formulas} of the form
$$ \exists x_1,\dots,x_n \, (\psi_1 \wedge \cdots  \wedge \psi_n)$$
where $\psi_1,\dots,\psi_n$ are atomic
$\tau$-formulas, i.e., they are of the form
$R(y_1,\dots,y_n)$ or of the form $y=y'$
where the variables 
might be from $\{x_1,\dots,x_n\}$;  otherwise they are called \emph{free}. 
We write $\phi(z_1,\dots,z_n)$ if the free variables of $\phi$ are contained in $\{z_1,\dots,z_n\}$. 
A formula without free variables is called a \emph{sentence}. 
\red{A formula is called \emph{existential positive} if it can be constructed from atomic formulas using conjunction, disjunction, and existential quantification; it is easy to see that every existential positive formula is equivalent to a disjunction of primitive positive formulas.} 

Let $\phi$ be a primitive positive $\tau$-formula without conjuncts of the form $y=y'$ and written in the unnested form presented above. Then the \emph{canonical database} of $\phi$ is the $\tau$-structure
$\bA$ whose elements are the variables of $\phi$, and such that for every $R \in \tau$ we have $(a_1,\dots,a_n) \in R^{\bA}$ if and only if $R(a_1,\dots,a_n)$ is a conjunct of $\phi$. 
We will apply the notion
of canonical database also to primitive positive formulas in general, by first rewriting them into unnested form and then applying the definition above. Since the rewriting might require 
that some of the existentially quantified variables are renamed, the resulting canonical database is not uniquely defined; but since we usually consider structures up to isomorphism, this should not cause  confusions. 
Also note that 
the information which variable is existentially quantified and which variable is free is lost in the passage from a primitive positive formula to the canonical database. 
The following is straightforward and well-known. 

\begin{proposition}[See, e.g.,~\cite{ChandraMerlin}]
Let $\bA$ and $\bB$ be two structures. 
The following are equivalent. 
\begin{itemize}
\item $\bA$ has a homomorphism to $\bB$. 
\item $\bB \models \exists \bar a. \phi$ where $\phi$ is the canonical query for $\bA$ and $\bar a$ lists the elements of $\bA$. 
\end{itemize}
\end{proposition}

\subsection{The finite-domain dichotomy theorem}
\label{sect:ua}
We will use an important result from universal-algebra, Theorem~\ref{thm:fin-ua} below; each of the equivalent items in this theorem will be used later in this article. 

\new{ A relation $R\subseteq B^m$ is \emph{preserved} by an operation $f\colon B^k\to B$ if for all tuples $\tuple a^1,\dots,\tuple a^k\in R$,
 the tuple $f(\tuple a^1,\dots,\tuple a^k)$ obtained by applying $f$ componentwise is also in $R$.
A \emph{polymorphism} of a structure $\bB$ is an operation that preserves all the relations of $\bB$;
alternatively, one can see a polymorphism of $\bB$ as a homomorphism from $\bB^k$ (a finite direct power of $\bB$) to $\bB$. }
For every $i,j \in {\mathbb N}$, $i \leq k$, the projection $\pi_i^k \colon B^k \to B$ given 
by $\pi^k_i(x_1,\dots,x_k) := x_i$ is a polymorphism.
\nomenclature[009]{$\pi_i^k$}{the $i$-th projection of arity $k$}
The set of all polymorphisms of $\bB$ is denoted by $\Pol(\bB)$; this
set forms a \emph{function clone},
 \nomenclature[010]{$\Pol(\bB)$}{the set of polymorphisms of $\bB$}
 i.e., it is a set of operations on the set $B$ that is closed under composition and contains the projections.
The clone on the set $\{0,1\}$ that only contains the projection
operations is denoted by $\Projs$.
 \nomenclature[013]{$\mathscr{B}, \mathscr{C}$}{function clones}
 \nomenclature[014]{$\Projs$}{the function clone of projections on a 2
   element set}
\red{A map $\xi \colon {\mathscr B} \to {\mathscr C}$ 
between two clones ${\mathscr B}$ and $\mathscr C$ that preserves the arities
is called 
\begin{itemize}
\item a \emph{clone homomorphism} if
$\xi(f(g_1,\dots,g_n)) 
= \xi(f)(\xi(g_1),\dots,\xi(g_n))$
for all $n$-ary operations $f \in \mathscr B$ 
and all 
$k$-ary operations $g_1,\dots, g_n \in {\mathscr B}$. 
\item a \emph{minor-preserving map}
if $\xi(f(\pi^k_{i_1},\dots, \pi^k_{i_n})) = \xi(f)(\pi^k_{i_1},\dots, \pi^k_{i_n})$ for all $n$-ary operations $f \in \mathscr B$, $k \in {\mathbb N}$, and  
$i_1,\dots,i_n \leq k$. 
\end{itemize}}

\begin{theorem}[\cite{Cyclic,JBK,wonderland}]\label{thm:fin-ua}
Let $\bB$ be a finite structure. 
Then the following are equivalent. 
\begin{enumerate}
\item $\bB$ has no polymorphism $s$ of arity 6 which is \emph{Siggers},
i.e., satisfies 
$$\forall x,y,z. \, s(x,y,x,z,y,z)) = s(y,x,z,x,z,y) \, .$$
\item $\bB$ has no polymorphism $f$ of arity $k \geq 2$ which is \emph{cyclic}, i.e., satisfies
$$\forall x_1,\dots,x_k. \, f(x_1,\dots,x_k) = f(x_2,\dots,x_k,x_1) \, .$$
\item There exists a minor-preserving map from $\Pol(\bB)$ to $\Pr$. 
\end{enumerate}
\end{theorem} 

It is known that if a finite structure $\bB$
satisfies the equivalent items from Theorem~\ref{thm:fin-ua},
then $\Csp(\bB)$ is NP-hard~\cite{JBK}.
Otherwise, we have the following recent result. 

\begin{theorem}[Finite-domain tractability theorem~\cite{BulatovFVConjecture,ZhukFVConjecture}]\label{thm:dichotomy}
Let $\bB$ be a finite structure with finite relational signature which does not satisfy the equivalent statements from Theorem~\ref{thm:fin-ua}. Then 
$\Csp(\bB)$ is in P. 
\end{theorem}

\subsection{Countable categoricity}
\label{sect:cat}
As we shall see below, MMSNP sentences describe problems that are finite unions of CSPs of countable structures.
Moreover these structures satisfy a strong property from model theory called \emph{$\omega$-categoricity}. 
A structure $\bB$ is  \emph{$\omega$-categorical} if all countable models of the first-order theory of $\bB$ are isomorphic. 

An \emph{endomorphism} of $\bB$ is a homomorphism from $\bB$ to $\bB$. 
The set of all endomorphisms of $\bB$, denoted by $\End(\bB)$, is a
transformation monoid with respect to composition $\circ$.
 \nomenclature[011]{$\End(\bB)$.}{the set of endomorphisms of $\bB$}
An \emph{automorphism} of $\bB$ is a bijective endomorphism $e$ of $\bB$ such that $e^{-1}$ is also an endomorphism of $\bB$. 
The set of all automorphisms of $\bB$, denoted by $\Aut(\bB)$, forms a
permutation group with respect to composition.
 \nomenclature[012]{$\Aut(\bB)$}{the set of automorphisms of $\bB$}
A structure $\bB$ is called \emph{homogeneous} if every isomorphism between finite substructures of $\bB$ can be extended to an automorphism of $\bB$. 
Homogeneous structures with finite relational signature are $\omega$-categorical; this is a straightforward
consequence of Theorem~\ref{thm:ryll} below. 
A permutation group $G$ on a countably infinite set $B$ is called \emph{oligomorphic} 
if for every $n \geq 1$ there are finitely many orbits of $n$-tuples on $B^n$ (with respect to the componentwise action of $G$ on $B^n$; this is often left implicit in the following).

\begin{theorem}\label{thm:ryll}
A countable structure $\bB$
is $\omega$-categorical if
and only if $G := \Aut(\bB)$ 
is oligomorphic. In an $\omega$-categorical structure, the orbits 
of the componentwise action of $G$
on $B^n$ 
are first-order definable in $\bB$. 
\end{theorem}

A finite or countably infinite 
$\omega$-categorical structure 
$\bB$ is called a \emph{core}
if all endomorphisms of $\bB$
are embeddings, and it is called
\emph{model-complete} 
if all embeddings of $\bB$ into $\bB$ preserve all first-order formulas. 
We say that two structures $\bB$ and $\bC$ are \emph{homomorphically equivalent} if there are homomorphisms $\bB\to\bC$ and $\bC\to\bB$.
\begin{theorem}[\cite{Cores-Journal}]
\label{thm:gen-mc-core}
Every $\omega$-categorical structure $\bB$ is homomorphically equivalent to a
model-complete core $\bC$, which 
is up to isomorphism unique, 
$\omega$-categorical, and embeds into $\bB$.
\end{theorem}

The set of all maps from $B \to B$  
carries a natural topology, the \emph{topology of pointwise convergence}, which is the product topology on $B^B$ where $B$ is taken to be discrete. We write $\overline{\mathscr S}$ for the closure of $\mathscr S$ with respect to this topology. 
\nomenclature[015]{$\overline{\mathscr S}$}{the closure of $\mathscr S$
  in the pointwise convergence topology}

\begin{proposition}[\cite{Cores-Journal}]
\label{prop:mc-core}
For a countable $\omega$-categorical structure $\bB$,
the following are equivalent.
\begin{itemize}
\item $\bB$ is a model-complete core;
\item the orbits of tuples 
of the componentwise action of $\Aut(\bB)$ are primitive positive definable in $\bB$;
\item $\End(\bB) = \overline{\Aut(\bB)}$. 
\end{itemize}
\end{proposition}

A map $\xi$ from a set of operations ${\mathscr B}$ on a set $B$ to a set of operations
${\mathscr C}$ on a set $C$ 
is \emph{uniformly continuous}\footnote{There is indeed a natural uniformity on the set of all operations on a set $B$ that 
induces the topology that we have introduced earlier; but we do not need this further and refer to~\cite{wonderland}.}
 if and only if for all $n \geq 1$ and all
finite $C' \subseteq C$ there exists a finite $B' \subseteq B$ such that whenever two $n$-ary functions $f, g \in \mathscr B$
agree on $B'$, then $\xi(f)$
 and $\xi(g)$ agree on $C'$.

\subsection{The infinite-domain dichotomy conjecture}

\new{It is known that for an arbitrary $\omega$-categorical structure $\bB$, the existence of a uniformly continuous minor-preserving map $\Pol(\bB)\to\Projs$
implies that $\Csp(\bB)$ is NP-hard (see~\cite{wonderland}, also stated in Theorem~\ref{thm:wonderland} below).
However this property cannot characterize NP-hardness for CSPs of $\omega$-categorical structures,
as there are $\omega$-categorical structures $\bB$ (even homogeneous digraphs) such that $\Pol(\bB)$ does not have uniformly continuous
minor-preserving map to $\Projs$  but such that $\Csp(\bB)$ is even undecidable~\cite{BodirskyNesetrilJLC}.} 
So to generalise the finite-domain tractability theorem we 
consider a subclass of the class of all
$\omega$-categorical structures, 
namely structures that are homogeneous and finitely bounded. 
More generally, we also consider \emph{first-order reducts} of such structures, i.e., structures $\bB$ 
with the same domain as a homogeneous finitely bounded structure $\bC$ such that all relations of $\bB$ are first-order definable over $\bC$. 
For such structures, Bodirsky and Pinsker~\cite{BPP-projective-homomorphisms} conjectured a complexity dichotomy
which can be rephrased as follows using the main result from~\cite{BKOPP}:

\begin{conjecture}[Infinite-domain tractability conjecture]\label{conj:inf-dichotomy}
Let $\bB$ be a first-order reduct of a finitely bounded homogeneous structure with finite relational signature. 
If there is no uniformly continuous minor-preserving map from $\Pol(\bB)$ to $\Projs$, then $\Csp(\bB)$ is in P.
\end{conjecture}

\section{MMSNP}
\label{sect:mmsnp}
In this section we introduce the logic MMSNP. 
Then we explain the 
 connection between MMSNP and infinite-domain CSPs:
we first syntactically characterise those
 MMSNP sentences that describe CSPs,
 by introducing the logic \emph{connected MMSNP},  
 and then we show that the dichotomy 
 for MMSNP and the dichotomy for connected MMSNP are equivalent (Section~\ref{sect:connected}). 

Let $\tau$ be a relational signature (we also refer to $\tau$ as the \emph{input signature}). 
\emph{SNP} is a syntactically restricted fragment of existential second order logic.
A sentence in SNP is of the form
$\exists P_1,\dots,P_n. \, \phi$ 
where $P_1,\dots,P_n$ are \emph{predicates} (i.e., relation symbols) and
$\phi$ is a \emph{universal} first-order-sentence
over the signature $\tau \cup \{P_1,\dots,P_n\}$; \red{we sometimes refer to $\phi$ as the \emph{first-order part of $\Phi$}}. 
\emph{Monotone Monadic SNP without inequality}, MMSNP, is the popular
restriction thereof which consists of sentences 
$\Phi$ of the form
$$\exists P_1,\dots,P_n \, \forall \bar x \bigwedge_i
\lnot\bigl(\alpha_i \land
\beta_i \bigr),$$
where $P_1,\dots,P_n$ are \emph{monadic} (i.e., unary) relation symbols not in $\tau$, where $\bar x$ is a tuple of first-order variables, and
for every negated conjunct:
\begin{itemize}
\item $\alpha_i$ consists of a conjunction of atomic formulas 
  involving relation symbols from $\tau$ and variables from
  $\bar x$; and 
\item $\beta_i$ consists of a conjunction 
of atomic formulas or negated
  atomic formulas involving relation symbols from $P_1,\dots,P_n$ and variables from $\bar x$.
\end{itemize}
Notice that the equality symbol is not allowed in MMSNP sentences.

Every MMSNP $\tau$-sentence
 describes a computational problem:
the input consists of a finite $\tau$-structure $\bA$, and the question is whether $\bA \models \Phi$, i.e., 
whether the sentence $\Phi$ is true in $\bA$. We sometimes identify MMSNP with the class of all computational problems described by MMSNP sentences. 
\nomenclature[0001]{$\tau$}{typically the input signature of an MMSNP sentence}

\subsection{Connected MMSNP}
\label{sect:connected}
\red{A conjunction $\phi$ of atomic formulas with at least one variable}
is called \emph{connected}
if the conjuncts of $\phi$ cannot be partitioned
into two non-empty sets of conjuncts with disjoint sets of variables, 
and \emph{disconnected} otherwise. 
Note that a primitive positive formula $\phi$ without equality conjuncts is connected if and only if the \emph{Gaifman graph}\footnote{The \emph{Gaifman graph} of a relational structure $\bA$ is the undirected graph with vertex set $A$ which contains an edge between $u,v \in A$ if and only if $u$ and $v$ both appear in a tuple contained in a relation of $\bA$.} of the canonical database of $\phi$ is connected in the graph theoretic sense. 
A connected conjunction $\phi$ of atomic formulas 
is called \emph{biconnected}
if the conjuncts of $\phi$ cannot be partitioned
into two non-empty sets of conjuncts
that only share one common variable. 
Note that formulas with only one variable might not be biconnected, e.g., the
formula $R_1(x) \wedge R_2(x)$ is not biconnected. 
An MMSNP $\tau$-sentence $\Phi$ is called \emph{connected} 
(or \emph{biconnected}) 
if for each conjunct $\neg (\alpha \wedge \beta)$ 
of $\Phi$ where $\alpha$ is a conjunction of 
$\tau$-formulas and $\beta$ is a conjunction of
unary formulas, the formula $\alpha$ is connected (or \emph{biconnected}, respectively). 

\begin{proposition}[Corollary 1.4.15 in~\cite{Bodirsky-HDR}]
\label{prop:csp-con}
An MMSNP sentence $\Phi$ describes a CSP if and only if $\Phi$ is logically equivalent to a connected MMSNP sentence. 
\end{proposition}

The following is implicitly contained in~\cite{FederVardi}; also see Section 6 of~\cite{MadelaineStewartSicomp}. 

\begin{proposition}
\label{prop:connected}
Let $\Phi$ be an MMSNP sentence. Then 
$\Phi$ is logically equivalent to 
a finite disjunction of
connected MMSNP sentences; these connected MMSNP sentences can be effectively computed 
from $\Phi$. 
\end{proposition}
\begin{proof}
Let $P_1,\dots,P_k$ be the existential monadic predicates in $\Phi$, and let $\tau$ be the input signature of $\Phi$. 
Suppose that $\Phi$ has a conjunct $\neg (\alpha \wedge \beta)$ where $\alpha$ is a disconnected 
conjunction of atomic $\tau$-formulas 
and $\beta$ contains unary predicates
only. Suppose that $\alpha$ 
is equivalent to $\alpha_1 \vee \alpha_2$ 
for non-empty formulas $\alpha_1$ and $\alpha_2$. 
Let $\Phi_1$ be the MMSNP sentence obtained from 
$\Phi$ by replacing $\alpha$ by $\alpha_1$,
and let $\Phi_2$ be the MMSNP 
sentence obtained from 
$\Phi$ by replacing $\alpha$ by $\alpha_2$.
It is then straightforward to check that
every finite $(\tau \cup \{P_1,\dots,P_k\})$-structure 
$\bA$ we have that $\bA$ satisfies the first-order part of
$\Phi$ if and only if $\bA$ satisfies the first-order part of $\Phi_1$ or
the first-order part of $\Phi_2$.
Iterating this process for each 
disconnected clause of $\phi$, 
we eventually arrive at a finite 
disjunction of connected
MMSNP sentences.
\end{proof}

It is well-known that the complexity classification for MMSNP can be reduced
to the complexity classification for \emph{connected} MMSNP; we add the simple proof for the convenience of the reader. 

\begin{proposition}\label{prop:connected-compl}
Let $\Phi$ be an MMSNP $\tau$-sentence which is logically equivalent to
$\Phi_1 \vee \cdots \vee \Phi_k$ for
 connected MMSNP $\tau$-sentences
 $\Phi_1,\dots,\Phi_k$ where $k$ is smallest possible. 
Then $\Phi$ is in P if each of $\Phi_1,\dots,\Phi_n$ is in P. If one of the $\Phi_i$ is NP-hard, then so is $\Phi$.
\end{proposition}
\begin{proof} 
If each $\Phi_i$ can be decided in polynomial time by an algorithm $A_i$, then it is clear
that $\Phi$ can be solved in polynomial time by running all of the algorithms $A_1,\dots,A_k$ on the input, and accepting if one of the algorithms accepts. 

Otherwise, if one of the $\Phi_i$ describes an NP-complete problem, then $\Phi_i$ can be reduced to $\Phi$ as follows. 
Since $k$ was chosen to be minimal, there exists a 
$\tau$-structure $\bB$ such that $\bB$ satisfies $\Phi_i$, 
but does not satisfy $\Phi_j$ for all $j \leq n$ that are
distinct from $i$, since otherwise we could have removed
$\Phi_i$ from the disjunction $\Phi_1 \vee \cdots \vee \Phi_k$ without affecting the equivalence of the disjunction to $\Phi$. 
We claim that  $\bA \uplus \bB$
satisfies $\Phi$ if and only if $\bA$ satisfies $\Phi_i$.
First suppose that $\bA$ satisfies $\Phi_i$. 
Since $\bB$ also satisfies $\Phi_i$ by choice of $\bB$,
and since $\Phi_i$ is closed under disjoint unions,
we have that $\bA \uplus \bB$ satisfies $\Phi_i$ as well.
The statement follows since $\Phi_i$ is a disjunct of $\Phi$.

For the opposite direction, 
suppose that $\bA \uplus \bB$ satisfies $\Phi$. 
Since $\bB$ does not satisfy $\Phi_j$
for all $j$ distinct from $i$, $\bA  \uplus \bB$ does not satisfy 
$\Phi_j$ as well, by monotonicity
of $\Phi_j$. Hence, $\bA \uplus \bB$ must satisfy $\Phi_i$. 
By the monotonicity of $\Phi_i$, it follows that $\bA$ satisfies $\Phi_i$. Since $\bA \uplus \bB$ is for fixed $\bB$ clearly 
computable from $\bA$ in linear time
this concludes our reduction from $\Phi_i$ to $\Phi$. 
\end{proof}

\section{Normal Forms}
\label{sect:nf}
We recall and adapt a normal form for MMSNP sentences that was initially proposed by Feder
and Vardi in~\cite{FederVardiSTOC,FederVardi} and later extended
in~\cite{MadelaineStewartSicomp}.
The normal form has been invented 
by Feder and Vardi to show that for every
connected MMSNP sentence $\Phi$ there is a polynomial-time equivalent 
finite-domain CSP. In their proof, the reduction from an MMSNP sentence to the corresponding finite-domain CSP is straightforward, but the reduction from the finite-domain CSP to $\Phi$ is tricky: it uses the fact that hard finite-domain CSPs are already hard when restricted to high-girth instances. The fact that
MMSNP sentences in normal form
are \emph{biconnected} is then the key to reduce
high-girth instances to the problem described by $\Phi$.

In our work, the purpose of the normal form is the reduction of the classification problem to MMSNP sentences that are \emph{precoloured} in a sense that will be made precise in Section~\ref{sect:precoloured}, which is later important to apply the universal-algebraic approach. 
Moreover, we describe a new \emph{strong normal form} 
that is based on recolourings introduced by Madelaine~\cite{Madelaine}. 
Recolourings have been applied by Madelaine to study the computational problem whether one MMSNP sentence implies another. In our context, the importance
of strong normal forms is that the templates that we construct
for MMSNP sentences in strong normal form, expanded with the inequality relation $\neq$, 
are model-complete cores (Theorem~\ref{thm:mc-core}). Let us mention that in order
to get this result, the biconnectivity of the MMSNP sentences in normal form is essential (e.g, the proof of Theorem~\ref{thm:mc-core} uses
Corollary~\ref{cor:1homogeneous}, which uses Lemma~\ref{lem:1homogeneous},
which uses Lemma~\ref{lem:nf}, which crucially uses biconnectivity of $\Phi$).

\subsection{The normal form for MMSNP}
\label{sect:normal-form}
Every connected MMSNP sentence can be 
rewritten to a connected MMSNP sentence of a
very particular shape, and this shape will be
crucial for the results that we prove in the
following sections. The following definition 
has its origins in~\cite{FederVardi}; also see~\cite{MadelaineStewartSicomp}. 

\begin{definition}
Let $\Phi$ be an MMSNP sentence 
where $M_1,\dots,M_n$, for $n \geq 1$, are the 
existentially quantified predicates (also called the \emph{colours} in the following). 
Then $\Phi$ is said to be in \emph{normal form} 
if it is connected and 
\begin{enumerate}
\item \label{item:prop-col-1} (Every vertex has a colour)
the first conjunct of $\Phi$ is $$\neg \big (\neg M_1(x) \wedge \cdots \wedge \neg M_n(x)\big )$$ 
for some variable $x$; 
\item \label{item:proper-col-2} (Every vertex has at most one colour) $\Phi$ contains the conjunct $$\neg \big (M_i(x) \wedge M_j(x) \big )$$ for all distinct $i,j \in \{1,\dots,n\}$;
\item \label{item:complete-col} (Clauses are fully coloured) \blue{for each conjunct $\neg \phi$ 
of $\Phi$ except the first, and for 
each 
variable $x$} that appears in $\phi$, 
there is an $i \leq n$ such that $\phi$ has a literal
of the form $M_i(x)$;
\item \label{item:biconn} (Clauses are biconnected) for all conjuncts $\neg \phi$ of $\Phi$
except for the ones from item 1 and 2, the formula $\phi$ is biconnected;
\item \label{item:hom} (Small clauses are explicit) any $(\tau \cup \{M_1,\dots,M_n\})$-structure $\bA$ with at most
$k$ elements satisfies the first-order part of $\Phi$ 
if $\bA$ satisfies all conjuncts of $\Phi$ 
with at most $k$ variables. 
\end{enumerate}
\end{definition}
\red{Note that if $\Phi$ is in normal form and $\neg \phi$ is a conjunct of $\Phi$ such that $\phi$ contains a negative literal $\neg M_i(x)$, for some variable $x$, then either we can remove $\neg M_i(x)$ from $\phi$ and obtain an equivalent formula, or we can remove $\neg \phi$ from $\Phi$ and obtain an equivalent formula. 
The reason is that $\phi$ must also contain a conjunct of the form $M_j(x)$ because of item~\ref{item:complete-col},
and if $i \neq j$ then we are in the first case, 
and if $i=j$ then we are in the second case. }
We illustrate item~\ref{item:biconn} and item~\ref{item:hom} in this definition with the following examples. 

\begin{example}\label{ex:c5}
\red{We present an example of a connected MMSNP sentence $\Phi$ which does not satisfy item~\ref{item:hom} from the definition of normal forms.} The formula $\Phi$ given by 
$$ \forall a,b,c,d,e. \, \neg \big (E(a,b) \wedge E(b,c) \wedge E(c,d) \wedge E(d,e) \wedge E(e,a) \big)$$
which is in fact a first-order formula. 
The canonical database of $$E(x_1,x_2)
\wedge E(x_2,x_3) \wedge E(x_3,x_4) 
\wedge E(x_4,x_3) \wedge E(x_3,x_1)$$
has only four elements, does not 
satisfy $\Phi$, but \red{satisfies all conjuncts of $\Phi$ with at most four variables, because the only conjunct of $\Phi$ has five variables}. 
However, $\Phi$ is logically equivalent to the following
MMSNP formula, and it can be checked that this
formula is in normal form. 
\begin{align*}
\exists M_1 \forall x_0,\dots,x_4 \big (\neg (\neg M_1(x_0)) \wedge & \neg (\bigwedge_{0 \leq i \leq 4} M_1(x_i) \wedge E(x_i,x_{i+1 \mmod 5})) \\
\wedge & \neg (\bigwedge_{0 \leq i \leq 2} M_1(x_i) \wedge E(x_i,x_{i+1 \mmod 3})) \\
\wedge & \neg (M_1(x_0) \wedge E(x_0,x_0))
\big).  
\end{align*}
\end{example}

Adding clauses to an MMSNP sentence
to obtain an equivalent sentence that satisfies
item 5 can make a biconnected sentence not biconnected, as we see in the following
example. 

\begin{example}\label{ex:p3}
Let $\Phi$ be the following biconnected MMSNP sentence.
$$  \forall a,b,c,d. \, \neg \big (E(a,b) \wedge E(b,d) \wedge E(a,c) \wedge E(c,d)\big) $$
\red{It can be checked that $\Phi$ does not satisfy
item 5} and in fact is equivalent to 
$$\forall a,b,d. \, \neg \big (E(a,b) \wedge E(b,d)\big)$$
which is not biconnected.
\end{example}

\begin{lemma}\label{lem:nf-existence}
Every connected MMSNP sentence $\Phi$ 
is equivalent to an MMSNP sentence $\Psi$
in normal form, and $\Psi$ can be computed 
from $\Phi$. 
\end{lemma}
\begin{proof}
We transform $\Phi$ in several steps
(their order is important).

\paragraph*{1: Biconnected clauses}
Suppose that $\Phi$ contains a conjunct $\neg \phi$
such that $\phi$ is not biconnected, i.e.,
$\phi$ can be written as $\phi_1(x,\bar y) \wedge \phi_2(x,\bar z)$ for tuples of variables
$\bar y$ and $\bar z$ with disjoint sets of variables, and where $\phi_1$ and $\phi_2$ are conjunctions of atomic formulas. 
Then we introduce a new existentially quantified predicate
$P$, 
and replace $\neg \phi$ by $\neg (\phi_1(x,\bar y) \wedge P(x)) \wedge \neg (\phi_2(x,\bar z) \wedge \neg P(x))$.
Repeating this step, 
we can establish item~\ref{item:biconn} in the definition of normal forms. 

\paragraph*{2: Making implicit small clauses explicit}
Let 
$\neg \phi(x_1,\dots,x_n)$ be a conjunct of $\Phi$ that is not the first conjunct. 
Let $x$ be a variable that
does not appear among $x_1,\dots,x_n$, 
and consider the formula $\phi(y_1,\dots,y_n)$ where
$y_i$ is either $x_i$ or $x$, and suppose that
$y_i = y_j = x$ for at least two different $i,j \leq n$. 
If $\phi(y_1,\dots,y_n)$ is biconnected,
then add $\neg \phi(y_1,\dots,y_n)$ to $\Phi$. 
Otherwise, $\phi(y_1,\dots,y_n)$ can be written
as $\phi_1(x,\bar z_1) \wedge \phi_2(x,\bar z_2)$. 
We then apply the procedure from 
step 1 with the formula $\neg \phi(y_1,\dots,y_n)$. 
In this way we can produce an equivalent MMSNP 
sentence that still satisfies
item~\ref{item:biconn} (biconnected clauses). 
When we repeat this in all possible ways
the procedure eventually terminates, and we 
claim
that the resulting sentence $\Psi$ satisfies
additionally item~\ref{item:hom}. 
To see this, let $\bA$ be a $(\tau \cup \{M_1,\dots,M_n\})$-structure with at most
$k$ elements which does not satisfy some conjunct
$\neg \phi$ of $\Phi$. Pick the conjunct
$\neg \phi$ from $\Phi$ with the least number of variables and this property.
\blue{Then there are 
$a_1,\dots,a_l \in A$ such that $\bA$ 
satisfies $\phi(a_1,\dots,a_l)$. If $l \leq k$,
we are done. Otherwise, 
there must be $i,j \leq l$ such that $a_i = a_j$. 
If the conjunct 
$\neg \phi(y_1,\dots,x_{i-1},x,x_{i+1},\dots,x_{j-1},x,x_{j+1},\dots,y_l)$ is biconnected, it has been added to $\Phi$, and it has less variables than $\phi$,
a contradiction.} Otherwise, our procedure did split the conjunct, and inductively we see that a clause that it not satisfied by $\bA$ and has less variables than $\phi$ has been added to $\Phi$.

\paragraph*{3: Predicates as colours}
Next, we want to ensure the 
 property that $\Phi$
contains for each pair of distinct existentially quantified monadic predicates $M_i,M_j$ the negated
conjunct $$\lnot \big(M_i(x) \land M_j(x)\big), $$ 
and if $M_1,\dots,M_c$ are all the existentially quantified predicates, then $\Phi$ contains the negated conjunct 
$$\lnot \big (\lnot M_1(x)\land
\cdots \wedge  \lnot M_c(x) \big ) .$$
We may transform every MMNSP sentence into an
equivalent MMSNP sentence of this form, 
via the addition of further monadic
predicates ($2^n$ predicates starting from $n$ monadic
predicates); \red{this step is standard and we refer to~\cite{FederVardi,MadelaineStewartSicomp} instead}. If $n=0$ then $\Phi$ was a first-order
formula; in this case, to have a unified treatment of all cases, we introduce a single existentially quantified predicate $M_1$, too.

\paragraph*{4: Fully coloured clauses}
Finally, if $\neg \phi$ is a conjunct of $\Phi$
and $x$ a variable from $\phi$ such that 
$x$ does not appear in any literal of the form $M_i(x)$
in $\phi$, then we replace $\neg \phi$ by
the conjuncts $$\neg (\phi \wedge M_1(x)) \wedge \cdots \wedge \neg (\phi \wedge M_n(x)) .$$
We do this for all conjuncts of $\Phi$ and all such variables, and obtain an MMSNP sentence that
finally satisfies all the items from the definition
of normal forms. 
\end{proof}

\begin{example}\label{ex:p3-nf}
We revisit an MMSNP sentence from Example~\ref{ex:p3},  
$$ \forall a,b,c. \, \neg \big(E(a,b) \wedge E(b,c) \big) \, .$$
An equivalent MMSNP sentence $\Psi$ in normal form is 
\begin{align*}
 \exists M_1,M_2 \, \forall x,y \, \big ( & \neg(\neg M_1(x) \wedge \neg M_2(x)) \\
\wedge \, &  \neg (M_1(x) \wedge M_2(x)) \\
\wedge \,& \neg (M_1(x) 
\wedge E(x,x)) \\
\wedge \,& \neg (M_2(x) \wedge E(x,x)) \\
\wedge \,& \neg (M_1(x) \wedge M_1(y)
\wedge E(x,y)) \\
\wedge \,& \neg (M_2(x) \wedge M_2(y)
\wedge E(x,y)) \\
\wedge \,& \neg (M_2(x) \wedge M_1(y) \wedge E(x,y) \big) .
\end{align*}
\red{This formula has been obtained by first applying step 1 once. In this example, there are two new clauses from step 2, namely $P(x) \wedge E(x,x)$
and $\neg P(x) \wedge E(x,x)$. In step 3, two monadic predicates $M_1$ and $M_2$ are introduced, because only once predicate was introduced in step 1. The first two clauses of $\Psi$ come from step 3 as well, and the other clauses are modified to get rid of the predicate $P$. 
Finally, step 4 produces the final three conjuncts of $\Psi$.}
\end{example}

The following lemma states a key property
that we have achieved with our normal form (in particular, we use the biconnectivity assumption). 

\begin{lemma}\label{lem:nf}
Let \red{$\theta$} be the first-order part of an MMSNP $\tau$-sentence in normal form with colour set $\sigma$ 
and let $\psi_1(x,\bar y)$ and $\psi_2(x, \bar z)$ be
two conjunctions of atomic $(\tau \cup \sigma)$-formulas
such that
\begin{itemize}
\item $\bar y$ and $\bar z$
are vectors of disjoint sets of variables;
\item 
the canonical databases of $\psi_1$
and of $\psi_2$ satisfy \red{$\theta$};
\item the canonical database $\bA$ of $\psi_1(x,\bar y) \wedge \psi_2(x,\bar z)$ does not
satisfy \red{$\theta$}.
\end{itemize}
Then $\psi_1$ must contain a literal $M_i(x)$ and
$\psi_2$ must contain a literal $M_j(x)$ for distinct colours
$M_i$ and $M_j$ of \red{$\theta$}. 
\end{lemma}
\begin{proof}
First observe that all vertices of $\bA$ must be coloured
since all vertices of the canonical databases of $\psi_1$
and of $\psi_2$ are coloured (because they satisfy \red{$\theta$}). Therefore, since $\bA$ does not satisfy \red{$\theta$}, there is a conjunct \red{$\neg \phi$ of $\theta$ 
and $a_1,\dots,a_l \in A$
such that $\bA \models \phi(a_1,\dots,a_l)$.
Pick the conjunct such that $l$ is minimal. 
Since both the canonical database
of $\psi_1$ and of $\psi_2$ satisfy $\theta$, not all of $a_1,\dots,a_l$ can lie in the canonical database
of $\psi_1$, or in the canonical database of $\psi_2$. 
If $\phi$ is of the form $M_i(x) \wedge M_j(x)$ for $i \neq j$ then we are done. 
Otherwise, since $\phi$ is biconnected, there are 
$i,j \leq n$ such that
$a_i = a_j = x$. In this case, the structure $\bA'$ induced 
by $a_1,\dots,a_l$ in $\bA$ has strictly
less than $l$ elements. 
Since $\Phi$ is in normal form, and since
$\bA'$ does not satisfy $\theta$, by item~\ref{item:hom}
in the definition of normal forms 
there must be a conjunct $\neg \phi'$ of $\theta$ with
at most $|A'|$ variables such
that $\phi'$ holds in $\bA'$. This contradicts the choice
of $\phi$.}
\end{proof}

{A way to rephrase Lemma~\ref{lem:nf} in terms of structures is as follows:}
\begin{corollary}\label{cor:nf}
	Let $\phi$ be the first-order part of an MMSNP sentence in normal form with colour set $\sigma$.
	Let $\bA,\bA'\models\phi$ and let $a\in \bA$ and $b\in \bA'$ be vertices of the same colour.
	Then the structure obtained by taking the disjoint union of $\bA$ and $\bA'$ and gluing them on $a$ and $b$ satisfies $\phi$.
\end{corollary}

\subsection{Templates for sentences in normal form}
\label{sect:templates}
\nomenclature[0002]{$\sigma$}{typically the colours
  (existentially quantified monadic predicates) of an
  MMSNP sentence in normal form}
Let $\Phi$ be an MMSNP $\tau$-sentence in normal form. Let $\sigma$
be the set of colours of $\Phi$. 
We will now construct an $\omega$-categorical $(\tau \cup
\sigma)$-structure $\bC_\Phi$ for an MMSNP sentence $\Phi$ in normal
form;
this structure will have several important 
properties:
\begin{enumerate}
\item a structure $\bA$ satisfies $\Phi$ 
if and only if $\bA$ maps homomorphically
to $\bC_\Phi^\tau$, i.e., the CSP of $\bC_\Phi^\tau$ is the problem described by $\Phi$; 
\item $\bC_\Phi$ has no algebraicity (see below for a definition);
\item the colours of $\Phi$ are in bijective correspondence to the orbits of $\bC_\Phi$;
\item $(\bC_\Phi,\neq)$ is a model-complete core; 
\item if $\Phi$ is furthermore in \emph{strong normal form} (to be introduced in Section~\ref{sect:snf})
then even $(\bC^\tau_\Phi,\neq)$ is a model-complete core.
\end{enumerate} 

If $\Phi$ is an MMSNP sentence in normal form with input signature $\tau$ and second-order predicates $\sigma$, one can view each negated conjunct (save the first one)
as a $(\tau\cup\sigma)$-structure that represents an obstruction to the satisfiability of $\Phi$.

\begin{definition}
Let $\Phi$ be an 
MMSNP $\tau$-sentence in normal form.
The \emph{coloured obstruction set}
 for $\Phi$ is the set $\mathcal F$ of all 
canonical databases for formulas $\phi$ such that
$\neg \phi$ is a conjunct of $\Phi$, except for the first conjunct.
\end{definition}

A structure $\bB$ \emph{does not have algebraicity} if for all first-order formulas $\phi$ with free variables 
$x_0,x_1,\dots,x_n$, 
and all elements $a_1,\dots,a_n$
of $\bB$ the set $$\{ x \mid \bB \models \phi(x,a_1,\dots,a_n)\}$$
is either infinite or contained in $\{a_1,\dots,a_n\}$; otherwise, we say that
the structure \emph{has algebraicity}. 
It is well-known that a homogeneous structure $\bA$ 
has no algebraicity if and only if its age has 
\emph{strong amalgamation} \red{(see, e.g.,~\cite{Oligo})}, 
i.e., if for any two finite substructures 
$\bB_1$ and $\bB_2$ of $\bA$ there exists a substructure $\bC$ of $\bA$ and embeddings
$e_1 \colon \bB_1 \hookrightarrow \bC$
and 
$e_2 \colon \bB_2 \hookrightarrow \bC$
such that $|B_1 \cap B_2| = |e_1(B_1) \cap e_2(B_2)|$.

\begin{theorem}[Theorem 4 in~\cite{CherlinShelahShi}]\label{thm:css}
Let $\mathcal F$ be a finite set of finite connected $\tau$-structures. 
Then there exists a countable model-complete $\tau$-structure $\bB^{\ind}_{\mathcal F}$ 
such that $\Age(\bB^{\ind}_{\mathcal F}) = \Forb^{\hom}({\mathcal F})$. 
The structure $\bB^{\ind}_{\mathcal F}$ is $\omega$-categorical, has no algebraicity, and is unique up to isomorphism.
\end{theorem}
\nomenclature[016]{$\bB^{\ind}_{\mathcal F}$}{a structure of age $\Forb^{\hom}({\mathcal F})$}

Theorem~\ref{thm:css} has the following variant in the category
of injective homomorphisms. 

\begin{theorem}\label{thm:css-new}
Let $\mathcal F$ be a finite set of finite connected $\tau$-structures. 
Then there exists a $\tau$-structure $\bB^{\hom}_{\mathcal F}$ such that 
\begin{itemize}
\item 
a finite $\tau$-structure $\bA$ homomorphically and injectively maps to $\bB^{\hom}_{\mathcal F}$
if and only if $\bA \in \Forb^{\hom}({\mathcal F})$;
\item $(\bB^{\hom}_{\mathcal F};\neq)$ is a model-complete core.  
\end{itemize}
The structure $\bB^{\hom}_{\mathcal F}$ is $\omega$-categorical, has no algebraicity, and is unique up to isomorphism.
\end{theorem}
\nomenclature[017]{$\bB^{\hom}_{\mathcal F}$}{a substructure of $\bB^{\ind}_{\mathcal F}$
}
\begin{proof}
Let $(\bB^{\hom}_{\mathcal F},\neq)$ be the model-complete core of $(\bB^{\ind}_{\mathcal F},\neq)$; by
Theorem~\ref{thm:gen-mc-core}
the structure 
$(\bB^{\hom}_{\mathcal F};\neq)$
is unique up to isomorphism,
and $\omega$-categorical.
Let $\bA$ be a finite $\tau$-structure. 
If $\bA \in \Forb^{\hom}({\mathcal F})$,
then $\bA$ embeds 
into $\bB^{\ind}_{\mathcal F}$
by Theorem~\ref{thm:css},
and since $(\bB^{\ind}_{\mathcal F},\neq)$ is homomorphically equivalent to
$(\bB^{\hom}_{\mathcal F},\neq)$, there is an injective homomorphism from $\bA$ to $\bB^{\hom}_{\mathcal F}$. 
These reverse implication can be shown similarly, 
and this shows the first item. 

For proving that $\bB^{\hom}_{\mathcal F}$ has no algebraicity, let
$\phi(x_0,x_1,\dots,x_n)$ be a first-order $\tau$-formula 
and $b_1,\dots,b_n$ 
be elements of $\bB_{\mathcal F}^{\hom}$. 
By Theorem~\ref{thm:gen-mc-core}
we can assume that 
$\bB_{\mathcal F}^{\hom}$
is a substructure of $\bB_{\mathcal F}^{\ind}$. 
Since $(\bB_{\mathcal F}^{\hom},\neq)$ is a model-complete core, the formula $\phi$ is equivalent to an
existential positive $(\tau \cup \{\neq\})$-formula $\psi$ over 
$(\bB_{\mathcal F}^{\hom},\neq)$. Suppose that the set 
$S := \{ x \mid \bB_{\mathcal F}^{\hom} \models \psi(x,b_1,\dots,b_n)\}$ contains an element $b_0 \notin \{b_1,\dots,b_n\}$. Then
$$b_0 \in T := \{x \mid \bB^{\ind}_{\mathcal F} \models \psi(x,b_1,\dots,b_n)\} \setminus \{b_1,\dots,b_n\}$$ 
and since $\bB^{\ind}_{\mathcal F}$ does not have algebraicity, the set $T$
must be infinite. Let $h$ be a homomorphism from 
$(\bB^{\ind}_{\mathcal F},\neq)$ to $(\bB^{\hom}_{\mathcal F},\neq)$. Since $h$ preserves $\neq$ we have that $h(T)$ is infinite, and since $h$ 
preserves the existential positive formula $\psi$ we have $h(T) \subseteq S$, which proves that
$S$ is infinite.  
\end{proof}

\red{The proof of Theorem~\ref{thm:css-new} shows that there is an injective homomorphism from $\bB^{\ind}_{\mathcal F}$ to $\bB^{\hom}_{\mathcal F}$ and vice versa; 
however, the two structures might not be isomorphic, as we see in the following example.}

\begin{example}\label{ex:difference-ind-hom}
The structure $\bB^{\hom}_{\mathcal F}$ might be
isomorphic to the structure $\bB^{\ind}_{\mathcal F}$:
it is for example easy to verify that for ${\mathcal F} := \{K_3\}$ the structure $\bB^{\ind}_{\mathcal F}$ 
is a model-complete core \red{(model-completeness follows from homogeneity, and for verifying that $\bB^{\ind}_{\mathcal F}$ is a core it suffices to find primitive positive definitions of the relations defined by $x \neq y$ and by $\neg E(x,y)$)}, and therefore
isomorphic to $\bB^{\hom}_{\mathcal F}$. 

In general, however, the two structures are not isomorphic. 
Consider for example the signature
$\tau = \{E\}$ for $E$ binary and 
$\mathcal F := \{L\}$ where $L := (\{x\};\{(x,x)\})$, i.e., $L$ is the canonical database of $E(x,x)$.
Then $\bB^{\ind}_{\mathcal F}$ is the \emph{random graph} (i.e., every finite simple graph embeds into $\bB^{\ind}_{\mathcal F}$),
but $\bB^{\hom}_{\mathcal F}$ is the countably infinite clique.
\end{example}

\begin{definition}\label{def:C-Phi}
Let $\Phi$ be an MMSNP $\tau$-sentence in normal form and let $\mathcal F$ be the coloured obstruction set of $\Phi$. Then 
$\bC_\Phi$ denotes the substructure of $\bB^{\hom}_{\mathcal F}$ 
induced by the coloured elements of 
$\bB^{\hom}_{\mathcal F}$.
\end{definition}
\nomenclature[018]{$\bC_\Phi$}{a substructure of $\bB^{\hom}_{\mathcal F}$}
The $\tau$-reduct $\bC^\tau_\Phi$ of the structure $\bC_\Phi$ that we constructed
for an MMSNP sentence $\Phi$ in normal form
is indeed a template for the CSP described by $\Phi$.
\nomenclature[019]{$\bC^\tau_\Phi$}{the $\tau$-reduct of the structure
  $\bC_\Phi$, a template for $\Phi$ viewed as a CSP}

\begin{lemma}\label{lem:c-basic}
Let $\Phi$ be an MMSNP $\tau$-sentence in normal form and let $\bA$ be a $\tau$-structure. 
Then the following are equivalent.
\begin{enumerate}
\item[(1)] $\bA \models \Phi$;
\item[(2)] $\bA$ homomorphically and injectively maps to $\bC^\tau_\Phi$;
\item[(3)] $\bA$ maps homomorphically to 
$\bC^\tau_\Phi$. 
\end{enumerate}
\end{lemma}
\begin{proof}
Let \red{$\sigma$} be the colour set and let $\mathcal F$
be the coloured obstruction set  of $\Phi$. 
$(1) \Rightarrow (2)$. If $\bA$ satisfies $\Phi$ it has
a $(\tau \cup \sigma)$-expansion $\bA'$ such
that no structure in ${\mathcal F}$ maps homomorphically to $\bA'$. 
\blue{So $\bA'$ homomorphically and injectively maps to $\bB^{\hom}_{\mathcal F}$
by Theorem~\ref{thm:css-new}.} 
Moreover,  
every element of $\bA'$ is contained in one
predicate from $\sigma$ (because of the first conjunct of $\Phi$) and hence the image
of the embedding must lie in $\bC_\Phi$. 

$(2) \Rightarrow (3)$ is trivial. For $(3) \Rightarrow (1)$, let $h$ be the homomorphism from $\bA$ to $\bC^\tau_\Phi$. Expand $\bA$ to a $(\tau \cup \sigma)$-structure $\bA'$ by colouring each element $a \in A$ by the colour of $h(a)$ in $\bC_\Phi$;
then there is no homomorphism from a structure
$\bF \in \mathcal F$ to $\bA'$, since the composition
of such a homomorphism with $h$ would give
a homomorphism from $\bF$ to $\bB^{\ind}_{\mathcal F}$, a contradiction. The expansion
$\bA'$ also satisfies the first conjunct of $\Phi$,
and hence $\bA \models \Phi'$. 
\end{proof}

In the following we prove that $\bC_\Phi$ indeed has
the properties that we announced at the beginning
of this section. 
We start with some remarkable properties of the structure $\bB^{\ind}_{\mathcal F}$ (Section~\ref{sect:css})
and continue with properties of $\bC_{\Phi}$ (Section~\ref{sect:our-templates}).

\subsubsection{Properties of Cherlin-Shelah-Shi structures}\label{sect:css}

An existential formula is called \emph{primitive}
if it does not contain disjunctions. 

\begin{lemma}\label{lem:define-orbits}
For every $k \in {\mathbb N}$,
the orbits of $k$-tuples in $\bB^{\ind}_{\mathcal F}$ 
can be defined by $\phi_1 \wedge \phi_2$ 
where $\phi_1$ is a primitive positive formula and $\phi_2$ is a conjunction of negated atomic formulas.
\end{lemma}
\begin{proof}
It suffices to prove the statement for $k$-tuples
$\bar a$ with pairwise distinct entries. 
       Since $\bB^{\ind}_{\mathcal F}$ is $\omega$-categorical and
model-complete, there is an existential definition $\phi(\tuple x)$ of the orbit of $\tuple a$ in
$\bB^{\ind}_{\mathcal F}$ \red{(apply Proposition~\ref{prop:mc-core} to the expansion of $\bB^{\ind}_{\mathcal F}$ by all relations defined by negations of atomic formulas)}.
       Since $\phi$ defines an orbit of $k$-tuples
       it can be chosen
to be primitive. Moreover, since $\bar a$ is 
        a tuple with pairwise distinct entries, $\phi$  can be chosen to be without conjuncts of the form $x=y$ (it is impossible that both $x$ and $y$ are among the free variables $x_1,\dots,x_n$; if one of the variables is existentially quantified, we can replace all occurrences of it by the other variable and obtain an equivalent formula).
       Let $\phi_1$ be the primitive positive formula 
obtained from $\phi$ by deleting all the negated conjuncts. Let $\phi_2$ be conjunction of all
negated atomic formulas that hold on $\bar a$. 
Clearly, $\phi$ implies $\phi_1 \wedge \phi_2$. 

Let $\bar b$ be a tuple that satisfies
$\phi_1 \wedge \phi_2$; we have to show that $\bar b$ satisfies $\phi$. 
Let $\psi(x_1,\dots,x_n)$ be the existential definition of the orbit of $\bar b$. Again, we may assume that $\psi$ is disjunction-free
and free of literals of the form $x=y$.  
Let $\psi_1$ 
be the formula obtained from $\psi$ by dropping negated conjuncts. 
Let $\bA$ be 
the canonical database of $\phi_1 \wedge \psi_1$ \blue{(which is well-defined since both $\phi_1$ and $\psi_1$ are primitive positive and do not involve literals of the form $x=y$)}. 
We have $\bB^{\ind}_{\mathcal F} \models
\phi_1(\bar b) \wedge \psi_1(\bar b)$, so $\bA$ does not homomorphically embed any structure from ${\mathcal F}$. By definition of 
$\bB^{\ind}_{\mathcal F}$ (Theorem~\ref{thm:css}), there exists an embedding $e$ 
of $\bA$ into $\bB^{\ind}_{\mathcal F}$. 
Then $e$ provides witnesses for
the existentially quantified variables in $\phi \wedge \psi$ showing that $\bB^{\ind}_{\mathcal F} \models (\phi \wedge \psi)(e(x_1),\dots,e(x_n))$ because for those witnesses the negated conjuncts will also be satisfied. Hence, 
$\phi$ and $\psi$ define the same orbit
of $n$-tuples. In particular, $t$ satisfies $\phi$ which is what we wanted to show.
\end{proof}

When $\bB$ is a structure, we write 
$\bB^*$ for the expansion of $\bB$ by
all primitive positive formulas. 
\nomenclature[020]{$\bB^*$}{expansion of $\bB$ by all primitive positive formulas}

\begin{corollary}\label{cor:homogeneous-exp}
The structure $(\bB_{\mathcal F}^{\ind})^*$ is homogeneous. 
\end{corollary}
\begin{proof}
Let $\bar a$, $\bar b$ be two $k$-tuples 
of elements of $(\bB_{\mathcal F}^{\ind})^*$
such that the map that sends $a_i$ to $b_i$, for $i \in \{1,\dots,k\}$, is an isomorphism between
the substructures induced by $\{a_1,\dots,a_n\}$
and by $\{b_1,\dots,b_n\}$ in $(\bB_{\mathcal F}^{\ind})^*$. Then $\bar a$ and $\bar b$ satisfy 
in particular the same negated atomic formulas,
and they also satisfy the same primitive positive
formulas in $\bB_{\mathcal F}^{\ind}$ since
$\alpha$ must preserve the relations that we have introduced for these formulas in 
$(\bB_{\mathcal F}^{\ind})^*$. The statement now follows from Lemma~\ref{lem:define-orbits}. 
\end{proof}

\begin{definition}
A relational structure $\bB$ is said to
be \emph{1-homogeneous} 
if it has the property that when $a,b \in B$ satisfy the same unary relations in $\bB$, then there exists an
automorphism of $\bB$ that maps $a$ to $b$. 
\end{definition}

\begin{lemma}\label{lem:1homogeneous}
Let $\Phi$ be an MMSNP sentence in normal form
with coloured obstruction set ${\mathcal F}$. 
Then $\bB_{\mathcal F}^{\ind}$
is 1-homogeneous. 
\end{lemma}

\begin{proof} 
Let $x_1$ and $x_2$ be two elements that
induce isomorphic 1-element substructures of  
$\bB^{\ind}_{\Phi}$.
Since $\bB^{\ind}_{\mathcal F}$ is
model-complete, the orbit of $x_i$, for $i = 1$ and $i=2$, 
has a primitive definition $\psi_i$ in
$\bB^{\ind}_{\mathcal F}$. 
Pick elements for the existentially quantified
variables in $\psi_i$ that witness the truth of
$\psi_i(x_i)$, 
and let $\psi_i'$ be the canonical
query of the structure induced by $x_i$ and those elements in $\bB^{\ind}_{\Phi}$.

Suppose for contradiction that 
$x_1$ and $x_2$ are in different orbits
of $\bB^{\ind}_{\mathcal F}$. 
This means that $\psi_1(x) \wedge \psi_2(x)$,
and therefore also $\psi_1'(x) \wedge \psi_2'(x)$, 
 is unsatisfiable in 
the structure $\bB^{\ind}_{\mathcal F}$. 
Since $x_1$ and $x_2$ induce isomorphic 1-element substructures, the contrapositive of Lemma~\ref{lem:nf} shows that 
already the canonical database of $\psi_1'$ or of $\psi_2'$ does not satisfy the first-order part of
$\Phi$, a contradiction. 
\end{proof}

\subsubsection{Properties of our templates for MMSNP}\label{sect:our-templates}
Some properties that we have derived for
$\bB_{\mathcal F}^{\ind}$ transfer via $\bB_{\mathcal F}^{\hom}$ to $\bC_\Phi$.

\begin{lemma}\label{lem:bhom-1homogeneous}
Let $\Phi$ be an MMSNP sentence in normal form
with coloured obstruction set ${\mathcal F}$. 
Then $\bB_{\mathcal F}^{\hom}$ 
is 1-homogeneous. 
\end{lemma}
\begin{proof} 
We already know that $\bB_{\mathcal F}^{\ind}$
is 1-homogeneous. Let \blue{$f$} be an injective homomorphism from $\bB_{\mathcal F}^{\ind}$ to 
$\bB_{\mathcal F}^{\hom}$ and 
$g$ an injective homomorphism from 
$\bB_{\mathcal F}^{\hom}$ to $\bB_{\mathcal F}^{\ind}$.
Let $u$ and $v$ be two elements of
$\bB_{\mathcal F}^{\hom}$ that induce isomorphic 1-element substructures. Then $g(u)$ and $g(v)$
must induce isomorphic 1-element substructures, too,
since otherwise the injection $e := f \circ g$ would not preserve all first-order formulas, in contradiction to the assumption that
$(\bB_{\mathcal F}^{\hom},\neq)$ is a model-complete core. By the 1-homogeneity of $\bB_{\mathcal F}^{\ind}$ \blue{(Lemma~\ref{lem:1homogeneous})} there exists $\alpha \in \Aut(\bB_{\mathcal F}^{\ind})$
such that $\alpha(g(u)) = g(v)$. The mapping $e' :=  f \circ \alpha \circ g$ is an endomorphism of
$(\bB_{\mathcal F}^{\hom},\neq)$, and since
$(\bB_{\mathcal F}^{\hom},\neq)$ is a model-complete core there exists $\beta \in \Aut(\bB_{\mathcal F}^{\hom},\neq)$ such that $\beta(u) = e'(u)$. 
There also exists a $\gamma \in \Aut(\bB_{\mathcal F}^{\hom},\neq)$ such that $\gamma(u) = e(v)$. 
Then \begin{align*}
\gamma^{-1}(\beta(u)) & = \gamma^{-1}(f(\alpha(g(u))))\\
& = \gamma^{-1}(f(g(v))) \\
& = \gamma^{-1}(e(v)) = v
\end{align*}
and so $u$ and $v$ are in the same orbit of
$\Aut(\bB_{\mathcal F}^{\hom})$. 
\end{proof}

\begin{corollary}\label{cor:1homogeneous}
Let $\Phi$ be an MMSNP sentence in normal form. 
Then $\bC_\Phi$ is 1-homogeneous.
\end{corollary}
\begin{proof}
Let ${\mathcal F}$ be
the coloured obstruction set for $\Phi$.  
Recall that $\bC_\Phi$ is a substructure of
$\bB^{\hom}_{\mathcal F}$. 
Let $x$ and $y$ be two elements of 
$\bC_{\Phi}$ that
induce isomorphic 1-element substructures.
By Lemma~\ref{lem:1homogeneous}, 
$x$ and $y$ lie in the same orbit of 
$\bB^{\hom}_{\mathcal F}$. 
When $x$ and $y$
are in the same orbit of $\bB^{\hom}_{\mathcal F}$,
they are clearly also in the same orbit of $\bC_\Phi$
since automorphisms of $\bB^{\hom}_{\mathcal F}$
respect the domain of $\bC_\Phi$. 
\end{proof}

\begin{lemma}\label{lem:one-element-substructs}
Let $\Phi$ be in normal form with colours 
$M_1,\dots,M_n$. 
Let $a$ and $b$ be two elements of
$\bC_\Phi$ that induce non-isomorphic one-element
structures in $\bC_\Phi$. Then there are distinct
$i,j \in \{1,\dots,n\}$ such that $\bC_\Phi \models M_i(a) \wedge M_j(b)$. 
\end{lemma}

\begin{proof}
By definition of $\bC_\Phi$ there are 
$i,j \in \{1,\dots,n\}$ such that 
$M_i(a)$ and $M_j(b)$.
Let ${\mathcal F}$ be the coloured obstruction set for $\Phi$. 
Since $(\bB_{\mathcal F}^{\hom},\neq)$ is a model-complete core, 
there is a primitive positive definition $\psi_1(x)$  
of the orbit of $a$ in $(\bB_{\mathcal F}^{\hom},\neq)$,
and similarly a primitive positive
definition $\psi_2(x)$ of the orbit of $b$ 
in 
$(\bB_{\mathcal F}^{\hom},\neq)$. 
Pick witnesses for the existentially quantified
variables that show that $\psi_1(a)$ and $\psi_2(b)$  hold, and 
let $\psi'_1(x)$ and $\psi'_2(x)$ be the
primitive positive formulas
in the \red{signature} of 
$\bB_{\mathcal F}^{\hom}$
that we obtain from $\psi_1$ and $\psi_2$
by 
\begin{enumerate}
\item dropping the conjuncts that involve the symbol $\neq$, and 
\item adding conjuncts of the form $M(x)$ for every
existentially quantified variable, where $M$ is the colour of the witness that we picked above. 
\end{enumerate} 
Clearly, the canonical databases of 
$\psi'_1$ and of
$\psi'_2$ satisfy the first-order part $\phi$ of $\Phi$.
We claim that the canonical database of $\psi'_1(x) \wedge \psi'_2(x)$ does not satisfy $\phi$.
Then Lemma~\ref{lem:nf} implies that $i \neq j$ and we are done.

To show the claim, suppose for contradiction that 
$\psi'_1(x) \wedge \psi'_2(x)$ is satisfiable. 
Then the canonical
database of this formula maps homomorphically to 
$\bB_{\mathcal F}^{\hom}$,
and by the first item of Theorem~\ref{thm:css-new} also map homomorphically  and 
injectively 
to $\bB_{\mathcal F}^{\hom}$. 
Hence, the formula
$\psi_1(x) \wedge \psi_2(x)$ is satisfiable
as well (any injective homomorphism gives a satisfying assignment). 
But $\psi_1(x) \wedge \psi_2(x)$ cannot
be satisfiable 
in $(\bB_{\mathcal F}^{\hom},\neq)$ because $a$ and $b$
must lie in different orbits of $\bB_{\mathcal F}^{\hom}$.
\end{proof}

Note that Lemma~\ref{lem:one-element-substructs}
would be false if instead of $\bB^{\hom}_{\mathcal F}$
we would have used $\bB^{\ind}_{\mathcal F}$
in the definition of $\bC_\Phi$, as shown by the
following example. 

\begin{example}
Let $\tau$ be the signature that only contains
the two unary predicates $P$ and $Q$. 
Let $\Phi$ be the MMSNP $\tau$-sentence in normal form with \red{no existentially quantified predicates and} an empty coloured obstruction set $\mathcal F$.
Then $\bB^{\ind}_{\mathcal F}$ 
would have 
four orbits, but just one colour, so there are 
vertices of the same colour that lie in different orbits. 
But $\bB^{\hom}_{\mathcal F}$ has only one orbit,
since all elements of $\bB^{\hom}_{\mathcal F}$ must
lie both in $P$ and in $Q$. 
\end{example}

\blue{The previous two lemmas jointly imply the following, which will become important in later sections.}

\begin{corollary}\label{cor:colors-are-orbits}
\blue{Let $\Phi$ be in normal form. Then the colours
of $\Phi$ denote the orbits of $\Aut(\bC_\Phi)$.}
\end{corollary}

\blue{The final goal of this section is to prove that
for MMSNP sentences $\Phi$ in normal form
the structure $(\bC_\Phi,\neq)$ is a model-complete core. To this end, we need the following.}

\begin{lemma}\label{lem:non-col}
Let $\Phi$ be an MMSNP $\tau$-sentence in normal form and $\mathcal F$ be the coloured obstruction set for $\Phi$. Let
$\bar a$ be a $k$-tuple of elements of 
$\bB^{\hom}_{\mathcal F}$ which has
an entry $a_i$ that does not satisfy
the first conjunct of $\Phi$. 
Then $\bB^{\hom}_{\mathcal F}
\models R(\bar a)$ for every 
$R \in \tau$ of arity $k$.
\end{lemma}
\begin{proof}
Let $\bB$ be the structure
obtained from
$\bB^{\hom}_{\mathcal F}$
by adding $\bar a$ to $R \in \tau$.
We claim that $\bB$ maps homomorphically to 
$\bB^{\hom}_{\mathcal F}$. By
$\omega$-categoricity of 
$\bB^{\hom}_{\mathcal F}$, it suffices to prove that every finite substructure $\bB'$ of the countable structure $\bB$ maps homomorphically
to $\bB^{\hom}_{\mathcal F}$. 
No structure from $\mathcal F$ maps homomorphically to $\bB'$, since 
\begin{itemize}
\item coloured obstructions from conjuncts as in item 2 of the definition of normal forms are satisfied by $\bB$ since 
$\bB^{\hom}_{\mathcal F}$ satisfies the conjunct, and $\bB^{\hom}_{\mathcal F}$ and $\bB$ coincide with respect to the unary relations; 
\item all other coloured obstructions cannot map to $\bB$ 
since they are fully coloured (item 3 of the definition of normal forms) 
and the element $a_i$ is by assumption not coloured. 
\end{itemize}
Therefore $\bB'$ maps homomorphically to 
$\bB^{\hom}_{\mathcal F}$ by
the first item in the definition of $\bB^{\hom}_{\mathcal F}$ from Theorem~\ref{thm:css-new}.
 Since the identity is a homomorphism from $\bB^{\hom}_{\mathcal F}$ to $\bB$, and $\bB^{\hom}_{\mathcal F}$ is a model-complete core,
we therefore must have that $\bB^{\hom}_{\mathcal F}
\models R(\bar a)$. 
\end{proof}

\begin{lemma}\label{lem:C-phi-mc-core}
Let $\Phi$ be an MMSNP $\tau$-sentence in normal form. Then $(\bC_\Phi,\neq)$ is 
a model-complete core.
\end{lemma} 
\begin{proof}
Let $M_1,\dots,M_n$ be the colours of $\Phi$,
and let ${\mathcal F}$
be the coloured obstruction set for $\Phi$. 
Let $e$ be an endomorphism of $\bC_\Phi$
and let $\bar b$ be a tuple of elements
of $\bC_\Phi$. We
have to show that there exists an automorphism $\beta$ of $\bC_\Phi$ such that $\beta(\bar b) = e(\bar b)$ \red{(see, e.g.,~\cite{Bodirsky-HDR})}. 
We extend $e$ to all elements of 
$\bB^{\hom}_{\mathcal F}$ by setting $e(a) := a$ for all uncoloured elements $a$ of $\bB^{\hom}_{\mathcal F}$, and verify that 
the resulting map $e'$ is an endomorphism of
$\bB^{\hom}_{\mathcal F}$. 
Clearly, $e'$ preserves $M_i$ for all $i \leq n$. Let $R \in \tau$, and
let $\bar a$ be such that 
$\bB^{\hom}_{\mathcal F} \models R(\bar a)$. 
If all entries of $\bar a$ are elements
of $\bC_\Phi$ then 
$\bB^{\hom}_{\mathcal F}
\models R(e'(\bar a))$ since $e'(\bar a) = e(\bar a)$ and $e'$ is an endomorphism. 
On the other hand, if
$\bar a$ has an entry $a_i$
which is not in $\bC_\Phi$,
then $\bB^{\hom}_{\mathcal F}
\models R(e'(\bar a))$ by Lemma~\ref{lem:non-col}. 
Since $(\bB^{\hom}_{\mathcal F},\neq)$ 
is a model-complete core
there exists an 
$\alpha \in \Aut(\bB^{\hom}_{\mathcal F})$ 
such that $\alpha(\bar b) = e(\bar b)$. 
The restriction $\beta$ of $\alpha$ to 
$\bC_\Phi$ is an automorphism of
$\bC_\Phi$ with the desired property. 
\end{proof}

\ignore{
From the fact that $(\bC_\Phi,\neq)$
is a model-complete core we can derive a consequence that will be used often in later sections.

\begin{lemma}
Let $\bA$ be a substructure of $\bC_\Phi$ and let $B$ be a structure containing $A$ 
such that $B$ satisfies $\Phi$. Then there exists an injective homomorphism $\bB->\bC_\Phi$ that is an embedding when restricted to A.
\end{lemma}
}

\subsection{Recolourings}
Let $\Phi_1$ and $\Phi_2$ be two MMSNP $\tau$-sentences
in normal form with colour sets $\sigma_1$
and $\sigma_2$, respectively. 
For $r \colon \sigma_1 \to \sigma_2$ and a $(\tau \cup \sigma_1)$-structure $\bA$ we write $r(\bA)$ for the structure
obtained from $\bA$ by renaming each predicate %
\newblue{$P \in \sigma_1$ to $r(P) \in \sigma_2$}. 
\begin{definition}
A \emph{recolouring (from $\Phi_1$ to $\Phi_2$)} is given by a function 
$r \colon \sigma_1 \to \sigma_2$ such that 
for every $(\tau \cup \sigma_1)$-structure $\bA$,
if a coloured
obstruction of $\Phi_2$ maps homomorphically to $r(\bA)$, 
 then a coloured obstruction of $\Phi_1$
 maps homomorphically to $\bA$. 
A recolouring $r \colon \sigma_1 \to \sigma_2$ is said to be \emph{proper} 
if $r$ is non-injective.
\end{definition}
\nomenclature[021]{$r \colon \sigma_1 \to \sigma_2$}{function between colour sets, e.g., a recolouring}
\nomenclature[022]{$r(\bA)$}{the structure obtained from $\bA$ by renaming
  each colour according to $r$}

\begin{example} \label{ex:nf1}
Consider the MMSNP sentence $\Phi$ given by
$$ \exists M_1,M_2 \; \forall x \, \big(  (M_1(x) \vee M_2(x)) \wedge
(\neg M_1(x) \vee \neg M_2(x))  \big ) $$
and note that this sentence is in normal form. 
There is a proper recolouring $r$ from $\Phi$ to $\Phi$, e.g., the map given by $r(M_1) = r(M_2) = M_1$. 
\end{example}

\begin{lemma}\label{lem:recolouring}
Let $\Phi_1$ and $\Phi_2$ be MMSNP $\tau$-sentences
in normal form. 
If $r$ is a recolouring from $\Phi_1$ to $\Phi_2$,
then every $\tau$-structure that satisfies
$\Phi_1$ also satisfies $\Phi_2$. 
\end{lemma}
\begin{proof}
Let $\tau$ be the signature of $\Phi_1$ and $\Phi_2$, 
and let $\sigma_1$ be the existentially quantified predicates of $\Phi_1$.  
Let $\bA$ be a finite model of $\Phi_1$.  
We have to show that $\bA \models \Phi_2$. 
Let $\sigma_1$ be the existentially quantified predicates
of $\Phi_1$. 
Let $\bA'$ be the $(\tau \cup \sigma_1)$-expansion of $\bA$ witnessing the truth of
$\Phi_1$ in $\bA$. Since $r$ is a recolouring,
the structure $r(\bA')$ does not embed any coloured
obstruction of $\Phi_2$, hence $\bA \models \Phi_2$. 
\end{proof}
 
 We mention that this lemma has a converse,
 as we will see in Theorem~\ref{thm:recol}.

\begin{example}\label{ex:nf2}
Consider the MMSNP $\{E\}$-sentence 
$$ \exists P \; \forall x,y \, \neg \big(\neg P(x) \wedge E(x,y) \wedge \neg P(y) \big)$$
It is not yet in normal form; an equivalent MMSNP
sentence $\Phi$ 
in normal form is 
\begin{align*}
 \exists M_1,M_2 \; \forall x,y \, \big(& \neg (\neg M_1(x) \wedge \neg M_2(x)) \; \wedge \\
& \neg (M_1(x) \wedge M_2(x)) \wedge \neg (M_1(x) \wedge E(x,y) \wedge M_1(y)) \big)
\end{align*} 
A proper recolouring from $\Phi$ to $\Phi$ is given
by $r(M_1)=r(M_2)=M_2$. 
To verify that $r$ is indeed a recolouring, consider 
the conjunct $\neg \phi_1 = \neg (M_1(x) \wedge E(x,y) \wedge M_1(y))$: if $\bB_1$ is 
the canonical database 
of $\phi_1$ then there does not exist any $(\tau \cup \sigma_1)$-structure $\bA$ such that $r(\bA) = \bB_1$. 
For the conjunct 
$\neg \phi_2 = \neg (M_1(x) \wedge M_2(x))$, 
if $\bB_2$ is the canonical database of 
$\phi_2$, there is again no $(\tau \cup \sigma_1)$-structure $\bA$ such that $r(\bA) = \bB_2$. 

In contrast, the map given by $r(M_1)=r(M_2)=M_1$ is \emph{not} a recolouring: consider the canonical database $\bA$
of the formula $M_1(x) \wedge E(x,y) \wedge M_2(y)$. 
It satisfies the quantifier-free part of $\Phi$,
but $r(\bA)$ is isomorphic to the canonical database 
of $\phi = (M_1(x) \wedge E(x,y) \wedge M_1(y))$, 
and $\neg \phi$ is a conjunct of $\Phi$. 
\end{example}

\begin{lemma}\label{lem:recol-dec}
Given two MMSNP sentences $\Phi_1$ and $\Phi_2$ in normal form,
one can effectively decide whether there exists a recolouring from $\Phi_1$ to $\Phi_2$. 
\end{lemma}
\begin{proof} 
In order to check whether a given map from $\sigma_1 \to \sigma_2$ 
is a recolouring, it suffices to check the recolouring
condition for $(\tau \cup \sigma_1)$-structures of size at most $|\Phi_2|$. 
 \end{proof}
 
\subsection{The strong normal form}
\label{sect:snf}
An MMSNP sentence $\Phi$ 
is defined to be in 
\emph{strong normal form} 
if it is in normal form and 
there is no proper recolouring
from $\Phi$ to $\Phi$. 

\begin{example}
The MMSNP sentence $\Psi$ from
Example~\ref{ex:p3-nf} is not only in normal form, but even in strong normal form. 
\end{example}

\begin{example}\label{ex:snf1}
Example~\ref{ex:nf1} was in normal form, but not
in strong normal form. An equivalent formula
in strong normal form is
$$ \exists M_1 \forall x. \neg (\neg M_1(x)) .$$
\end{example}

\begin{example}\label{ex:snf2}
The sentence 
\begin{align} \exists M_1 \forall x,y \,  \big( \neg (\neg M_1(x)) \wedge \neg (M_1(x) \wedge E(x,y) \wedge M_1(y)) \big ).
\label{eq:snf2} 
\end{align}
is a strong normal form for the sentence from 
Example~\ref{ex:nf2}. 
\end{example}

\begin{theorem}
\label{thm:snf-existence}
For every connected MMSNP sentence $\Phi$ 
there exists an equivalent connected MMSNP 
$\Psi$ in strong normal form, and $\Psi$ can
be effectively computed from $\Phi$. 
\end{theorem}
\begin{proof}
By Lemma~\ref{lem:nf-existence}, we
can assume that $\Phi$ is already given
in normal form; let $\sigma$ be the colours of $\Phi$. 
To compute a strong normal form for $\Phi$ 
we exhaustively check for proper recolourings
from $\Phi$ to $\Phi$ (see Lemma~\ref{lem:recol-dec}). 

 If there is no such recolouring
we are done. Otherwise, let $r$ be such a proper recolouring. Let $\Psi$ be the MMSNP sentence obtained from $\Phi$ by performing the following
for each colour $M$ not in the image of $r$:
\begin{enumerate}
\item drop all conjuncts $\neg \phi$ of $\Phi$ 
such that $M$ appears positively in $\phi$,
\item remove the literal in which $M$ appears negatively from the first conjunct of $\Phi$, and 
\item remove $M$ from the existential quantifier prefix of $\Phi$.
\end{enumerate}
(Step 1 and 2 amount to replacing $M$ by \emph{false}.) 
Since the identity map is clearly a recolouring from
$\Psi$ to $\Phi$, Lemma~\ref{lem:recolouring} implies
that $\Psi$ is equivalent to $\Phi$. 
We now repeat the procedure with $\Psi$ instead of
$\Phi$. 
Since $\Psi$ has fewer existential predicates than $\Phi$
 this procedure must eventually terminate with an MMSNP sentence in strong normal form that is equivalent to the sentence we started with. 
\end{proof}

\section{Recolouring and Containment}
The following result has
already been announced in~\cite{MadelaineCP10}. 

\begin{theorem}[Recolouring Theorem]\label{thm:recol}
Let $\Phi_1$ and $\Phi_2$ be two MMSNP sentences in normal form. 
Then the following are equivalent: 
\begin{enumerate}
\item All finite $\tau$-structures that satisfy
$\Phi_1$ also satisfy $\Phi_2$;
\item $\Phi_1$ has a recolouring to $\Phi_2$;
\item All $\tau$-structures that satisfy
$\Phi_1$ also satisfy $\Phi_2$.
\end{enumerate}
\end{theorem}

\red{Since deciding the existence of a recolouring from $\Phi_1$ to $\Phi_2$ is decidable by Lemma~\ref{lem:recol-dec}, we have the following consequence which has been foreseen by Feder and Vardi~\cite{FederVardi}. 
\begin{corollary}
For given MMSNP sentences $\Phi_1$ and $\Phi_2$, the task to decide whether $\Phi_1$ implies $\Phi_2$ is decidable. 
\end{corollary}}

The implication from 3.\ to 1.\ is trivial.
For the converse implication, 
let $\bA$ be a $\tau$-structure that satisfies $\Phi_1$. Clearly, all finite substructures of $\bA$ satisfy $\Phi_1$,
so by 1.\ they also satisfy $\Phi_2$. 
We now use the well-known
fact that a structure satisfies an SNP sentence if and only if all finite substructures satisfy the sentence (see Lemma~9 in~\cite{FederVardiNegation}),
and obtain that $\bA$ satisfies $\Phi_2$. 

The implication from 2.\ to 3.\ is Lemma~\ref{lem:recolouring}. 
The proof of the implication from 3.\ to 2.\ requires some tools that we present in the next sections. 

\subsection{Ramsey theory}
\label{sect:ramsey}
Let $\tau$ be a relational signature, and
$\mathcal F$ a finite set of finite $\tau$-structures.
\red{Let ${\bB \choose \bA}$ be the set of all
embeddings of $\bA$ into $\bB$. If $e \colon A \to B$ and $f \colon B \to C$ are functions we write $f \circ e \colon A \to C$ for the composition of the two functions;
we also use this notation set-wise, i.e., if $S$ is a set of functions from $B$ to $C$ then we write
$e \circ S$ for $\{e \circ f \mid f \in S\}$.  
For $\tau$-structures $\bA,\bB,\bC$ and $r \in {\mathbb N}$ we write $\bC \to (\bB)^\bA_r$ if for every
$\chi \colon {\bC \choose \bA} \to \{1,\dots,r\}$ 
there exists an $e \in {\bC \choose \bB}$ such that
$|\chi(e \circ {\bB \choose \bA})| \leq 1$.} 
\nomenclature[023]{${\bB \choose \bA}$}{the set of all embeddings of $\bA$ into $\bB$}
\nomenclature[024]{$\bC \to (\bB)^\bA_r$}{the Ramsey partition arrow}

\begin{definition}[\blue{see, e.g.,~\cite{BodirskyRamsey}}]
We say that a homogeneous
structure $\bC$
is \emph{Ramsey} if $\bC \to (\bB)^\bA_r$ holds
for every $r \in \mN$
and all finite substructures
$\bA,\bB$ of $\bC$.
An $\omega$-categorical structure
is Ramsey if its (homogeneous) expansion by all first-order definable relations is Ramsey. 
\end{definition}

A recent result of Hubi\v{c}ka and \Nesetril\ (Theorem~\ref{thm:HN-Ramsey} below) asserts that a certain homogeneous structure $\HN$ that can be associated to a finite set of finite $\tau$-structures ${\mathcal F}$ is Ramsey.
The structure $\HN$ has first been described in~\cite{Hubicka-Nesetril-ForbiddenHomomorphisms}. 
Since a homogeneous structure is uniquely up to isomorphism given by
its age, it suffices to specify the age of $\HN$.
\nomenclature[025]{$\HN$}{The Hubi\v{c}ka-\Nesetril{} homogeneous
  Ramsey structure}
Let $m$ be the size of the largest structure in ${\mathcal F}$ (with respect to the number of its elements). 
Let ${\mathcal P}$ be the class of all structures
in $\Forb^{\hom}({\mathcal F})$ that have been
expanded by all relations defined by connected primitive positive formulas with 
at most $m$ variables and with at least one free variable.
Then the class of all substructures of structures
in $\mathcal P$ is an amalgamation class, and $\HN$ is its \Fresse-limit.

Theorem~\ref{thm:HN-Ramsey} states
that $\HN$ 
has a homogeneous order expansion 
${(\HN,<)}$ 
which is Ramsey. We will see that this gives Ramsey order 
expansions of $\bB^{\ind}_{\mathcal F}$,
$\bB^{\hom}_{\mathcal F}$, and 
$\bC_\Phi$, too. 
We need the following folklore result (see e.g. Section 2.7 in~\cite{Oligo} \red{or~\cite{BodirskyGreinerCombinations}}):

\begin{lemma}\label{lem:stardef}
Let $\bB_1$ and $\bB_2$ be
two $\omega$-categorical structures without
algebraicity and disjoint signatures $\tau_1$
and $\tau_2$. Then there exists an 
up to isomorphism unique $(\tau_1 \cup \tau_2)$-structure $\bB =: \bB_1 * \bB_2$, called the \red{generic combination of $\bB_1$ and $\bB_2$}, such that
\begin{enumerate}
\item $\bB^{\tau_i}$ is isomorphic to $\bB_i$ for $i =1$ and $i=2$; 
\item $\overline{\Aut(\bB^{\tau_1}) \circ \Aut(\bB^{\tau_2})} = \overline{\Aut(\bB^{\tau_2}) \circ \Aut(\bB^{\tau_1})} = \End(\bB;\neq)$; 
\item for finite tuples $\bar a, \bar b$ of elements of $\bB$ there exists an automorphism of $\bB$ that maps $\bar a$ to $\bar b$ if and only if there 
exist automorphisms of $\bB^{\tau_1}$
and of $\bB^{\tau_2}$ that 
map $\bar a$ to $\bar b$.
\end{enumerate}
The structure $\bB$ is $\omega$-categorical and has no algebraicity. 
\end{lemma}
\nomenclature[026]{$\bB_1 * \bB_2$}{
the generic combination of $\bB_1$ and $\bB_2$, 
see Lemma~\ref{lem:stardef}}

\begin{theorem}[implied by Theorem 2.1 in~\cite{Hubicka-Nesetril-All-Those}]
\label{thm:HN-Ramsey}
For every finite sets of finite connected $\tau$-structures ${\mathcal F}$ 
the structure $(\HN,<) := \HN * ({\mathbb Q};<)$ is Ramsey.
\end{theorem}

Theorem 2.1 in~\cite{Hubicka-Nesetril-All-Those} is stronger and the terminology is different than here, but for the convenience of the reader we state Theorem 2.1 in Appendix~\ref{sect:hn} in full detail and explain how to specialise it in order to derive
Theorem~\ref{thm:HN-Ramsey}. 
From Theorem~\ref{thm:HN-Ramsey} we will deduce that several
other structures that are important later
are Ramsey, too. For this, we use the
following lemma.

\begin{lemma}\label{lem:mc-core-join}
\red{Let $\bB_1$ and $\bB_2$ be
two $\omega$-categorical structures without
algebraicity and suppose that the model-complete cores $\bC_1$ and $\bC_2$
of $\bB_1$ and $\bB_2$ also do not have algebraicity. Then the model-complete core of $\bB_1 * \bB_2$ 
is isomorphic to $\bC_1 * \bC_2$.}
\end{lemma}

\begin{proof}
We first show that 
$\bC_1 * \bC_2$ is a model-complete core. 
By Theorem~3.6.11 in~\cite{Bodirsky-HDR}, 
being a model-complete core is equivalent 
to the existence of a homogeneous expansion by relations with an existential positive
definition whose complement also has an existential positive definition. By item 3 in Lemma~\ref{lem:stardef}, the expansion of 
$\bC_1 * \bC_2$ by all relations with a first-order definition $\phi$ in $\bC_1$ 
or in $\bC_2$ is homogeneous.
Since $\bC_i$ is a model-complete core, 
the formulas $\phi$ and $\neg \phi$ are equivalent
to existential positive formulas over $\bC_i$. 
Hence, when we expand $\bC_1 * \bC_2$ by all
relations with an existential positive definition in 
either $\bC_1$ or in $\bC_2$, the resulting structure is homogeneous. Therefore, $\bC_1 * \bC_2$ is a model-complete core. 

To show that there exists a homomorphism
from $\bC_1 * \bC_2$ to $\bB := \bB_1 * \bB_2$, let $\bA$
be a finite substructure of 
$\bC_1 * \bC_2$. 
Then $\bA^{\tau_i}$ has a homomorphism $h_i$ to $\bB^{\tau_i}$, for $i=1$ and $i=2$. 
By item 2.\ in Lemma~\ref{lem:stardef}  there exist $\alpha_1 \in \Aut(\bB^{\tau_1})$ and $\alpha_2 \in \Aut(\bB^{\tau_2})$ 
such that $\alpha_1 \circ h_1 = \alpha_2 \circ h_2 := h$. 
The map $h$ is
a homomorphism from $\bA$ to $\bB_1 * \bB_2$. The existence of a homomorphism 
from $\bC_1 * \bC_2$ to $\bB_1 * \bB_2$ now follows by compactness. 
The existence of a homomorphism from 
$\bB_1 * \bB_2$ to $\bC_1 * \bC_2$
can be shown analogously. 
\end{proof} 

\begin{corollary}\label{cor:ramsey}
For all finite sets of finite connected $\tau$-structures ${\mathcal F}$ 
the structures $\bB^{\ind}_{\mathcal F} * ({\mathbb Q};<)$ and $\bB^{\hom}_{\mathcal F} * ({\mathbb Q};<)$ are Ramsey.
\end{corollary}
\begin{proof}
\blue{By Theorem~\ref{thm:HN-Ramsey}, the
structure $\HN * ({\mathbb Q};<)$ is Ramsey. Let $\bD$ be the
$\tau$-reduct of this structure.  
Note that $(\bD,\neq) * ({\mathbb Q};<)$ is Ramsey, too, since it has the same automorphism group
as $\HN * ({\mathbb Q};<)$. The model-complete core of $(\bD,\neq)$
equals $(\bB^{\hom}_{\mathcal F},\neq)$
and the structure $({\mathbb Q};<)$ already is a model-complete core.
So by Lemma~\ref{lem:mc-core-join},
the model-complete core of $(\bD,\neq) * ({\mathbb Q};<)$ is 
$(\bB^{\hom}_{\mathcal F},\neq) * ({\mathbb Q};<)$. }
Theorem~3.18 of~\cite{BodirskyRamsey}
states that the model-complete core 
of an $\omega$-categorical Ramsey structure
is again Ramsey. So 
$(\bB^{\hom}_{\mathcal F},\neq) * ({\mathbb Q};<)$ is Ramsey, and therefore also 
$\bB^{\hom}_{\mathcal F} * ({\mathbb Q};<)$. 
\blue{The statement for $\bB^{\ind}_{\mathcal F} * ({\mathbb Q};<)$ can be shown similarly,
using Theorem~3.15 in~\cite{BodirskyRamsey}  instead of Theorem~3.18 in~\cite{BodirskyRamsey}.}
\end{proof}

\begin{definition}\label{def:cphi<}
We write 
\begin{itemize}
\item $(\bB^{\ind}_{\mathcal F},<)$
for the expansion of $\bB^{\ind}_{\mathcal F}$
 isomorphic to $\bB^{\ind}_{\mathcal F} * ({\mathbb Q};<)$;
\item 
$(\bB^{\hom}_{\mathcal F},<)$
for the expansion of $\bB^{\hom}_{\mathcal F}$
 isomorphic to $\bB^{\hom}_{\mathcal F} * ({\mathbb Q};<)$; 
\item $(\bC_\Phi,<)$ for the substructure of $(\bB^{\hom}_{\mathcal F},<)$ induced by the elements
that satisfy the first conjunct of $\Phi$. 
In other words: we obtain
$(\bC_\Phi,<)$ from $(\bB^{\hom}_{\mathcal F},<)$ by
removing all uncoloured vertices. Note that $(\bC_\Phi,<)$ is indeed an expansion 
of $\bC_\Phi$.
\end{itemize}
\end{definition}
\nomenclature[027]{$(\bB^{\ind}_{\mathcal F},<)$}{a shortcut for $\bB^{\ind}_{\mathcal F} * ({\mathbb Q};<)$}
\nomenclature[028]{$(\bB^{\hom}_{\mathcal F},<)$}{a shortcut for $\bB^{\hom}_{\mathcal F} * ({\mathbb Q};<)$}
\nomenclature[029]{$(\bC_\Phi,<)$}{the substructure of $(\bB^{\hom}_{\mathcal F},<)$ induced by the coloured vertices}

\begin{corollary}\label{cor:ramsey-cphi}
Let $\Phi$ be an MMSNP sentence
in normal form. 
Then $(\bC_\Phi,<)$
is Ramsey. 
\end{corollary} 
\begin{proof}
Kechris, Pestov, and Todorcevic~\cite{Topo-Dynamics} proved that 
an ordered $\omega$-categorical 
structure $\bB$ is Ramsey if
and only if the automorphism
group of $\bB$ is extremely amenable.
Hence, $\Aut(\bB^{\hom}_{\mathcal F},<)$ is extremely amenable by Corollary~\ref{cor:ramsey}. 

Moreover, by Lemma 6.18 in~\cite{Topo-Dynamics}, if $\pi \colon G \to H$
is a continuous homomorphism between
topological groups whose image is
dense, and $G$ is extremely amenable,
then so is $H$. 
Thus, it suffices to prove that
there exists a continuous 
homomorphism from $\Aut(\bB^{\hom}_{\mathcal F},<)$ to $\Aut(\bC_\Phi,<)$
whose image is dense in $\Aut(\bC_\Phi,<)$, 
because in this case 
the backwards direction of the KPT connection implies
that the structure $(\bC_\Phi,<)$
is Ramsey. 
The restriction map from 
$\Aut(\bB^{\hom}_{\mathcal F},<)$ to 
$\Aut(\bC_\Phi,<)$ clearly is a continuous homomorphism. We show that its image is dense.
Let $\bar a$ be an $n$-tuple 
of elements of $(\bC_\Phi,<)$
and $\alpha \in \Aut(\bC_\Phi,<)$. 
We have to show that there exists a 
$\beta \in \Aut(\bB^{\hom}_{\mathcal F},<)$ such that
$\beta(\bar a) = \alpha(\bar a)$. 
Let $\sigma$ 
be the colours of $\Phi$.
By the third item of Lemma~\ref{lem:stardef}, 
the expansion of 
$\bB := (\bB^{\hom}_{\mathcal F},<)$
by all relations that are first-order definable
in %
$\bB^{\tau \cup \sigma} = \bB^{\hom}_{\mathcal F}$ 
and by all relations that are first-order definable in 
$\bB^{<}$ 
 is homogeneous. 
 Since by the 
homogeneity of $({\mathbb Q};<)$ the tuples
$\bar a$ and $\alpha(\bar a)$ 
satisfy the same first-order $\{<\}$-formulas in $\bB^{<}$,
it suffices to show that 
$\alpha$ can be extended to an injective endomorphism
of $\bB^{\hom}_{\mathcal F}$. 
This can be done by setting \red{$\alpha$ to be equal to $\beta$} for all
elements of $\bB^{\hom}_{\mathcal F}$ that are 
not elements of $(\bC_\Phi,<)$, 
as in the proof of Lemma~\ref{lem:C-phi-mc-core}. 
\end{proof}

\begin{lemma}\label{lem:1homogeneous-order}
Let $\Phi$ be an MMSNP sentence in normal form. Then the structure $(\bC_\Phi,<)$ is 1-homogeneous. 
\end{lemma}
\begin{proof}
Let $x$ and $y$ be elements of $(\bC_\Phi,<)$
that induce isomorphic 1-element substructures.  
Recall that $(\bC_\Phi,<)$ 
is a substructure of 
$(\bB^{\hom}_{\mathcal F},<) = \bB^{\hom}_{\mathcal F} * ({\mathbb Q};<)$, 
which is an expansion of 
$\bB^{\hom}_{\mathcal F}$. 
By Lemma~\ref{lem:1homogeneous}, there exists
an automorphism of $\bB^{\hom}_{\mathcal F}$
which maps $x$ to $y$. 
Moreover, there is an automorphism
of $({\mathbb Q};<)$ mapping $x$ to $y$, so by item 3 of Lemma~\ref{lem:stardef} there exists 
an automorphism of 
$(\bB^{\hom}_{\mathcal F},<)$ mapping $x$ to $y$. 
This automorphism $\alpha$ must preserve 
the elements of $(\bC_\Phi,<)$, and hence
the restriction of $\alpha$ to $(\bC_\Phi,<)$
is an automorphism of $(\bC_\Phi,<)$, which maps
$x$ to $y$, showing 1-homogeneity of 
$(\bC_\Phi,<)$.
\end{proof}

\subsection{Canonical functions}
\label{sect:can}
Let $\bA$ and $\bB$ be two structures. 
We call a function $f \colon A \to B$  \emph{canonical (from $\bA$ to $\bB$)} if
for every $m \in {\mathbb N}$ and all 
$x,y \in A^m$,
if $x$ and $y$ lie in the same orbit with respect to the componentwise action of $\Aut(\bA)$ on $A^m$
then $f(x)$ and $f(y)$ lie in 
the same orbit of the componentwise action of 
$\Aut(\bB)$ on $B^m$. In other words, 
$f$ induces a function from the orbits
of $m$-tuples of $\Aut(\bA)$ to the orbits
of $m$-tuples of $\Aut(\bB)$.

\begin{theorem}[\cite{BodPin-CanonicalFunctions}]\label{thm:can}
Let $\bA$ be a countable homogeneous $\tau$-structure \red{which  is Ramsey} and let $\bB$ be $\omega$-categorical.
Then for any function $f \colon A \to B$
there exists a function in 
$$\overline{ \{ \beta \circ f \circ \alpha \mid \alpha \in \Aut(\bA), \beta \in \Aut(\bB)\} }$$
that is canonical from $\bA$ to $\bB$. 
In particular, if there exists a homomorphism from
$\bA$ to $\bB$ then there also
exists a canonical homomorphism from $\bA$ to $\bB$. 
\end{theorem}

The following lemma explains how
homomorphisms from $\bC^\tau_\Phi$ to $\bC_\Psi^\tau$ that are 
canonical from $(\bC_\Phi,<)$ to $(\bC_\Psi,<)$ give rise to recolourings. 

\begin{lemma}\label{lem:can-recol}
Let $\Phi$ and $\Psi$ be two MMSNP sentences in normal form
and $h$ a homomorphism
from $\bC^\tau_\Phi$ to $\bC_\Psi^\tau$ which is canonical from 
$(\bC_\Phi,<)$ to $(\bC_{\Psi},<)$.
\blue{Then the map $r$ from the colours of $\Phi$ to the colours of $\Psi$ that takes $M$ to the colour
of $h(x)$ for some $x \in M$ is well-defined 
and a recolouring from $\Phi$ to $\Psi$.}
\end{lemma}
\begin{proof}
\blue{Let $\sigma$ be the colours of $\Phi$. 
To show that $r$ is well-defined, let $a,b$ 
be elements of $(\bC_\Phi,<)$ of the same color.
By Lemma~\ref{lem:one-element-substructs},
$a$ and $b$ must induce \red{isomorphic} 1-element
substructure of $\bC_\Phi$, and hence also of
$(\bC_\Phi,<)$. Since $(\bC_\Phi,<)$ is 1-homogeneous by
Lemma~\ref{lem:1homogeneous-order},
$a$ and $b$ lie in the same orbit of $\Aut(\bC_\Phi,<)$. The canonicity of
$h$ then implies that the images $h(a)$ 
and $h(b)$ lie in the same
orbit of $\Aut(\bC_{\Psi},<)$, and in particular 
they must have the same color in $\bC_\Psi$. 
Hence, $r$ is well-defined.}

Let $\bA$ be a $(\tau \cup \sigma)$-structure
and suppose that no 
coloured obstruction of $\Phi$ homomorphically maps to $\bA$. 
Then there exists a homomorphism $g$ from 
$\bA$ to $\bC_\Phi$. 
By the canonicity of $h$ 
and the definition of $r$, the map $h \circ g$
is a homomorphism from $r(\bA)$ to $\bC_\Psi$. 
Hence, $r$ is a recolouring from $\Phi$ to $\Psi$. 
\end{proof}

\red{Ramsey theory will be also be applied
to functions of higher arity. }
Let $\bA$ and $\bB$ be two structures. 
A function $f \colon A^k \to B$ is called 
\begin{itemize}
\item \emph{1-canonical from $\bA$ to $\bB$} if for all $a_1,\dots,a_k\in A$,
the orbit of $f(a_1,\dots,a_k)$ with respect to $\Aut(\bB)$ only depends on the orbits of $a_1,\dots,a_k$ with respect to $\Aut(\bA)$,
\item \emph{canonical from $\bA$ to $\bB$} if for all $\tuple t_1, \dots, \tuple t_k \in A^m$
the orbit of $f(\tuple t_1,\dots,\tuple t_k)$ with respect to $\Aut(\bB)$ 
only depends on the orbits of $\tuple t_1,\dots,\tuple t_k$ with respect to $\Aut(\bA)$.
\item \emph{diagonally canonical from $\bA$ to $\bB$} if  for all $\tuple t_1,\dots,\tuple t_k \in A^m$
the orbit of $f(\tuple t_1,\dots,\tuple t_k)$ with respect to $\Aut(\bB)$ 
only depends on the orbit of the $km$-tuple $(\tuple t_1,\dots,\tuple t_k)$
with respect to $\Aut(\bA)$.
\end{itemize}

\begin{theorem}[\cite{BodPin-CanonicalFunctions}]\label{thm:can-3}
Let $\bA$ be a countable homogeneous
$\tau$-structure whose age is Ramsey and let $\bB$ be $\omega$-categorical. 
Then for any function $f \colon A^k \to B$
there exists a function in 
$$\overline{ \{ \beta \circ f \circ (\alpha_1,\dots,\alpha_k) \mid \alpha_1,\dots,\alpha_k \in \Aut(\bA),  \beta \in \Aut(\bB)\} }$$
that is canonical from $\bA$ to $\bB$.
\end{theorem}

The following already follows from Theorem~\ref{thm:can}. 

\begin{theorem}\label{thm:can-4}
Let $\bA$ be a countable homogeneous
$\tau$-structure whose age is Ramsey and let
$\bB$ be $\omega$-categorical. 
Then for any function $f \colon A^k \to B$
there exists a function in 
$$\overline{ \{ \beta \circ f \circ (\alpha,\dots,\alpha) \mid \alpha \in \Aut(\bA), \beta \in \Aut(\bB)\} }$$
that is diagonally canonical from $\bA$ to $\bB$. 
\end{theorem}

We say that $f \colon B^k \to B$ is \emph{canonical with respect to $\bB$} if it is canonical
from $\bB$ to $\bB$.
Note that the set of all polymorphisms of $\bB$ 
that are canonical with respect to $\bB$ forms a clone. 

\subsection{Proof of the recolouring theorem}
We will show the following in cyclic order;
this clearly implies Theorem~\ref{thm:recol} and Theorem~\ref{thm:containment-intro} from the introduction. \red{The result plays the same role for the logic MMSNP as the theorem of Chandra and Merlin~\cite{ChandraMerlin} plays for conjunctive queries.} 

\begin{theorem}[Recolouring Theorem, full version]\label{thm:recol-full}
Let $\Phi$ and $\Psi$ be two MMSNP sentences in %
normal form. 
Then the following are equivalent: 
\begin{enumerate}
\item All finite $\tau$-structures that satisfy
$\Phi$ also satisfy $\Psi$;
\item $\bC^\tau_{\Phi}$ homomorphically maps to $\bC^\tau_{\Psi}$;
\item There exists a homomorphism $h$ from $\bC^\tau_{\Phi}$ to $\bC^\tau_{\Psi}$ which is
canonical as a map from
$(\bC_\Phi,<)$ to $(\bC_{\Psi},<)$; 
\item $\Phi$ has a recolouring to $\Psi$;
\item All $\tau$-structures that satisfy
$\Phi$ also satisfy $\Psi$.
\end{enumerate}
\end{theorem}
\begin{proof}
$1. \Rightarrow 2.$: Observe that by assumption, all finite substructures of
$\bC^\tau_\Phi$ homomorphically map to $\bC^\tau_\Psi$.
Since $\bC_\Psi$ is $\omega$-categorical and $\bC_\Phi$ is countable, this implies the statement. 

$2. \Rightarrow 3.$: by Corollary~\ref{cor:ramsey-cphi}, the $\omega$-categorical structure 
$(\bC_\Phi,<)$ is Ramsey. 
Hence, the implication 
is a direct consequence of 
Theorem~\ref{thm:can}. 

$3. \Rightarrow 4.$ is Lemma~\ref{lem:can-recol}. 

$4. \Rightarrow 5.$ is Lemma~\ref{lem:recolouring}. 

$5. \Rightarrow 1.$ is trivial. 
\end{proof}

\subsection{Strong normal forms and cores}
\label{sect:mccores}
In this section we prove that if $\Phi$ is an MMSNP sentence in strong normal form with input 
signature $\tau$, then the structure $(\bC^\tau_\Phi,\neq)$ is a model-complete core. In the proof, the following binary relation $O$ on $\bC_\Phi$ is important:
 $O(x,y)$ expresses that $x$ and $y$ lie in the same orbit
of $\Aut(\bC_\Phi)$. This relation is clearly first-order definable in $\bC_\Phi$, but actually it is also first-order definable in 
$\bC_\Phi^\tau$, and therefore even existentially positively definable in 
$(\bC_\Phi^\tau,\neq)$ since 
$(\bC^\tau_\Phi,\neq)$ is a model-complete core.

\begin{theorem}\label{thm:mc-core}
Let $\Phi$ be an MMSNP sentence in strong normal form and with input signature $\tau$.
Then $(\bC^\tau_\Phi,\neq)$ is a model-complete core.
\end{theorem}
\begin{proof}
Let $\bC$ be the model-complete core of $(\bC^\tau_\Phi,\neq)$, and let 
$h$ %
be a homomorphism from $(\bC^\tau_\Phi,\neq)$
to $\bC$. Since $\bC$ is isomorphic to a substructure
of $(\bC^\tau_\Phi,\neq)$ we can assume in the following that $\bC$ equals such a substructure. 
It suffices to show that $h$ preserves
all first-order formulas. 
By Corollary~\ref{cor:ramsey-cphi}, 
the structure 
$(\bC_\Phi,<)$ is Ramsey. 
By Theorem~\ref{thm:can}, there exists a function 
$$g \in \overline{\{\beta \circ h \circ \alpha \mid \alpha \in \Aut(\bC_\Phi,<),\beta \in \Aut(\bC) \}}$$
which is canonical as a function from
 $(\bC_\Phi,<)$ to $\bC$, and an endomorphism of
 $(\bC^\tau_\Phi,\neq)$ (recall that $\bC$ is a substructure of $(\bC^\tau_\Phi,\neq)$). 
 
 We first consider the case that 
 the range of $g$ is contained
in the union of a proper subset of the set of all orbits of $\bC_\Phi$. 
By Lemma~\ref{lem:can-recol},
the canonical $g \in \End(\bC^\tau_\Phi,\neq)$ 
induces a recolouring from 
$\Phi$ to $\Phi$, which is proper because the orbits of $\bC_\Phi$ are in bijective corrspondance with the 1-element substructures of $\bC_\Phi$ by Corollary~\ref{cor:colors-are-orbits}.
This is in contradiction to the assumption
that $\Phi$ is in strong normal form. 

Otherwise, if the map induced
by $g$ on the colours of $\Phi$
is injective, then in particular
the relation $O$ is preserved by $g$,
and for sufficiently large $n$ 
the map $g^n \in \End(\bC^\tau_\Phi,\neq)$ 
preserves the orbits of $\bC_\Phi$.  
Hence, $g^n$ is an endomorphism of 
$\bC_\Phi$, and since $(\bC_\Phi,\neq)$
is a model-complete core by Lemma~\ref{lem:C-phi-mc-core}, the function 
$g^n$ preserves all first-order formulas over
$\bC_\Phi$. Hence, $g$ and $g^{n-1} \in \End(\bC^\tau_\Phi,\neq)$ 
\emph{locally invert each other} 
in the sense of~\cite{Bodirsky-HDR}, and $g \in \overline{\Aut(\bC^\tau_\Phi)}$ by Corollary 3.4.13 in~\cite{Bodirsky-HDR}. So $g$ preserves all first-order $\tau$-formulas. This shows in particular that 
$\bC$ and $(\bC^\tau_\Phi,\neq)$ have the same
first-order theory, and are isomorphic by $\omega$-categoricity. We conclude that $(\bC^\tau_\Phi,\neq)$ is a model-complete core.
\end{proof}

We give an example that shows that the assumption that $\Phi$ is in \emph{strong} normal form in Theorem~\ref{thm:mc-core} is necessary. 
 
\begin{example}
Consider again the MMSNP sentence 
$$\exists P \; \forall x,y. \, \neg \big(\neg P(x) \wedge E(x,y) \wedge \neg P(y) \big)$$ from Example~\ref{ex:nf2}; as we have observed, it is not in strong normal form. And indeed, the domain 
of $(\bC^\tau_\Phi,\neq)$ consists of two countably infinite sets such there are no edges within the first set, and otherwise all edges are present. Clearly, this structure is not a model-complete core since there are endomorphisms whose range does not contain any element from the first set. 
\end{example}

The following corollary shows that, in some sense, a description of the model-complete core of an MMSNP template can be computed algorithmically (via the strong normal form and Theorem~\ref{thm:snf-existence}). The corollary is
not needed in the further course of the paper; however, we want to state it here since we find this a good explanation for the concepts introduced so far. 

\begin{corollary}\label{cor:snf-mc-core}
Let $\Phi$ be an MMSNP sentence in normal form,
and let $(\bB,\neq)$ be the model-complete core of $(\bC^\tau_\Phi,\neq)$. 
Then there exists an 
MMSNP sentence $\Psi$ in strong normal form
such that
$\bC^\tau_{\Psi}$ is 
isomorphic to $\bB$. 
\end{corollary}
\begin{proof}
Let $\Psi$ be a strong normal form for $\Phi$, 
which exists 
due to Theorem~\ref{thm:snf-existence} in Section~\ref{sect:nf}. Since $\Phi$ and 
$\Psi$ are logically equivalent, 
all finite structures that satisfy $\Phi$
have an injective homomorphism to 
$\bC^\tau_\Psi$. By compactness,
also $(\bB,\neq)$ homomorphically and injectively maps
to $(\bC^\tau_\Psi,\neq)$. The existence
of an injective homomorphism from 
$(\bC^\tau_\Psi,\neq)$ to $(\bB,\neq)$ can be
shown analogously. 
By Theorem~\ref{thm:mc-core},
the structure $(\bC^\tau_\Psi,\neq)$ is a model-complete core. 
Hence, $(\bC_\Psi,\neq)$ 
is the model-complete core 
of $(\bC_\Phi,\neq)$. 
The statement now follows from
the uniqueness of the model-complete core up to isomorphism. 
\end{proof}

The following example 
shows that $O$ introduced above 
is in general 
not primitive positive definable \blue{in $(\bC_\Phi,\neq)$}. 

\begin{example}
Let $R$ and $B$ be two unary relation symbols. 
The first-order $\{R,B\}$-formula
\begin{align*}
\neg (R(x) \wedge B(y)) \\
\end{align*}
can be transformed \blue{into} an MMSNP sentence
$\Phi$ in normal form. Note that $\bC_\Phi^2$
does not embed into $\bC_\Phi$, since 
in $\bC_\Phi$ every element is either in $R$ or in $B$,
but some
elements of $\bC_\Phi^2$ are in neither $R$ nor in $B$. Let $a,b \in R^{\bC_\Phi}$ and $c \in B^{\bC_\Phi}$. Any mapping that preserves $R$ and $B$ is a homomorphism from $\bC_\Phi^2$ to $\bC_\Phi$,
so there exists an injective binary polymorphism $f$
of $\bC_\Phi$ 
that maps 
$(a,c)$ to $R^{\bC_\Phi}$ and $(b,c)$ to 
$B^{\bC_\Phi}$. But $(a,b) \in O$ and $(c,c) \in O$,
and $(f(a,c),f(b,c)) \notin O$, so $O$ is not preserved by $f$, and $O$ is not primitive positive
definable in $(\bC_\Phi,\neq)$.  
\end{example}

            \tikzset{every loop/.style={},
           every node/.style={minimum size=8pt,inner sep=0,outer
sep=0,circle, draw,thick},
           bv/.style={rectangle,fill=teal}, rv/.style={fill=magenta}}

\section{Precoloured MMSNP}
\label{sect:precoloured}
An MMSNP $\tau$-sentence $\Phi$ in normal form is called
\emph{precoloured} if, informally, for each colour
of $\Phi$ 
there is a corresponding unary relation symbol in $\tau$ that forces elements to have this colour. 
In this section we show that every MMSNP
sentence is polynomial-time equivalent
to a precoloured MMSNP sentence;
this answers a question posed in~\cite{LutzWolter}. 
We first formally introduce precoloured MMSNP and state some basic properties in Section~\ref{sect:precol-basics}. 
We then prove a stronger result than the complexity statement above:  we show that the Bodirsky-Pinsker tractability conjecture
is true for CSPs in MMSNP if and only if it is true
for CSPs in precoloured MMSNP (Theorem~\ref{thm:precol}).
In order to prove this stronger result we relate in Section~\ref{sect:neq} the algebraic properties of $\bC^\tau_\Phi$
with the algebraic properties of the expansion of $\bC^\tau_\Phi$ by the inequality relation $\neq$.
The main results are stated in Section~\ref{sect:standard-precol}. 
In Section~\ref{sect:precol-proof} we complete the proofs of the results in this section.

\subsection{Basic properties of precoloured MMSNP} 
\label{sect:precol-basics}
Formally, an MMSNP $\tau$-sentence 
$\Phi$ is 
 \emph{precoloured} if it is in normal form  and for every colour  
$M$ of $\Phi$ there exists a unary symbol $P_M \in \tau$
such that for every colour $M'$ of $\Phi$ 
which is distinct from $M$ 
the formula $\Phi$ contains 
the conjunct $\neg (P_M(x) \wedge M'(x))$. 

\begin{lemma}\label{lem:precol-snf}
Every precoloured MMSNP sentence
is in strong normal form.
\end{lemma} 
\begin{proof}
Let $\Phi$ be a precoloured MMSNP
sentence with colour set $\sigma$. 
We will show that every 
recolouring $r \colon \sigma \to \sigma$ 
of $\Phi$ must 
be the identity. 
Let $M \in \sigma$, and let $A$ be
the canonical database of 
$P_M(x) \wedge M(x)$. 
Note that $A$ does not homomorphically 
embed any coloured
obstruction of $\Phi$. 
But if $M' := r(M) \neq M$, then 
$r(A)$ homomorphically embeds 
the canonical database
of $P_M(x) \wedge M'(x)$,
in contradiction to the assumption that
$r$ is a recolouring. 
Hence, $r(M) = M$ for all $M \in \sigma$. 
\end{proof}

\blue{Finally, we prove an important property that will be used in Section~\ref{sect:alg}: the colours in a precoloured MMSNP sentence $\Phi$
denote (all) the orbits of $\Aut(\bC_\Phi^\tau)$.}

\begin{lemma}\label{lem:precol-idemp}
Let $\Phi$ be a precoloured MMSNP
sentence. Then for each colour $M$, the symbol $P_M$ and $M$ both interpret the same orbit of $\Aut(\bC_\Phi) = \Aut(\bC^\tau_\Phi)$, and each orbit is denoted by some colour $M$ of $\Phi$. 
\end{lemma}
\begin{proof}
By Lemma~\ref{lem:C-phi-mc-core}
the structure $(\bC_\Phi;\neq)$ is a model-complete core. Note that the $\omega$-categorical structures 
$(\bC_\Phi;\neq,M)$ and 
$(\bC_\Phi;\neq,P_M)$ have the same
CSP, and hence they are homomorphically equivalent. 
The fact that $\omega$-categorical model-complete cores are up to isomorphism unique then implies that $M$ and $P_M$ have the same interpretation in $\bC_\Phi$. \blue{
Since $\Phi$ is in particular in normal form,
Corollary~\ref{cor:colors-are-orbits} 
states that
$M$ and $P_M$ denote an
orbit of $\Aut(\bC_\Phi) = \Aut(\bC_\Phi^\tau)$, and that each orbit of $\Aut(\bC^\Phi)$ is denoted by some colour of $\bC^\Phi$.}
\end{proof} 

\subsection{Primitive positive constructions}
\label{sect:pp-constructions}
In order to reduce to the case of precoloured MMSNP sentences, we need the following relational counterpart of minor-preserving maps.
A \emph{pp-power} of $\bB$ is a structure with domain $B^d$, for $d \in {\mathbb N}$, whose $k$-ary relations are primitive positive definable when viewed as $dk$-ary relations over $\bB$.
Let ${\mathcal C}$ be a class of structures. 
We write 
\begin{itemize}
\item $\Ho({\mathcal C})$ for the class
of all structures that are homomorphically equivalent to structures in ${\mathcal C}$. 
\item $\PP({\mathcal C})$ for the class
of all structures obtained from structures in ${\mathcal C}$ by taking pp-powers.  
\end{itemize}
A structure $\bC$ is said to have a \emph{pp-construction} over $\bB$
if it can be obtained from $\{\bB\}$ 
by repeated applications of $\Ho$ and 
$\PP$. %
\nomenclature[030]{$\Ho({\mathcal C})$}
{the class of structures that are homomorphically equivalent to a structure from ${\mathcal C}$}
\nomenclature[031]{$\PP({\mathcal C})$}
{the class of pp-powers of structures from 
${\mathcal C}$}

\begin{lemma}[\cite{wonderland}]
\label{lem:pp-reduce}
Let $\bB$ be a relational structure with a finite relational signature. 
Then the structures with a pp-construction 
over $\bB$ are precisely the structures
in $\Ho(\PP(\{\bB\}))$. 
If $\bC \in \Ho(\PP(\{\bB\}))$
then there is a polynomial-time reduction
from $\Csp(\bC)$ to $\Csp(\bB)$. 
\end{lemma}

\begin{theorem}[\cite{wonderland}]\label{thm:wonderland}
Let $\bB$ be an $\omega$-categorical structure. Then the following are equivalent. 
\begin{itemize}
\item $\Ho(\PP(\bB))$ contains the complete graph on 3 vertices. 
\item $\Pol(\bB)$ has a uniformly continuous minor-preserving map to $\Pr$. 
\end{itemize} 
\end{theorem} 

We will also need the following consequence of results from~\cite{wonderland}.
 
\begin{proposition}\label{prop:uch1}
Let $\bB$ and $\bC$ be at most countable $\omega$-categorical structures  that are homomorphically equivalent. Then there is a uniformly continuous minor-preserving map from $\Pol(\bB)$ to $\Pol(\bC)$. 
\end{proposition}
\begin{proof}
Let $h_1\colon\bC\to\bB$ and $h_2\colon \bB\to\bC$ be homomorphisms.
Let $\xi\colon\Pol(\bB)\to\Pol(\bC)$ be the minor-preserving map $f\mapsto h_2 f(h_1,\dots,h_1)$.
It is clearly uniformly continuous: for any finite $C' \subseteq C$, if two $n$-ary functions $f, g \in \mathscr B$
agree on the finite set $h_1(C') \subseteq B$, then $\xi(f)$ and $\xi(g)$ agree on $C'$. 
\end{proof}

If $\bB$ is an $\omega$-categorical model-complete core, then any expansion of $\bB$ by finitely many unary singleton relations is pp-constructible in $\bB$~\cite{wonderland}.

\subsection{Adding inequality}
\label{sect:neq}
Let $\Phi$ be an MMSNP sentence in normal form with input signature $\tau$. We first show that adding the inequality relation to $\bC^\tau_\Phi$ does not increase the complexity of its CSP.  

\begin{proposition}\label{prop:neq}
$\Csp(\bC^\tau_\Phi)$ and $\Csp(\bC^\tau_\Phi,\neq)$ are polynomial-time equivalent.
\end{proposition}
\begin{proof}
If a given instance of $\Csp(\bC^\tau_\Phi,\neq)$,
viewed as a primitive positive sentence, 
contains conjuncts of the form $x \neq x$, then the 
instance is unsatisfiable. Otherwise, we only consider
the constraints using relations from $\tau$, 
and let $\bA$ be the canonical database of those constraints. If $\bA$ has no homomorphism to $\bC^\tau_\Phi$ then the instance is unsatisfiable. Otherwise, by Lemma~\ref{lem:c-basic} there is an 
injective homomorphism from $\bA$ to $\bC^\tau_\Phi$.
The injectivity implies that the homomorphism also 
satisfies all the inequality constraints, so we
have a polynomial-time reduction from $\Csp(\bC^\tau_\Phi,\neq)$ to $\Csp(\bC^\tau_\Phi)$. 
\end{proof}

We would now like to prove that 
$\bC^\tau_\Phi$ satisfies Conjecture~\ref{conj:inf-dichotomy}
if and only if $(\bC^\tau_\Phi,\neq)$ does. 
We do not know whether there exists a uniformly continuous minor-preserving map
$\Pol(\bC^\tau_\Phi)\to\Pol(\bC^\tau_\Phi,\neq)$ but we can prove the following, which turns out to be sufficient.

\begin{proposition}\label{prop:neq-is-algebraic}
    There exists a uniformly continuous minor-preserving map $\Pol(\bC^\tau_\Phi)\to\Projs$ if, and only if,
    there exists a uniformly continuous minor-preserving map $\Pol(\bC^\tau_\Phi,\neq)\to\Projs$.
\end{proposition}

\red{A formula is called \emph{equality free}  if it does not use the equality symbol = of first order logic.} 
In the proof of this \red{proposition}, we need the following fact.
 
\begin{lemma}\label{lem:construction-of-injective-maps}
   Let $\bA$ be any structure that has a homomorphism $g$ to
$\bB^{\ind}_{\mathcal F}$.
   Then there exists an injective homomorphism $h\colon\bA\to\bB^{\ind}_{\mathcal F}$
    such that for all tuples $\tuple a$ from $\bA$
    and all equality-free existential formulas $\phi$,
    if $\phi(g(\tuple a))$ holds in $\bB^{\ind}_{\mathcal F}$,  then $\phi(h(\tuple a))$ also holds in $\bB^{\ind}_{\mathcal F}$.
   Moreover, for all tuples $\tuple a,\tuple b$ with pairwise distinct entries from $\bA$,
   if $g(\tuple a)$ and $g(\tuple b)$ lie in the same orbit in $\Aut(\bB^{\ind}_{\mathcal F})$ then $h(\tuple a)$ and $h(\tuple b)$ lie in the same orbit in $\Aut(\bB^{\ind}_{\mathcal F})$.
    \end{lemma}
\begin{proof}
   Assume first that $\bA$ is finite with domain $A$.
   Build a new structure $\bA'$ as follows.
   For every $\tuple a$ in $\bA$ and existential formula $\phi(\tuple
x):=\exists y_1,\dots,y_s. \psi(\tuple x,\tuple y)$ such that
   $\bB^{\ind}_{\mathcal F}\models \phi(g(\tuple a))$ holds, pick
elements $b_1,\dots,b_s$ of $\bB^{\ind}_{\mathcal F}$ such that $\bB^{\ind}_{\mathcal F}\models\psi(g(\tuple
a),b_1,\dots,b_s)$.
   Let $A'$ be the set consisting of $A$ 
   as well as new elements $a'_1,\dots,a'_s$,  
   and define $g(a'_i):=b_i$.
   Let $\bA'$ be the $(\tau\cup\sigma)$-structure on $A'$ obtained by pulling back the
relations from the structure induced by $g(A')$ in $\bB^{\ind}_{\mathcal F}$.
   We therefore have that $g$ is a homomorphism $\bA' \to
\bB^{\ind}_{\mathcal F}$.
   It follows that there exists an embedding $h\colon \bA' \to
\bB^{\ind}_{\mathcal F}$.

We prove the first part of the statement.
Let $\phi(\tuple x):=\exists y_1,\dots,y_s. \psi(\tuple x,\tuple y)$
be an equality-free existential formula. 
   Assume that $\bB^{\ind}_{\mathcal F}\models \phi(g(\tuple a))$.
    By construction and the fact that $\phi$ is equality free, this is equivalent to $\bA'\models \psi(\tuple
a,a'_1,\dots,a'_s)$ for some elements $a'_1,\dots,a'_s\in A'$.
  Since $h$ is an embedding, this implies
   $\bB^{\ind}_{\mathcal F}\models \exists y_1,\dots,y_s. \psi(h(\tuple a),\tuple y)$, i.e., $\phi(h(\tuple a))$ holds in $\bB^{\ind}_{\mathcal F}$.

   We now prove the second part of the statement.
   Let $\tuple a,\tuple b$ be tuples with pairwise injective entries from $\bA$
   such that $g(\tuple a)$ and $g(\tuple b)$ are in the same orbit in $\bB^{\ind}_{\mathcal F}$.
   Since $\bB^{\ind}_{\mathcal F}$ is $\omega$-categorical and by Theorem~\ref{thm:ryll}, the orbit of
the tuple $g(\tuple a)$ has a first-order definition $\phi(\tuple x)$.
   Since $\bB^{\ind}_{\mathcal F}$ is model-complete and $\phi$ defines an orbit, we can assume
that $\phi$ is existential without disjunctions, of the form $\exists
y_1,\dots,y_s \, \big(\psi_1(\tuple x,\tuple y))\land\psi_2(\tuple x)\big)$
   with $\psi_1$ equality free and quantifier free, and $\psi_2$ a conjunction of literals of the form $x_i\neq x_j$. 
   Since $h$ is injective and the tuples $\tuple a$ and $\tuple b$ are injective, $\psi_2(h(\tuple a))$ and $\psi_2(h(\tuple b))$
   hold.
    Moreover, since $\psi_1$ is equality free, the previous paragraph gives us that both $\exists y_1,\dots,y_s. \psi_1(h(\tuple a),\tuple y)$
    and $\exists y_1,\dots,y_s. \psi(h(\tuple b),\tuple y)$ hold.
    \blue{Therefore, $h(\tuple a)$ and $h(\tuple b)$ lie in the same orbit of $\Aut(\bB^{\ind}_{\mathcal F})$.}

	In case $\bA$ is infinite, it suffices to apply a compactness argument using the statement for finite substructures of $\bA$.
\end{proof}

\begin{proof}[Proof of Proposition~\ref{prop:neq-is-algebraic}]
    Let $K_3$ be the clique on $\{R,G,B\}$.
    We prove that $K_3$ is pp-constructible in $\bC^\tau_\Phi$ if,
    and only if, it is pp-constructible in $(\bC^\tau_\Phi,\neq)$. By Theorem~\ref{thm:wonderland}, this is equivalent to the statement in Proposition~\ref{prop:neq-is-algebraic}.
    Suppose then that $K_3$ is homomorphically equivalent to a pp-power of $(\bC^\tau_\Phi,\neq)$,
    and let $\bA$ be such a pp-power with minimal exponent $d$.
    Let $\phi_E(\tuple x,\tuple y)$ be the defining primitive positive formula of the edge relation of $\bA$.
    
   \red{We first show that $\phi_E$ does not contain conjuncts of the form $x=y$. One may suppose that an existentially quantified variables $x$ does not appear 
    in such a conjunct of $\phi_E$ since otherwise one could simply substitute all occurrences of $x$ by $y$ and then erase $x$ from the formula (or, if $x$ and $y$ are the same variable, drop the conjunct).} 
    Suppose that $x_i=x_j$ is in $\phi_E$ for some $i\neq j$, and for ease of notation assume that $j=d$.
    Then one could obtain a pp-power $\bA'$ with exponent $d-1$ by existentially quantifying $x_d$ and $y_d$.
    Simply observe that $\bA$ and $\bA'$ are homomorphically equivalent, with the homomorphisms being $h_1\colon (a_1,\dots,a_d)\mapsto (a_1,\dots,a_{d-1})$
    and $h_2\colon (a_1,\dots,a_{d-1})\mapsto (a_1,\dots,a_{d-1},a_i)$.
    Similarly, we obtain that $\phi_E$ does not contain $y_i=y_j$ for $i\neq j$.
    Suppose now that $\phi_E$ contains $x_i=y_j$. If $i=j$ (and again for simplicity, assume $i=j=d$), let $\bA'$ be the pp-power (with exponent $d-1$) obtained by existentially quantifying $x_d$ and $y_d$.
    Then $\bA'$ and $\bA$ are homomorphically equivalent, with the homomorphisms being $h_1\colon \bA\to\bA'$ as above and $h_3\colon (a_1,\dots,a_{d-1})\mapsto (a_1,\dots,a_{d-1},b)$,
    where $b$ is an arbitrary element of $\bC_\Phi$ that belongs to a $d$-tuple incident to an edge in $\bA$.
    Finally, consider the case that $\phi_E$ contains $x_i=y_j$ with $i\neq j$.
    Since there is a homomorphism from $K_3$ to $\bA$, there exist $\tuple a,\tuple b,\tuple c\in\bA$ such that $\phi_E$ holds for all 6 pairs of distinct tuples from $\{\tuple a,\tuple b,\tuple c\}$.
    In particular, $\phi_E(\tuple a,\tuple b)$, $\phi_E(\tuple c,\tuple b),$ and $\phi_E(\tuple c,\tuple a)$ hold in $(\bC_\Phi^\tau,\neq)$.
    We conclude that $a_i=b_j=c_i=a_j$, and similarly that $b_i=b_j$ and $c_i=c_j$.
    Therefore, one could add the literal $x_i=x_j$ to $\bA$ to obtain a subgraph of $\bA$ that still contains a triangle, and is therefore homomorphically equivalent to $K_3$.
    By the argument above, we could then find a suitable pp-power with a lower exponent, a contradiction to the minimality of $d$.

    Let $\psi_E$ be the formula $\phi_E$ where all conjuncts of the form $x=y$ have been removed (note that a literal $x\neq x$ cannot appear,
    for otherwise the edge relation of $\bA$ is empty, and $K_3$ would not have a homomorphism to $\bA$).
    Let $\bB$ be the structure defined by $\psi_E$ in $\bC^\tau_\Phi$. Observe that $\bB$ contains all the edges of $\bA$, so $\bB$ contains a triangle.

\medskip 
    \underline{Claim:} $\bB$ does not contain any loop.\\
    \textsc{Proof.} Suppose the contrary, and let $\tuple a\in B$ be such that $\bC^\tau_\Phi\models \psi_E(\tuple a,\tuple a)$.
    Let $D=\{b_1,\dots,b_d,c_1,\dots,c_d\}$ be a set with \new{$2d$ elements}.
    Let $g\colon b_i,c_i\mapsto a_i$ for all $i\in\{1,\dots,d\}$.
    Let $\bD$ be the $(\tau\cup\sigma)$-structure on $D$ obtained by pulling back the relations from the structure induced by $g(D)$ in $\bC_\Phi$.
    Note that all the elements of $\bD$ are coloured.
    By Lemma~\ref{lem:construction-of-injective-maps}, there is an injective homomorphism $g'\colon\bD\to\bB^{\ind}_{\mathcal F}$
    with the additional property that $g'(\tuple b)$ and $g'(\tuple c)$ are in the same orbit in $\bB^{\ind}_{\mathcal F}$, because $g(\tuple b)$
    and $g(\tuple c)$ are in the same orbit (they are actually equal).
    By composing with an appropriate $\alpha\in\Aut(\bB^{\ind}_{\mathcal F})$ we may assume that $g'(\tuple b)$ and $g'(\tuple c)$ are in the same
    orbit in $(\bB^{\ind}_{\mathcal F},<)$.
Compose with an injective homomorphism $h\colon\bB^{\ind}_{\mathcal F}\to\bB^{\hom}_{\mathcal F}$
    that is canonical from $(\bB^{\ind}_{\mathcal F},<)$ to $(\bB^{\hom}_{\mathcal F},<)$ to get an injective homomorphism $g''\colon\bD\to\bB^{\hom}_{\mathcal F}$
    such that $g''(\tuple b)$ and $g''(\tuple c)$ are in the same orbit in $(\bB^{\hom}_{\mathcal F},<)$.
    Note that all the elements of the image of $g''$ are coloured, because all the elements of $\bD$ are coloured. So the image of $g''$ lies in $\bC_\Phi$. We prove that $\phi_E(g''(\tuple b),g''(\tuple c))$ holds in $\bC^\tau_\Phi$. Indeed, $\bC^\tau_\Phi\models\psi_E(g(\tuple b),g(\tuple c))$.
    By Lemma~\ref{lem:construction-of-injective-maps} and since $\psi_E$ is equality free, $(g'(\tuple b),g'(\tuple c))$ satisfies $\psi_E$.
    This implies that $\bB^{\hom}_{\mathcal F}\models \psi_E(g''(\tuple b),g''(\tuple c))$ and by the injectivity of $g''$ we have
    that $\bC^\tau_\Phi\models\phi_E(g''(\tuple b),g''(\tuple c))$ holds. 
    
    Let $\chi\colon\bA\to K_3$ be a homomorphism.  By Corollary~\ref{cor:ramsey} and Theorem~\ref{thm:can-4}, we may assume that $\chi$ is diagonally canonical from $(\bC_\Phi,<)$ to $(K_3,R,G,B)$. 
    Since $\chi$ is \red{diagonally} canonical, we have that $\chi(g''(\tuple b)) = \chi(g''(\tuple c))$. This contradicts the fact that $\chi$ is a homomorphism $\bA \to K_3$.
    Therefore, $\bB$ has no loops. $\lozenge$
	\medskip
	
    We now prove that every finite substructure $\bS$ of $\bB$ has a homomorphism to $K_3$, which proves by compactness that $\bB$ has a homomorphism to $K_3$. Let $\tuple s^1=(s^1_1,\dots,s^1_d),\dots,\tuple s^K=(s^K_1,\dots,s^K_d)$ be a list of the elements of $\bS$. 
    Let $\theta(\tuple x^1,\dots,\tuple x^K)$ be the formula with $Kd$ free variables that is a conjunction of
        the formulas $\psi_E(\tuple x^i,\tuple x^j)$ for all $i,j\in\{1,\dots,K\}$ such that $\bC^\tau_\Phi\models\psi_E(\tuple s^i,\tuple s^j)$.
        \new{Since this formula is satisfiable in $\bC^\tau_\Phi$, it is also satisfiable in $\bC^\tau_\Phi$ by an injective assignment $g$.}
    Let $\tuple t^i:=(g(x^i_1),\dots,g(x^i_d))$.
    Let $\mathfrak T$ be the structure induced by $\{\tuple t^1,\dots,\tuple t^K\}$ in $\bB$.
    \red{The map that sends $\tuple s^i$ to $\tuple t^i$ is a homomorphism from $\bS$ to $\bT$.}
    If $\bC^\tau_\Phi\models \psi_E(\tuple t^i,\tuple t^j)$, then $i\neq j$ because $\bB$ has no loops.
    By the injectivity of $g$, it follows that $\bC^\tau_\Phi\models\phi_E(\tuple t^i,\tuple t^j)$.
    Therefore, $\mathfrak T$ is a subgraph of $\bA$, which homomorphically maps to $K_3$. \red{Composing homomorphisms,} we obtain a homomorphism $\bS\to K_3$.
 
    Thus, $K_3$ is homomorphically equivalent to a pp-power of $\bC^\tau_\Phi$.
\end{proof}

\subsection{The standard precolouration}
\label{sect:standard-precol}
Let  $\Phi$ be an MMSNP sentence in strong normal form
with colour set $\sigma$, and let $\Psi$
be the following precolored MMSNP sentence: 
we obtain $\Psi$ from $\Phi$ by adding 
for each $M \in \sigma$ a new input
predicate $P_M$ 
and adding the conjunct 
$\neg (P_M(x) \wedge M'(x))$ for each colour $M' \in \sigma \setminus \{M\}$.
We call this sentence the \emph{standard precolouration of $\Phi$}.

\begin{theorem}[Precolouring Theorem]\label{thm:precol}
\blue{Let $\Phi$ be an MMSNP sentence in strong normal form with input signature $\tau$.}
Let $\Psi$ be the standard precolouration of $\Phi$ and let $\rho$ be the input signature of $\Psi$.
Then:
\begin{itemize}
\item $\bC^{\rho}_\Psi$ is pp-constructible in $(\bC^\tau_\Phi,\neq)$,
\item $\bC^\tau_\Phi$ is pp-constructible in $\bC^{\rho}_\Psi$ (in fact, $\bC^\tau_\Phi$ is isomorphic to a reduct of $\bC^{\rho}_\Psi$).
\end{itemize}
Moreover, there exists a uniformly continuous minor-preserving map $\Pol(\bC_\Phi^{\tau})\to\Projs$ if,
and only if, there exists a uniformly continuous minor-preserving map $\Pol(\bC_\Psi^{\rho})\to\Projs$.
\end{theorem}

The proof of this theorem will be given in Section~\ref{sect:precol-proof}. 
We first point out that Theorem~\ref{thm:precolouring-intro} from the introduction is an immediate consequence. 

\begin{corollary}\label{cor:precol}
Let $\Phi$ be an MMSNP sentence in strong normal form and let $\Psi$ be its standard precoloration. 
Then $\Phi$ and $\Psi$ describe polynomial-time equivalent problems.
\end{corollary}
\begin{proof}
    It is clear that the problem described by $\Phi$ reduces to the problem described by $\Psi$.
    We now prove that there is a polynomial-time reduction in the other direction. Let $\tau$ and $\rho$ be the input signatures of $\Phi$ and $\Psi$.
    Since $\bC^{\rho}_\Psi$ is pp-constructible in $(\bC^\tau_\Phi,\neq)$ \blue{by Theorem~\ref{thm:precol}}, 
    we have that $\Csp(\bC^{\rho}_\Psi)$ reduces in polynomial-time to
    $\Csp(\bC^\tau_\Phi,\neq)$, by Lemma~\ref{lem:pp-reduce}. 
    Moreover, by Proposition~\ref{prop:neq}, 
    there is a polynomial-time reduction from 
    $\Csp(\bC^\tau_\Phi,\neq)$ to $\Csp(\bC^\tau_\Phi)$.
    Therefore, $\Csp(\bC^\rho_\Psi)$ reduces to $\Csp(\bC^\tau_\Phi)$.
\end{proof}

\subsection{Proof of the precolouring theorem}
\label{sect:precol-proof}
Let $\bA$ be a properly coloured $(\tau\cup\sigma)$-structure, i.e., every element appears in the interpretation of precisely one symbol from $\sigma$.
For an element $a\in A$, denote by $\bA[a\mapsto *]$ the structure obtained by uncolouring $a$.
For $M\in\sigma$ and a tuple $\tuple a$ of elements of $\bA$, denote by $\bA[\tuple a\mapsto M]$ the structure obtained by uncolouring the elements of $\tuple a$ 
and giving them the colour $M$.
\nomenclature[032]{$\bA[a\mapsto *]$}{the structure obtained from $\bA$ by
  uncolouring $a$}
\nomenclature[033]{$\bA[\tuple a\mapsto M]$}{the structure obtained from $\bA$ by uncolouring the elements of $\tuple a$ and giving them the colour $M$}
Let $C(\bA,a)$ be the subset of $\bC_\Phi$ containing all elements $c$
such that there exists a homomorphism $$h\colon\bA[a\mapsto
*]\to\bC_\Phi$$ that satisfies $h(a)=c$.
Note that $C(\bA,a)$ is, by 1-homogeneity of $\bC_\Phi$, a union of colours. So we can also see $C(\bA,a)$ as the union of $M^{\bC_\Phi}$ for $M\in\sigma$ such that $\bA[a\mapsto M]$ is $\mathcal F$-free,
where $\mathcal F$ is the coloured obstruction set of $\Phi$.
\nomenclature[034]{$C(\bA,a)$}{the subset of $\bC_\Phi$ that consists of colours $M$ of
  $\sigma$ such that $\bA[a\mapsto M]$ satisfies the first-order part
  of $\Phi$}

\begin{example}
{Consider the MMSNP sentence $\Phi$ whose obstructions are the two monochromatic triangles of Figure~\ref{fig:ex-precol}.
Let $\bA$ be the triangle with the blue square vertices and let $a$ be any vertex thereof. Then one sees that $C(\bA,a)$ consists precisely of the red round vertices of $\bC_\Phi$.
Indeed, any homomorphism from $\bA[a\mapsto *]$ to $\bC_\Phi$ has to map $a$ to a red vertex, because $\bC_\Phi$ does not contain blue triangles.}
\end{example}

\begin{lemma}\label{lem:colours-as-intersection}
    Suppose that $\Phi$ is in strong normal form and let $M$ be a colour of $\Phi$.
    Then $M^{\bC_\Phi} = \bigcap C(\mathfrak F, a)$ where the intersection ranges over all $\bF\in\mathcal F$ and $a\in F$ such that $M^{\bC_\Phi}\subseteq C(\bF,a)$.
\end{lemma}
\begin{proof}
    The left-to-right inclusion is immediate, by definition.
    We prove the other inclusion. To do this, it suffices to show that for every $M'\in\sigma\setminus\{M\}$,
    there exists $\mathfrak G\in\mathcal F$ and $b\in G$ such that $M^{\bC_\Phi}\subseteq C(\mathfrak G,b)$ but $(M')^{\bC_\Phi}\not\subseteq C(\mathfrak G,b)$.
    Let $r\colon\sigma\to\sigma$ be defined by $r(M)=M'$ and $r(N)=N$ for all $N\in\sigma\setminus\{M\}$.
    Since $\Phi$ is in strong normal form and $r$ is not surjective, $r$ cannot be a recolouring of $\Phi$.
    This means that there exists a $\mathcal F$-free structure $\bA$ and $\bF\in\mathcal F$ such that there exists a homomorphism
    $h\colon\bF\to r(\bA)$.
    Let $a_1,\dots,a_k$ be the elements of $\mathfrak F$ that are mapped to $M^{\bA}$ by $h$.
    In $r(\bA)$, these elements are in $M'$, so since $h$ is a homomorphism and $\bF$ is completely coloured, we have that $a_1,\dots,a_k\in (M')^\bF$.
    Moreover, since $\bA$ is $\mathcal F$-free, the structure $\bF[a_1,\dots,a_k\mapsto M]$ is $\mathcal F$-free.
    Let $0\leq j\leq k$ be minimal such that $\bF[a_1,\dots,a_j\mapsto M]$ is $\mathcal F$-free.
    Since $\bF\in\mathcal F$, we have $j\geq 1$.
    Let now $\mathfrak G\in\mathcal F$ be such that there exists $g\colon \mathfrak G\to\bF[a_1,\dots,a_{j-1}\mapsto M]$, which exists by minimality of $j$.
    Note that $a_j$ is in the image of $g$, otherwise $g$ would be a homomorphism $g\colon \mathfrak G\to\bF[a_1,\dots,a_j\mapsto M]$, in contradiction to the choice of $j$.
    Thus, let $b\in G$ be such that $g(b)=a_j$, and note that $b\in (M')^{\mathfrak G}$, so that $(M')^{\bC_\Phi}\not\subseteq C(\mathfrak G,b)$.
    Since $g$ is a homomorphism $\mathfrak G[b\mapsto M]\to \bF[a_1,\dots,a_j\mapsto M]$, the structure $\mathfrak G[b\mapsto M]$ is $\mathcal F$-free.
    This implies that $M^{\bC_\Phi}\subseteq C(\mathfrak G,b)$.
    We therefore found a $\mathfrak G\in\mathcal F$ and $b\in G$ such that $M^{\bC_\Phi}\subseteq C(\mathfrak G,b)$ but $(M')^{\bC_\Phi}\not\subseteq C(\mathfrak G,b)$.
\end{proof}

For $M\in\sigma$, let $P(M)$ be the set of pairs $(\bF,a)$ with $\bF\in\mathcal F$, $a\in F$, and such that $M^{\bC_\Phi}\subseteq C(\bF,a)$.
{For each colour $M$ and each $n\geq 0$, we define a tree-like $(\tau\cup\sigma)$-structure $\bT_{M,n}$ with a designated root $r$ and leaves as follows (the construction
is illustrated in Figure~\ref{fig:chi-formula}).
For $n=0$, we define $\bT_{M,0}$ to be the disjoint union of all structures $\bF[a_F\mapsto M]$ such that $(\bF,a_F)\in P(M)$, where we then identify all the vertices $a_F$ into a single vertex $r$.
The root of $\bT_{M,0}$ is defined to be $r$ and the other vertices are leaves.
For $n>0$, we define $\bT_{M,n}$ to be $\bT_{M,0}$, to which for each leaf $b$ of colour $M'$, we add a fresh copy of $\bT_{M',n-1}$ whose root we glue at $b$.
The leaves of $\bT_{M,n}$ are defined to be the leaves of all the added copies of $\bT_{M',n-1}$.
We let $\overline{\bT_{M,n}}$ be the structure obtained by uncolouring all the non-leaves of $\bT_{M,n}$.}

\begin{lemma}\label{lem:gadget-as-colour}
 Let $n \in {\mathbb N}$, $M\in\sigma$.
 There exists a homomorphism $\bT_{M,n}\to \bC_\Phi$, and every homomorphism $h\colon\overline{\bT_{M,n}}\to\bC_\Phi$ satisfies $h(r)\in M^{\bC_\Phi}$.
\end{lemma}
\begin{proof}
   The fact that there exists a homomorphism as in the statement follows immediately from Corollary~\ref{cor:nf},
   given that each $\bT_{M,n}$ is obtained by gluing structures that are $\mathcal F$-free on vertices of corresponding colours.
   
   We prove the second part of the statement by induction.
    Suppose $n=0$.
    We have $C(\bT_{M,0},r) \subseteq \bigcap_{(\bF,a)\in P(M)} C(\bF,a) = M^{\bC_\Phi}$, where the inclusion follows from the fact that $\bT_{M,0}$ contains a copy of $\bF[a\mapsto M]$
    and the equality follows from Lemma~\ref{lem:colours-as-intersection}. 
    Suppose now that the result is proved for some $n>0$. A homomorphism $h$ from $\overline{\bT_{M,n}}$ to $\bC_\Phi$ induces homomorphisms
    $\overline{\bT_{M',n-1}}\to\bC_\Phi$
    for all the copies of $\bT_{M',n-1}$ contained in $\bT_{M,n}$. Therefore, the induction hypothesis implies that for each such copy of $\bT_{M',n-1}$ with root $r$,
    we have $h(r)\in (M')^{\bC_\Phi}$.
    Since the leaves of $\bT_{M,0}$ are the roots of the copies of $\bT_{M',n-1}$,
     $h$ induces a homomorphism $\overline{\bT_{M,0}}\to \bC_\Phi$, which by the base case of the induction maps the root $r$ to $M^{\bC_\Phi}$, as required.
\end{proof}

\begin{figure}
\begin{center}
\begin{tikzpicture}
    \node[rv] (x) at (1,0) {};
    \node[bv] (y1) at (0.443,-1) {};
    \node[bv] (y2) at (1.557,-1) {};

    \draw (x) -- (y1) -- (y2)  -- (x);
    \end{tikzpicture}%
    \hspace{0.4cm}
   \begin{tikzpicture}
    \node[bv] (x) at (1,0) {};
    \node[rv] (y1) at (0.443,-1) {};
    \node[rv] (y2) at (1.557,-1) {};

    \draw (x) -- (y1) -- (y2)  -- (x);
    \end{tikzpicture}%
    \hspace{0.4cm}
\begin{tikzpicture}
    \node[rv] (x) at (1,0) {};
    \node[bv] (y1) at (0,-1) {};
    \node[bv] (y2) at (2,-1) {};
   
    \foreach \i in {0,...,3} 
        \node[rv] (z\i) at ({(\i-0.3)/1.3},-2) {};

    \draw (x) -- (y1) -- (y2)  -- (x);
    \draw (y1) -- (z0) -- (z1) -- (y1);
    \draw (y2) -- (z2) -- (z3) -- (y2);
    \end{tikzpicture}%
    \hspace{0.4cm}
\begin{tikzpicture}
    \node[label={above:$a$}] (x) at (1,0) {};
    \node[label={[label distance=5pt]left:$b$}] (y1) at (0,-1) {};
    \node[label={[label distance=5pt]right:$c$}] (y2) at (2,-1) {};
   
    \foreach \i in {0,...,3} 
        \node[rv] (z\i) at ({(\i-0.3)/1.3},-2) {};

    \draw (x) -- (y1) -- (y2)  -- (x);
    \draw (y1) -- (z0) -- (z1) -- (y1);
    \draw (y2) -- (z2) -- (z3) -- (y2);
    \end{tikzpicture}

\caption{Illustration of the construction of the structures $\bT_{M,1}$, $\bT_{B,1}$, $\bT_{M,2}$, and $\overline{\bT_{M,2}}$, in the case of the MMSNP sentence given by the obstructions in Figure~\ref{fig:ex-precol},
with monadic symbols $M$ and $B$ (depicted in magenta/round and blue/square).
The structures are shown with the leaves at the bottom and the root at the top.
    Note that if $\bF$ is the triangle with $B$-coloured vertices and $a$ is any vertex of this triangle then $P(M)= \{(\bF,a)\}$.
    Thus, $\bT_{M,1}$ is simply $\bF[a\mapsto M]$, with root $a$, and $\bT_{B,1}$ is constructed similarly by exchanging $B$ and $M$.
    By gluing two copies of $\bT_{B,1}$ at the leaves of $\bT_{M,1}$, we obtain $\bT_{M,2}$.
    Note that any homomorphism from $\overline{\bT_{M,2}}$ to $\bC_\Phi$ has to map $b$ and $c$ to $B^{\bC_\Phi}$, and therefore has to map $a$ to $M^{\bC_\Phi}$.
}
\label{fig:chi-formula}
\end{center}
\end{figure}

{Let $n>|\Phi|$. It is a consequence of Lemma~\ref{lem:gadget-as-colour}
 that for each $M\in\sigma$, the structure $\overline{\bT_{M,n}}$ has a homomorphism to $\bC_\Phi$.
 By Lemma~\ref{lem:c-basic}, it also injectively maps to $\bC_\Phi$. Let $c_1,\dots,c_k$ be the images of the leaves of $\overline{\bT_{M,n}}$ under this homomorphism.
We use different constant symbols for each colour $M$ and let $\tuple c$ be the list of all the constant used in this construction.
 Therefore, each leaf of each $\bT_{M,n}$ is associated with a unique element from $\bC_\Phi$ and a unique constant symbol.}

\begin{proposition}\label{prop:universality}
    Let $\bA$ be a finite $(\tau\cup\sigma)$-structure with a homomorphism to $\bC_\Phi$. Let $n>|\Phi|$.
    Let $\bA'$ be the structure obtained by gluing to every \red{$a \in M^{\bA}$} the structure $\bT_{M,n}$,
    followed by identifying the corresponding leaves of different copies of $\bT_{M,n}$.
    Then there is a homomorphism $h\colon \bA'\to\bC_\Phi$ that maps the leaves of $\bA'$ to $\tuple c$.
\end{proposition}

See Figure~\ref{fig:prop-universality} for an illustration.

\begin{proof}
	Without loss of generality we may assume that $\bA$ is fully coloured.
    Let $\psi(\tuple x)\land \theta(\tuple x)$ be a formula describing the orbit of the tuple $\tuple c$ in $\bB^{\ind}_{\mathcal F}$ where \blue{$\psi(\tuple x)$} is a primitive positive formula
    in the \red{signature} of $\bB^{\ind}_{\mathcal F}$ and $\theta(\tuple x)$ is a conjunction of negated atomic formulas (that such a formula exists is a consequence of 
    Lemma~\ref{lem:define-orbits}).
    Let $\tuple d$ be the tuple of all the leaves in $\bA'$.
    Let $\bA''$ be the $(\tau\cup\sigma)$-structure obtained 
   by gluing the canonical database of $\psi(\tuple x)$ to the elements of $\tuple d$. 
    We show that $\bA''$ has a homomorphism to $\bC_\Phi$ that takes each $d_i$ to $c_i$, for all $i$.
    
We first give an embedding of $\bA''$ into $\bB^{\ind}_{\mathcal F}$, by showing that $\bA''$ is $\mathcal F$-free.
    Suppose that some $\mathfrak F\in\mathcal F$ maps to $\bA''$ under some homomorphism $h$.
    Since $\Phi$ is in normal form, $\mathfrak F$ is connected and therefore the image of $h$ is a connected substructure of $\bA''$ of size at most $|\mathfrak F| < n$.
    Note that such a substructure of $\bA''$ can be of three types: contained in $\bA$, contained in $\bA''\setminus \bA$, or not biconnected.
    The image of $h$ cannot fall in the first case, since $\bA$ is fully coloured and maps to $\bC_\Phi$, and is therefore $\mathcal F$-free.
    In the second case, the image of $h$ would be contained in a gluing of several copies of $\bT_{M,n}$ for the same $M\in\sigma$, together with the canonical database of $\psi(\tuple x)$.
    Since such a structure homomorphically maps to a single copy of $\bT_{M,n}$ glued with the canonical database of $\psi(\tuple x)$,
    we get a contradiction to the fact that $\bT_{M,n}$ has a homomorphism to $\bC_\Phi$ (Lemma~\ref{lem:gadget-as-colour}) that additionally maps the leaves to $\tuple c$ (by the very choice of $\tuple c$).
    In the third case, we apply Corollary~\ref{cor:nf} and obtain a contradiction.
    Thus, $\bA''$ has an embedding $e$ to $\bB^{\ind}_{\mathcal F}$.
     \begin{figure}
     \begin{center}
\begin{tikzpicture}[vertex/.style = {minimum size=8pt,inner sep=0,outer sep=0,circle, draw,thick,fill=white}]
    
    \draw[dashed] (-2,0.3) ellipse (3cm and 0.5 cm);
    \node[rv,label={above:$x_j$}] (x) at (0.3,0) {};
    
    \node[bv] (y1) at (-0.9,-1) {};
    \node[bv] (y1) at (-0.9,-1) {};
    \node[rv] (z0) at (-1.4,-2) {};
    \node[rv] (z1) at (-0.5,-2) {};
     \node[rv] (z0) at (-1.4,-2) {};
    \node[rv] (z1) at (-0.5,-2) {};
    \node[bv] (y2) at (1.5,-1) {};
    \node[bv] (y2) at (1.5,-1) {};

     \node[rv] (z2) at (1,-2) {};
    \node[rv] (z3) at (2,-2) {};

    \draw (x) -- (y1) -- (y2)  -- (x);
    \draw (y1) -- (z0) -- (z1) -- (y1);
    \draw (y2) -- (z2) -- (z3) -- (y2);

    \node[rv,label={above:$x_i$}] (x2) at (-3,-0.2) {};
    \node[bv] (y12) at (-3.9,-1) {};
    \node[bv] (y22) at (-2.5,-1) {};
    \draw (x2) -- (y12) -- (y22) -- (x2);
    \draw (y12) edge (z0) edge (z1);
    \draw (y22) edge (z2) edge (z3);
    
     \end{tikzpicture} \end{center}
     \caption{Depiction of $\bA'$, in case $x_i$ and $x_j$ are two vertices of $\bA$ coloured by $M$.}
     \label{fig:prop-universality}
     \end{figure}
   By construction, $e(\tuple d)$ satisfies $\psi$ and also $\theta$ since $e$ is an embedding.
     It follows that $e(\tuple d)$ is in the same orbit as the elements $\tuple c$ from $\bC_\Phi$.
    Let $g$ be any injective homomorphism $\bB^{\ind}_{\mathcal F}\to\bC_\Phi$.
        The restriction of $g$ to $\bC_\Phi\subseteq\bB^{\ind}_{\mathcal F}$ is an embedding, since $(\bC_\Phi,\neq)$ is a model-complete core.
    Therefore, $(g\circ h)(\tuple d)$ and $\tuple c$ are in the same orbit, and without loss of generality we can assume that $(g\circ h)(\tuple d)=\tuple c$.
    In conclusion, $g\circ h$ is a homomorphism from $\bA'$ to $\bC_\Phi$ that maps $\tuple d$ to $\tuple c$.
\end{proof}
Finally, let \red{$\chi_M(r)$} be the primitive positive formula (in the expansion of $\tau$ by the constant symbols) that is obtained by:
\begin{itemize}
\item uncolouring all the vertices of $\bT_{M,n}$;
\item labelling each leaf of $\bT_{M,n}$ by the corresponding constant symbol;
\item taking the canonical query of the resulting structure;
\item \red{existentially quantifying all variables in the query except for the root $r$}. 
\end{itemize}

\begin{proof}[Proof of Theorem~\ref{thm:precol}]
We first show that $\bC^{\rho}_\Psi$ is pp-constructible in $(\bC^\tau_\Phi,\neq)$.
Let $\bD$ be the expansion with signature $\rho$ of the structure $\bC^\tau_\Phi$
such that for every color $M \in \sigma$ of $\Phi$
the symbol $P_M \in \rho$ denotes 
 the relation defined by $\chi_M(x)$.
Since $(\bC^\tau_\Phi,\neq)$ is a model-complete core and $\bD$ is primitive positive definable
in $\bC^\tau_\Phi$ after having added finitely many constants, we obtain that $\bD$ is pp-constructible from $(\bC^\tau_\Phi,\neq)$.
\blue{Hence, it suffices to show that $\bD$ and $\bC^{\rho}_\Psi$ are homomorphically equivalent.
We first show that $\bD$ satisfies $\Psi$. 
Consider the expansion of $\bD$ where $M \in \sigma$
denotes $M^{\bC_{\Phi}}$. This expansion satisfies
for distinct $M,M' \in \sigma$ the clause
$\forall x. \neg (P_M(x) \wedge M'(x))$ of $\Psi$ as a consequence of Lemma~\ref{lem:gadget-as-colour}.
The expansion clearly satisfies
all other conjuncts of $\Psi$.
Therefore, $\bD$ satisfies $\Psi$ and} we obtain a homomorphism $\bD \to \bC^{\rho}_\Psi$.
Conversely, Proposition~\ref{prop:universality} gives that every finite substructure of $\bC^{\rho}_\Psi$ has a homomorphism to $\bD$.
By the $\omega$-categoricity of $\bD$, we get a homomorphism from $\bC^{\rho}_\Psi$ to $\bD$.

To prove that $\bC^\tau_\Phi$ is pp-constructible in $\bC^{\rho}_\Psi$, it suffices to note that the structures $\bC^\tau_\Phi$ and $\bC^{\tau}_\Psi$ are isomorphic (since $(\bC^\tau_\Phi,\neq)$ and
$(\bC^{\tau}_\Psi,\neq)$ are model-complete cores and have the same CSP),
and that $\bC^\tau_\Psi$ is obtained from $\bC^{\rho}_\Psi$ by dropping the relations from $\rho\setminus\tau$, and is in particular a pp-power of $\bC^{\rho}_\Psi$.

{By Proposition~\ref{prop:uch1}, we obtain a uniformly continuous minor-preserving map from $\Pol(\bC^{\rho}_\Psi)$ to $\Pol(\bC^\tau_\Phi)$.
So if $\Pol(\bC^\tau_\Psi)\to\Projs$, we obtain by composing the two maps that $\Pol(\bC^\rho_\Psi)\to\Projs$.
Conversely, suppose that $\Pol(\bC^\rho_\Psi)\to\Projs$. From Proposition~\ref{prop:uch1} again, we have that $\Pol(\bC^\tau_\Phi,\neq)\to\Pol(\bC^\rho_\Psi)$ so that
by composition we obtain $\Pol(\bC^\tau_\Phi,\neq)\to\Projs$. Finally, we obtain by Proposition~\ref{prop:neq-is-algebraic} that $\Pol(\bC^\tau)\to\Projs$.}
\end{proof}

\section{An Algebraic Dichotomy for MMSNP}
\label{sect:alg}

We prove in this section that MMSNP exhibits a complexity dichotomy, that is, that every problem in MMSNP is in P or NP-complete.
Moreover, we show that the tractability border can be described in terms of minor-preserving maps to $\Projs$, thus confirming the general conjecture of Bodirsky and Pinsker
for the class of constraint satisfaction problems in MMSNP.

\begin{theorem}[Theorem~\ref{thm:main-intro} of the introduction]
   Let $\bB$ be an $\omega$-categorical structure such that $\Csp(\bB)$ is described by an MMSNP sentence.
Then one of the following holds:
\begin{itemize}
	\item There is a uniformly continuous minor-preserving map from $\Pol(\bB)$ to $\Projs$ and $\Csp(\bB)$ is NP-complete, or
	\item there is no such map and $\Csp(\bB)$ is in P.
\end{itemize}
\end{theorem}

We briefly describe the road to proving Theorem~\ref{thm:main-intro}.
In virtue of Theorem~\ref{thm:precol} and Corollary~\ref{cor:precol}, it suffices to focus on the case that $\Csp(\bB)$ is described by a precoloured MMSNP sentence.
For each precoloured sentence $\Phi$, we consider the structure $\bC^\tau_\Phi$ whose CSP is described by $\Phi$ and prove that the complexity of $\Csp(\bC^\tau_\Phi)$ and the existence of a uniformly continuous minor-preserving map $\Pol(\bC^\tau_\Phi)\to\Projs$ are determined
by the existence of a clone homomorphism $\scrC^\typ_1\to\Projs$, where $\scrC^\typ_1$ is a clone on a finite set that we introduce in the next subsection. 

From now on, we fix a precoloured MMSNP sentence $\Phi$ with coloured obstruction set $\mathcal F$, input signature $\tau$, and colour signature $\sigma$.
All the structures defined above will appear again in the proof below. Figure~\ref{fig:recap-structures} summarises the important properties of these structures.

\begin{figure}
\begin{center}
\resizebox{\textwidth}{!}{
\begin{tabular}{|c|c|c|l|}
\hline
 & Signature & Orbits of elements & Model-theoretic properties \\
\hline
$\HN$ (Sect.~\ref{sect:ramsey}) & $\rho$ &  & \parbox{0.3\linewidth}{Homogeneous}\\
$\bB^{\ind}_{\mathcal F}$ (Thm.~\ref{thm:css})& $\tau\cup\sigma$ &  & 1-homogeneous, model-complete\\
$\bB^{\hom}_{\mathcal F}$ (Thm.~\ref{thm:css-new}) & $\tau\cup\sigma$ & $\simeq\sigma$+\{uncoloured elements\}& Model-complete core (with $\neq$)\\
$\bC_\Phi$\ (Def.~\ref{def:C-Phi}) & $\tau\cup\sigma$ & $\simeq\sigma$ & Model-complete core (with $\neq$)\\
$\bC^\tau_\Phi$ & $\tau$ & $\simeq\sigma$ & Model-complete core (with $\neq$)\\
\hline
\end{tabular}}
\end{center}
\caption{A description of the different structures we defined, for a precoloured $\Phi$ in normal form with input signature $\tau$, colour signature $\sigma$, and coloured obstruction set $\mathcal F$.
The signature $\rho$ consists of $\tau\cup\sigma$ as well as a fresh relation symbol for each connected pp-formula in the signature $\tau\cup\sigma$ with at least 1 free variable and at most $m$ free variables,
where $m$ is the maximal size of a structure in $\mathcal F$.
The domains of the structures are ordered by inclusion, from bottom to top.
The age of $\HN$ consists of all the structures in $\Forb^{\hom}(\mathcal F)$ that have been expanded by all pp-definable relations of arity $\leq m$.
For each structure $\bA$ above, the expansion $\bA*(\mathbb Q,<)$ of $\bA$ by a ``random'' order has the Ramsey property.}
\label{fig:recap-structures}
\end{figure}
\newblue{%
Let $\scrC$ be any clone that consists of operations $f$ that are canonical with respect to $\bB$.  
Each $f\in\scrC$ induces a function on the orbits of a single element in $B$ with respect to $\Aut(\bB)$. We denote this function by $\xi^\typ_1(f)$.
Moreover, we write $\scrC^\typ_1$ for the clone of functions of the form $\xi^\typ_1(f)$, with $f\in\scrC$.
It can be checked  easily that $\xi^\typ_1\colon\scrC\to\scrC^\typ_1$
is a continuous clone homomorphism.
}%
\newblue{%
We state a consequence of the assumption that $\Phi$ is precoloured
and in normal form for $\scrC^\typ_1$. Note that in this context, the
orbits of $\Aut(\bC_\Phi)$ are in  one-to-one correspondence with the colours from $\Phi$ (by Corollary~\ref{cor:colors-are-orbits} since $\Phi$ is in normal form)\nomenclature[035]{$\scrC^\typ_1$}{the function clone over orbits induced
  by a canonical clone $\scrC$}. 
}
\begin{proposition}\label{prop:1-type-idempotent}
	Let $\Phi$ be a precoloured MMSNP sentence in normal form.
	Let $\scrC$ be the set of polymorphisms of $\bC^\tau_\Phi$ that are canonical with respect to $(\bC_\Phi,<)$.
	Then all functions in $\scrC^\typ_1$ are idempotent.
\end{proposition}
\begin{proof}
    Since $\Phi$ is precoloured and by Lemma~\ref{lem:precol-idemp}, the symbols $P_M\in\tau$ and $M\in\sigma$ have the same interpretation in $\bC_\Phi$.
    This implies that all polymorphisms of $\bC^\tau_\Phi$ (and in particular, the ones that are canonical with respect to $(\bC_\Phi,<)$) preserve the orbits of elements of $\bC^\tau_\Phi$.
    Therefore, every function in $\scrC^\typ_1$ is idempotent.
\end{proof}

\subsection{The tractable case}\label{sect:tractable-case}

In this section, we prove that $\Csp(\bC^\tau_\Phi)$ is polynomial-time tractable,
under the assumption that $\bC^\tau_\Phi$ has a polymorphism that is canonical with respect to $(\bC_\Phi,<)$ and whose behaviour on orbits of elements is Siggers
(or equivalently, if $\scrC$ is the clone of polymorphisms of $\bC^\tau_\Phi$ that are canonical with respect to $(\bC_\Phi,<)$,
under the assumption that $\scrC^\typ_1$ has no clone homomorphism to $\Projs$).
For that we use the recent solutions to the Feder-Vardi conjecture~\cite{BulatovFVConjecture,ZhukFVConjecture}.

Recall that $\Phi$ has coloured obstruction set $\mathcal F$, input signature $\tau$, and colour signature $\sigma$.
Without loss of generality, we can assume that every obstruction $\mathfrak F\in\mathcal F$ of size $n$ has domain $\{1,\dots,n\}$.
Given $\mathfrak F\in\mathcal F$ of size $n$, let \[R_{\mathfrak F}=\{(a_1,\dots,a_n)\in (C_\Phi)^n \mid \exists h\colon\mathfrak F^\tau\to\bC^\tau_\Phi : h(1,\dots,n)=(a_1,\dots,a_n)\}.\]
Note that every $R_{\mathfrak F}$ is preserved by all polymorphisms of $\bC^\tau_\Phi$.

Let $\bD$ be the structure with domain $\sigma$ and that has the relation
\[(R_{\mathfrak F})^{\bD}:=\{(M_1,\dots,M_n)\in \sigma^n \mid \exists h\colon\mathfrak F^\tau\to\bC^\tau_\Phi : h(i)\in M_i \text{ for all $i\in\{1,\dots,n\}$}\}\]
for all $\mathfrak F\in\mathcal F$ of size $n$.
Note that since $R_{\mathfrak F}$ is preserved by all polymorphisms of $\bC^\tau_\Phi$, the relation $(R_{\mathfrak F})^{\bD}$ is preserved by 
$\scrC^\typ_1$.

\begin{theorem}\label{thm:conclusion-tractable-case}
    If there is no clone homomorphism $\scrC^\typ_1\to\Projs$, then $\Csp(\bC^\tau_\Phi)$ is in P.
\end{theorem}
\begin{proof}
    Since there is no clone homomorphism $\scrC^\typ_1\to\Projs$ and since $\scrC^\typ_1$ is an idempotent clone on a finite set,
    we obtain that $\Csp(\bD)$ is in P by the Dichotomy Theorem (Theorem~\ref{thm:dichotomy}).
    We show that $\Csp(\bC^\tau_\Phi)$ reduces to $\Csp(\bD)$ in polynomial-time (Feder and Vardi gave a similar reduction from MMSNP to finite-domain CSPs;
    since our language is slightly different we give the proof in full detail).
    
    Let $\bA$ be an input to $\Csp(\bC^\tau_\Phi)$.
    Let $\bB$ be the structure with the same domain as $\bA$ and the same signature as $\bD$, where for each $\mathfrak F\in\mathcal F$ of size $n$
    and for each homomorphism $h\colon\mathfrak F^\tau\to\bA$, the tuple $(h(1),\dots,h(n))$ belongs to
    the interpretation of the relation symbol $R_{\mathfrak F}$ in $\bB$.
    It is clear that $\bB$ can be constructed in polynomial time, because there are only polynomially many such homomorphisms.
    We claim that $\bB$ homomorphically maps to $\bD$ if, and only if, $\bA$ homomorphically maps to $\bC^\tau_\Phi$.
    
    Let $f\colon \bB\to\bD$ be a homomorphism.
    Let $\bA'$ be the expansion of $\bA$ to the signature $\tau\cup\sigma$,
    where for each $M\in\sigma$ and each $a\in \bA$, we have $a\in M^{\bA'} \Leftrightarrow f(a)=M$.
    We prove that $\bA'$ is $\mathcal F$-free.
    Suppose that $h\colon\mathfrak F\to\bA'$ is a homomorphism, where $\mathfrak F\in\mathcal F$ has size $n$.
    Then $h$ is also a homomorphism from $\mathfrak F^\tau\to\bA$, so that $(h(1),\dots,h(n))\in (R_{\mathfrak F})^{\bB}$.
    Therefore, $f(h(1),\dots,h(n))\in (R_{\mathfrak F})^{\bD}$. Let $M_i:=f(h(i))$.
    By definition of $\bD$, there exists a homomorphism $g\colon\mathfrak F^\tau\to \bC^\tau_\Phi$
    such that $g(i)\in M_i$ for all $i\in\{1,\dots,n\}$.
    We claim that $g$ is a homomorphism $\mathfrak F\to\bC_\Phi$, a contradiction to the fact that $\mathfrak F\in\mathcal F$.
    Let $i\in\{1,\dots,n\}$. By definition of the colours of $\bA'$, if $i\in M$ for some $M\in\sigma$ then $h(i)\in M^{\bA'}$, so that $f(h(i)) = M$,
    i.e., $M=M_i$. Thus, $g$ is indeed a homomorphism $\mathfrak F\to\bC_\Phi$.
    
    Conversely, suppose that $f\colon\bA\to\bC^\tau_\Phi$ is a homomorphism, and let $\bA'$ be the colouration of $\bA$ obtained according to $f$.
    For $a\in B$, define $f'(a)$ to be the colour of $f(a)$ in $\bC_\Phi$.
    We claim that $f'$ is a homomorphism from 
    $\bB\to\bD$.
    Suppose that $(a_1,\dots,a_n)\in (R_{\mathfrak F})^{\bB}$.
    Then there exists a homomorphism $h\colon\mathfrak F^\tau\to \bA$ such that $h(i)=a_i$ for all $i$.
    By composing with $f$, we get a homomorphism from $\mathfrak F^\tau\to\bC^\tau_\Phi$.
    Let $M_i$ be the colour of $f(h(i))$ in $\bC_\Phi$.
    Then $(M_1,\dots,M_n)\in (R_{\mathfrak F})^{\bD}$.
    Note that by definition $M_i=f'(h(i))=f'(a_i)$, so that $(f'(a_1),\dots,f'(a_n))\in (R_{\mathfrak F})^{\bD}$ and $f'$ is a homomorphism from $\bB\to\bD$ as claimed.
\end{proof}

\subsection{Membership in FO}

We consider here the computational problem of deciding whether a given MMSNP sentence is equivalent to a first-order sentence (also called the \emph{FO-rewritability} problem).
This problem was proven to be decidable in~\cite{FeierKuusistoLutzRewritability} with a 2NExpTime complexity.
For finite-domain CSPs, membership in FO was proven decidable in~\cite{LLT}  (the problem is NP-complete in general and in P when restricted to input structures that are core).
Finally, a necessary and sufficient condition for membership in FO for $\omega$-categorical structures was provided in~\cite{BodHilsMartin-Journal}: for an $\omega$-categorical structure $\bA$,
$\Csp(\bA)$ is in FO iff there exists an $n\geq 1$ such that $\bA$ has an \emph{$n$-ary $1$-tolerant polymorphism}: a map $f\colon\bA^n\to\bA$
such that for each $k$-ary relation $R$ of $\bA$ and all tuples $\tuple a^1,\dots, \tuple a^n$ such that all but at most one are in $R$,
then $f(\tuple a^1,\dots,\tuple a^n)$ is in $R$. Note that this does not give a decision procedure for arbitrary $\omega$-categorical structures,
since a priori one does not have a bound on $n$. However, this criterion will be useful to prove the correctness of our algorithm.

\begin{theorem}\label{thm:membership-FO}
	Let $\Phi$ be an MMSNP sentence in normal form with colour set $\sigma$. Then $\Phi$ is equivalent to a first-order sentence if, and only if, $\Csp(\bD)$ is in FO,
	where $\bD$ is the structure with domain $\sigma$ constructed above.
	In particular, the FO-rewritability problem is decidable and is NP-complete for sentences in normal form and in P for sentences in strong normal form.
\end{theorem}
\begin{proof}
	Let $\tau$ be the input signature of $\Phi$.
	We saw in the proof of  Proposition~\ref{thm:conclusion-tractable-case}
	that $\Csp(\bC^\tau_\Phi)$ reduces to $\Csp(\bD)$. Moreover, one sees that this reduction is a first-order reduction. Hence, if $\Csp(\bD)$ is in FO then so is $\Csp(\bC^\tau_\Phi)$,
	so that $\Phi$ is equivalent to a first-order sentence.
	
	Conversely, suppose that $\Phi$ is equivalent to a first-order sentence, so that $\Csp(\bC^\tau_\Phi)$ is in FO.
	By the criterion mentioned above, $\bC^\tau_\Phi$ has a $1$-tolerant polymorphism $f$ from $(\bC^\tau_\Phi)^n$ to $\bC^\tau_\Phi$ for some $n$.
	Let $g$ be canonical with respect to $(\bC_\Phi,<)$ and in
	$$\overline{ \{ \beta \circ f \circ (\alpha_1,\dots,\alpha_n) \mid \alpha_1,\dots,\alpha_n \in \Aut(\bC_\Phi,<),  \beta \in \Aut(\bC_\Phi,<)\} },$$
	which exists by Theorem~\ref{thm:can-3}.
	Note that $g$ is again $1$-tolerant. Indeed, let $R$ be a relation of $\bC^\tau_\Phi$ of arity $k$ and let $\tuple a^1,\dots,\tuple a^n$ be $k$-tuples
	such that all but at most one are in $R$.
	There exist $\alpha_1,\dots,\alpha_n,\beta\in\Aut(\bC_\Phi,<)$ such that $g(\tuple a^1,\dots,\tuple a^n)=\beta f(\alpha_1(\tuple a^1),\dots,\alpha_n(\tuple a^n))$.
	All but at most one of $\alpha_1(\tuple a^1),\dots,\alpha_n(\tuple a^n)$ are in $R$, so that by applying $f$ we obtain a tuple in $R$.
	Composing with $\beta$ gives a tuple in $R$, so $g$ is $1$-tolerant.
	It is now easy to see that $\xi^\typ_1(g)$ is a $1$-tolerant polymorphism of $\bD$, so that $\Csp(\bD)$ is in FO.
	
	Given an MMSNP sentence $\Phi$ in normal form, one can then decide whether $\Phi$ is equivalent to a first-order formula
	by checking whether $\Csp(\bD)$ is in FO. Since $\bD$ is constructible in polynomial time, this gives an NP decision procedure.
	If $\Phi$ is furthermore assumed to be in strong normal form, then the FO-rewritability of $\Phi$ can even be decided in polynomial time.
	
	A decision procedure for the general FO-rewritability problem is then as follows: given $\Phi$, compute connected MMSNP sentences $\Phi_1,\dots,\Phi_k$
	such that $\Phi$ is equivalent to $\Phi_1\lor\dots\lor\Phi_k$ (which is possible by Proposition~\ref{prop:connected}),
	then compute a normal form $\Psi_k$ for each $k$ (which is possible by Lemma~\ref{lem:nf-existence}).
	For each $i\in\{1,\dots,k\}$, one can decide whether $\Psi_i$ is equivalent to a first-order formula by checking whether the corresponding finite-domain CSP
	is in FO. If it is the case for all $i$, then $\Phi$ is equivalent to a first-order formula.
	Otherwise, since the reduction in the proof of Proposition~\ref{prop:connected-compl} can be coded into FO,
	we obtain that $\Phi$ is not equivalent to a first-order formula.
\end{proof}

The complexity of our procedure to compute a normal form (given in Lemma~\ref{lem:nf-existence}) is not clear. If this procedure can be proved to run in doubly exponential time,
then the complexity of our decision procedure for the FO-rewritability problem would match the one given in~\cite{FeierKuusistoLutzRewritability}.

\subsection{The hard case}\label{sect:hard-case}

Let $\Phi$ be a precoloured MMSNP sentence and let $\scrC$ be the clone of polymorphisms of $\bC^\tau_\Phi$ that
are canonical with respect to $(\bC_\Phi,<)$.
In this section, we deal with the case that there exists a clone homomorphism $\xi\colon\scrC^\typ_1\to\Projs$, and prove that there exists a uniformly continuous
minor-preserving map $\Pol(\bC^\tau_\Phi)\to\Projs$.
There is a natural candidate for a minor-preserving map $\Pol(\bC^\tau_\Phi)\to\Projs$, which we describe now.
By Theorem~\ref{thm:can-3}, for every $f\in \Pol(\bC_\Phi)$ of arity $k$, the set 
$${
{\mathscr I}_f
:= \overline{\{a_0(f(a_1,\dots,a_k)) \mid a_0,a_1,\dots,a_k \in \Aut(\bC_\Phi,<)\} } }$$ has a non-empty
intersection with $\scrC$. Thus, a natural definition of a 
map $\phi$ from $\Pol(\bC_\Phi)$ to $\Projs$ is
given by $$\phi(f) := \xi(\xi^\typ_1(g)) \text{ where } g\in\scrC\cap {{\mathscr I}_f} .$$
This map is well-defined only if for every $g,h\in\scrC\cap {{\mathscr I}_f} $
 we have $\xi(\xi^\typ_1(g))=\xi(\xi^\typ_1(h))$.
We focus on proving that $\phi$ {(potentially after replacing $\xi$ with another clone homomorphism from $\scrC^\typ_1\to\Projs$)} is a well-defined uniformly continuous minor-preserving map in the following series of propositions. 

Let $\rho$ be a subset of $\sigma$ such that $\rho$ is preserved by $\scrC^\typ_1$ (we identify the relation symbols with the domain of $\scrC^\typ_1$).
Let $\Theta$ be an equivalence relation on $\rho$ that is preserved by $\scrC^\typ_1$ and with two equivalence classes $S,T\subseteq\rho$.
We call $\{S,T\}$ a \emph{subfactor} of $\scrC^\typ_1$.
The clone $\scrC^\typ_1$ naturally induces a clone on the two-element set $\{S,T\}$.
If this clone is (isomorphic to) the projection clone $\Projs$, then we call $\{S,T\}$ a \emph{trivial subfactor}.
The theory of finite idempotent algebras implies that $\scrC^\typ_1$ has a homomorphism to $\Projs$ if, and only if, $\scrC^\typ_1$
has a trivial subfactor $\{S,T\}$ (see~\cite[Proposition 4.14]{BulatovJeavons}).

In the following we write
$S^{\bC_\Phi}$ for $\bigcup_{R\in S} R^{\bC_\Phi}$ and $T^{\bC_\Phi}$ for $\bigcup_{R \in T} R^{\bC_\Phi}$. 
Note that if $\{S,T\}$ is a subfactor of $\scrC^\typ_1$, then $S^{\bC_\Phi}\cup T^{\bC_\Phi}$ 
is preserved by every operation in $\scrC$.

Let $X$ be a primitive positive definable subset of $\bC_\Phi$. A binary symmetric relation $N\subseteq X^2$ defines an undirected graph on $\sigma$: there is an edge between $M$ and $M'$
iff there exist $x\in M^{\bC_\Phi}$ and $y\in {M'}^{\bC_\Phi}$ such that $(x,y)\in N$.
If $N$ is primitive positive definable in $\bC_\Phi$,
we call the resulting graph on $\sigma$ a \emph{definable colour graph over $X$}.
In the following technical propositions, we prove that the existence of a trivial subfactor $\{S,T\}$ of $\scrC^\typ_1$
implies the existence of a definable colour graph  with an edge from $S$ to $T$ and without loops (Proposition~\ref{prop:case-2}).
Refining this even further, we show the existence of such a graph without any path of even length between $S$ and $T$ (Proposition~\ref{prop:connected-components-colour-graph}).

\begin{lemma}\label{lem:canonisation-free-orbit}
  For every pair of colours $R,B\in\sigma$, there are endomorphisms $e_1$ and $e_2$
    of $\bC_\Phi$ such that 
    for all $x,x' \in R^{\bC_\Phi}$
    and $y,y' \in B^{\bC_\Phi}$,
    the four pairs 
    \begin{align*}
    (e_1(x),e_2(y)), & \quad (e_1(x'),e_2(y')) \\
    (e_2(x),e_1(y)), & \quad (e_2(x'),e_1(y'))
    \end{align*} 
    are in the same orbit in $(\bC_\Phi,<)$. 
\end{lemma}

\begin{proof}%
We build the endomorphisms by compactness, showing that partial homomorphisms with the given properties exist {for every finite substructure $\mathfrak F$ of $\bC_\Phi$.}
        Let $\mathfrak G$ be the disjoint union of $2$ copies of $\mathfrak F$, with domain $F\times\{1,2\}$.
    We prepare a new structure $\mathfrak H$ which contains $\mathfrak G$ as a substructure.
    For all elements $x$ and $x'$ of $\mathfrak G$
    {of the same color}, take a fresh copy $\mathfrak G'$ of $\mathfrak G$ and add to $\mathfrak H$ this fresh copy, where the vertex corresponding to $x$ in $\mathfrak G'$
    is glued on top of the vertex corresponding to $x'$ in the original copy of $\mathfrak G$.
    This way, every two elements of the original $\mathfrak G$ that are in the same colour satisfy the same primitive positive formulas in $\mathfrak H$.
    It is also clear that $\mathfrak H$ is $\mathcal F$-free, since $\Phi$ is in normal form.
    Since $\mathfrak H$ is $\mathcal F$-free, the expansion $\mathfrak H^*$ of $\mathfrak H$ by all relations {with a primitive positive definition with at most $m$ variables} embeds into $\HN$ {(where $m$ denotes the size of the largest structure in ${\mathcal F}$)}. 

Let $<$ be any linear order on $\mathfrak G$ such that $(x,1)<(y,2)$ and $(x,2)<(y,1)$ for all $x\in R^{\mathfrak F}$ and $y\in B^{\mathfrak F}$.
Complete $<$ arbitrarily into a linear order on $\mathfrak H$, so that there exists an embedding $e$ of $(\mathfrak H^*,<)$ into $(\HN,<)$. By the homogeneity of $(\HN,<)$, the pairs 
    \begin{align*}
    (e(x,1),e(y,2)) & \quad (e(x',1),e(y',2)) \\
    (e(x,2),e(y,1)) & \quad (e(x',2),e(y',1))
    \end{align*} 
    are all in the same orbit in $(\HN,<)$,
    for all $x,x'\in R^{\mathfrak F}$ and $y,y'\in B^{\mathfrak F}$.
    Let $e'\colon \mathfrak G\to\bB^{\hom}_{\mathcal F}$ be obtained by composing $e$ with an injective homomorphism from the $(\tau\cup\sigma)$-reduct of $\HN$ to
    $\bB^{\hom}_{\mathcal F}$ that is canonical from $(\HN,<)$ to $(\bB^{\hom}_{\mathcal F},<)$ {(we use Theorem~\ref{thm:HN-Ramsey} and Theorem~\ref{thm:can})}. 
    Since all the vertices of $\mathfrak G$ are coloured, the image of $e'$ is included in $\bC_\Phi$.
    We obtain a homomorphism {$h$} from $\mathfrak G$ to $\bC_\Phi$ such that the given pairs are in the same orbit under $\Aut(\bC_\Phi,<)$.
    For $i\in\{1,2\}$, define the partial endomorphisms $e_i$ of $\bC_\Phi$ by {$x\mapsto h(x,i)$}.
    It is easy to check that these partial endomorphisms satisfy the required properties.
\end{proof}

\begin{proposition}\label{prop:case-2}
Let $\Phi$ be a precoloured MMSNP sentence and let $\scrC$ be the clone of polymorphisms of $\bC_\Phi$ that are canonical with respect to $(\bC_\Phi,<)$. Let $\{S,T\}$ be a trivial subfactor of $\scrC^\typ_1$ and let $X$ be a primitive positive definable subset of $\bC_\Phi$ such that $X\cap S^{\bC_\Phi}\neq\emptyset$
 and $X\cap T^{\bC_\Phi}\neq\emptyset$. 
Then there exists a loopless definable colour graph over $X$ containing an edge from $S$ to $T$.
\end{proposition}
\begin{proof}
	\new{We prove the result by contraposition, assuming that $\{S,T\}$ is a subfactor of $\scrC^\typ_1$ and
	that every definable colour graph over $X$ that contains an edge from $S$ to $T$ also contains a loop.}
        The crux of the proof is to show that this assumption implies the existence of a canonical polymorphism $\hat h$ of $\bC_\Phi$ such that for all $x,y\in X$
    	the equivalence $\hat h(x,y)\in S^{\bC_\Phi}\Leftrightarrow \hat h(y,x)\in S^{\bC_\Phi}$ holds.
	
       First, we show that for every finite subset $A$ of $\bC_\Phi$ 
        there exists a binary polymorphism $f$ of $\bC_\Phi$ such that the following property $(\dagger)$ holds for all $a,b\in A\cap X$:
    \begin{align*}
    f(a,b),f(b,a)\in S^{\bC_\Phi}\cup T^{\bC_\Phi} \text{ implies }
    (f(a,b)\in S^{\bC_\Phi}\Leftrightarrow f(b,a)\in S^{\bC_\Phi}). \quad (\dagger)
    \end{align*}
    For a binary polymorphism $f$ of $\bC_\Phi$  define $$C(f) := \{(a,b)\in A^2 \mid \exists\alpha\in\Aut(\bC_\Phi) : f(a,b)=\alpha f(b,a)\}.$$
    Let $f$ be such that $C(f)$ is maximal. Suppose that $f$ does not satisfy $(\dagger)$.
    This means that there exist $a,b\in A\cap X$ such that such that $f(a,b)\in S^{\bC_\Phi}$ and $f(b,a)\in T^{\bC_\Phi}$.
    Let $N$ be the smallest binary relation containing $(f(a,b),f(b,a)),(f(b,a),f(a,b))$ and being preserved by the polymorphisms of $\bC_\Phi$.
    Note that $N\subseteq X^2$, since $a$ and $b$ are in $X$ and $X$ is preserved by all the polymorphisms of $\bC_\Phi$.
    Since $\bC_\Phi$ is $\omega$-categorical, this relation has a pp-definition in $\bC_\Phi$~\cite[Theorem 5.1]{BodirskyNesetrilJLC}.
    Moreover, it is symmetric and $(f(a,b),f(b,a))\in N\cap (S^{\bC_\Phi}\times T^{\bC_\Phi})$. By hypothesis, the colour graph defined by $N$ contains a loop.
    This implies that there exist $g\in\Pol(\bC_\Phi)$ and $\alpha\in\Aut(\bC_\Phi)$ such that $g(f(a,b),f(b,a)) = \alpha g(f(b,a),f(a,b))$.
    Define $f'(x,y):= g(f(x,y),f(y,x))$ for all $x,y\in\bC_\Phi$.
    It is clear from the above that $(a,b)\in C(f')$.
    Moreover, we have $C(f)\subseteq C(f')$. Indeed, let $(a',b')\in C(f)$. Then $f(a',b'),f(b',a')$ are in the same orbit, and since $\Phi$ is precoloured,
    this implies that $f'(a',b')$ and $f'(b',a')$ are in the same orbit.
    This contradicts the maximality of $C(f)$, so that it must be the case that $f$ satisfies $(\dagger)$.
    
    Using a standard compactness argument (see e.g.\ the proof of Proposition 13 in \cite{Bodirsky-Mottet}), we obtain a binary polymorphism $f$ of $\bC_\Phi$ that satisfies $(\dagger)$ for all $a,b\in X$.

    Let $g$ be any polymorphism obtained by diagonally canonising $f$, using Theorem~\ref{thm:can-4}.
    We claim that $g$ still satisfies $(\dagger)$ on $X$. Indeed, let $a,b\in X$ and suppose that $g(a,b),g(b,a)\in S^{\bC_\Phi}\cup T^{\bC_\Phi}$.
    There exist $\alpha,\beta\in\Aut(\bC_\Phi)$ such that $g(a,b) = \alpha f(\beta a,\beta b)$ and $g(b,a)=\alpha f(\beta b,\beta a)$.
    Since $S^{\bC_\Phi}$ and $T^{\bC_\Phi}$ are union of colours, they are preserved by automorphisms of $\bC_\Phi$.
    We conclude that $f(\beta a,\beta b),f(\beta b,\beta a) \in S^{\bC_\Phi}\cup T^{\bC_\Phi}$.
    Since $f$ satisfies $(\dagger)$ on $X$, the equivalence $f(\beta a,\beta b)\in S^{\bC_\Phi}\Leftrightarrow f(\beta b,\beta a)\in S^{\bC_\Phi}$ holds.
    It follows that $g(a,b)\in S^{\bC_\Phi}\Leftrightarrow g(b,a)\in S^{\bC_\Phi}$, so that $g$ also satisfies $(\dagger)$ on $X$.

    Let $R\in S, B\in T$ be such that $R^{\bC_\Phi}\subseteq X$ and $B^{\bC_\Phi}\subseteq X$.
    Let $e_1,e_2$ be the endomorphisms of $\bC_\Phi$ given by Lemma~\ref{lem:canonisation-free-orbit}.
    Define $h(x,y):=g(e_1(x),e_2(y))$ for all $x,y\in\bC_\Phi$.
    Note that $h$ is 1-canonical on $R^\bC_\Phi\cup B^\bC_\Phi$: for $(a,b),(a',b')\in R^{\bC_\Phi}\times B^{\bC_\Phi}$,
    the pairs $(e_1(a),e_2(b))$ and {$(e_1(a'),e_2(b'))$} are in the same orbit of $(\bC_\Phi,<)$, according to Lemma~\ref{lem:canonisation-free-orbit}.
    Since $g$ is diagonally canonical, this implies that $h(a,b)$ and $h(a',b')$ are in the same orbit.
    Similarly, for $(a,b),(a',b')\in B^{\bC_\Phi}\times R^{\bC_\Phi}$, the pairs $(e_1(a),e_2(b))$ and $(e_1(a'),e_2(b'))$ are in the same orbit of $(\bC_\Phi,<)$.
    Moreover, $h$ satisfies $(\dagger)$ on $R^{\bC_\Phi}\cup B^{\bC_\Phi}$.
    Indeed, let $(a,b)\in R^{\bC_\Phi}\times B^{\bC_\Phi}$ be such that $h(a,b)$ and $h(b,a)$ are in $S^{\bC_\Phi}\cup T^{\bC_\Phi}$.
    Then $g(e_1(a),e_2(b))$ and $g(e_1(b),e_2(a))$ are in $S^{\bC_\Phi}\cup T^{\bC_\Phi}$.
    Since $g$ is diagonally canonical and $(e_1(b),e_2(a))$ and $(e_2(b),e_1(a))$ are in the same orbit,
    we have that also $g(e_2(b),e_1(a))$ is in $S^{\bC_\Phi}\cup T^{\bC_\Phi}$.
    By $(\dagger)$, we have $g(e_1(a),e_2(b))\in S^{\bC_\Phi}$ if, and only if, $g(e_2(b),e_1(a))\in S^{\bC_\Phi}$.
    By definition, this implies that $h(a,b)\in S^{\bC_\Phi}\Leftrightarrow h(b,a)\in S^{\bC_\Phi}$ holds.
    So $h$ satisfies $(\dagger)$ on $R^{\bC_\Phi}\cup B^{\bC_\Phi}$.

    Let now $\hat h$ be obtained by canonising $h$ with respect to $(\bC_\Phi,<)$.
    {Since $h$ was already 1-canonical on 
    $R^{\bC_\Phi}\cup B^{\bC_\Phi}$, 
    the restrictions of $\xi^\typ_1(h)$ and 
    $\xi^\typ_1(\hat h)$ to $\{R, B\}$ are equal.}
    This implies that $\hat h$ still satisfies  
    $(\dagger)$ 
    on $R^{\bC_\Phi}\cup B^{\bC_\Phi}$.
    By assumption, $S^{\bC_\phi}\cup T^{\bC_\Phi}$ is preserved by $\hat h$.
    This implies that for all $a\in R^{\bC_\Phi}, b\in B^{\bC_\Phi}$, we have that $\hat h(a,b)\in S^{\bC_\Phi}\Leftrightarrow \hat h(b,a)\in S^{\bC_\Phi}$.
    Finally, since the partition $\{S,T\}$ is preserved by $\hat h$ by assumption, for all $a,a'\in S^{\bC_\Phi}$ and $b,b'\in T^{\bC_\Phi}$
    we must have that $\hat h(a,b) \in S^{\bC_\Phi}$ iff $\hat h(a',b')\in S^{\bC_\Phi}$, and similarly for $\hat h(b,a)$ and $\hat h(b',a')$.
    This finishes the construction of $\hat h$.
   
   Note that the function induced by $\hat h$ on the subfactor $\{S,T\}$ is binary and commutative.
   \new{Therefore, the clone induced by $\scrC^\typ_1$ on $\{S,T\}$ contains an operation that is not a projection, and
   $\{S,T\}$ is not a trivial subfactor of $\scrC^\typ_1$.}
\end{proof}

\begin{proposition}\label{prop:connected-components-colour-graph}
	Let $\Phi$ be a precoloured MMSNP sentence and let $\scrC$ be the clone of polymorphisms of $\bC_\Phi$ that are canonical with respect to $(\bC_\Phi,<)$.
	Suppose that $\scrC^\typ_1$ has a trivial subfactor.
    Then there exist a trivial subfactor $\{S,T\}$ and a binary symmetric relation $N$ \blue{that is pp-definable in $\bC_\Phi$, defines a colour graph with an edge from $S$ to $T$, and does not contain a path of even length} between $S$ and $T$.
\end{proposition}
\begin{proof}
    Consider the set $\mathcal{S}$ of all triples $(S,T,N)$ such that:
    \begin{itemize}
        \item $\{S,T\}$ is a trivial subfactor of $\scrC^\typ_1$;
        \item $N$ is a binary symmetric relation pp-definable in $\bC_\Phi$ whose colour graph is loopless;
        \item if the colour graph defined by $N$ is bipartite, then $N$ contains an edge between $S$ and $T$.
    \end{itemize}
    By the assumption that $\scrC^\typ_1$ has a trivial subfactor and Proposition~\ref{prop:case-2} applied to $X:=\bC_\Phi$, the set $\mathcal S$ is nonempty.
    Pick a triple $(S,T,N)\in\mathcal S$ such that the \emph{support} of $N$ (i.e., the set of $x\in\bC_\Phi$ such that there exists $(x,y)\in N$ for some $y$)
    intersects the fewest number of orbits of $\bC_\Phi$. We show that $N$ then satisfies the conclusion of the proposition.

    We first claim that the colour graph defined by $N$ is bipartite.
    For the sake of contradiction, suppose that it is not the case.
    If $\ell$ is the smallest length of an odd cycle in the colour graph defined by $N$, then \red{the relation $N^{\circ(\ell-2)}$
    that contains all pairs linked by a directed path 
    with $\ell-2$ many edges from $N$}
     is again symmetric, pp-definable in $\bC_\Phi$,
    and such that the colour graph it defines is loopless and contains a triangle. We therefore assume without loss of generality that $\ell=3$.
    The set defined by the unary formula
     $$\phi(x):=\exists y,z,x' (N(x,y)\land N(y,z)\land N(z,x')\land x\sim x')$$ is
    preserved by $\scrC$, where the meaning of $x \sim x'$ is that $x$ and $x'$ are in the same orbit of $\Aut(\bC_\Phi)$, i.e., in the same colour of $\bC_\Phi$.
    The corresponding subset $\rho$ of $\sigma$ is therefore preserved by $\scrC^\typ_1$,
    and consists of the set of colours that belong to a triangle in the colour graph defined by $N$. In particular, $\rho$ is not empty.
    When restricted to $\rho$, the colour graph is therefore not bipartite and does not contain a loop.
    It follows from~\cite[Theorem 1]{BulatovHColoring} that there is a trivial subfactor $\{S',T'\}$ of $\scrC^\typ_1$
    such that $S'\cup T'\subseteq \rho$.
    Let now $R\in S'$, and let $N'(x,y)$ be defined by
    \[N(x,y)\land \exists z,z'\left(R(z)\land R(z')\land N^{\circ 2}(z,x)\land N^{\circ 2}(z',y)\right),\] which is a pp-definition in $\bC_\Phi$.
    Again, $N'\subseteq N$ is symmetric and the colour graph that it defines is loopless and contains a triangle, so that $(S',T',N')\in\mathcal{S}$.
    By \new{the} minimality of $(S,T,N)$, the support of $N'$ must be equal to the support of $N$.
    \new{In particular, every $B\in T'$ has a path of length $2$ to $R$ since it is in the support of $N$.}
    Pick an arbitrary $B\in T'$ and let $G$ be the midpoint of a path of length $2$ between $R$ and $B$. The situation is described in Figure~\ref{fig:case-1-N-relation}.
    Then the formula $\theta(x):= \exists y (G(y)\land N(x,y))$ defines a proper subset $Y$ of the support of $N$ (proper because the colour graph is loopless, which implies that
    no element in $G$ satisfies $\theta$), and this subset intersects $S'$ and $T'$.
    By Proposition~\ref{prop:case-2} applied to $X:=Y$, we obtain a new binary symmetric relation $M\subseteq Y^2$ whose colour graph is loopless
    and contains an edge between $S'$ and $T'$. In particular, $(S',T',M)\in\mathcal{S}$, a contradiction to the minimality of $(S,T,N)$.
    \begin{figure}
	\begin{center}
	\begin{tikzpicture}[every node/.style = {circle,fill,black,inner sep=2pt} ]
		\node[label={below right:$R$}] (r) at (0,0) {};
		\node[label={below left:$B$}] (b) at (-2,0) {};
		\node[label={left:$G$}] (g) at (-1,1.2) {};
		\node (p) at (-1,-1.5) {};
		\node (v) at (-1.5,-1.4) {};
		\node (w) at (0,-1.7) {};
		\node[above right= of w] (m) {};
		\node[above right = of r] (s) {};
		
		\draw (r) -- (g) -- (b);
		\draw (b) -- (v) -- (p) -- (b);
		\draw (g) -- (m) -- (s) -- (g);
		\draw (r) -- (v);
		\draw (r) -- (s);
		\draw (w) -- (g);
		\draw (w) to [out=-170,in=-90] (v);
		\draw (w) -- (r);
		
		\draw plot [smooth] coordinates { (2,-1) (-0.5,0) (1,2) (2,-1)};
		\draw (-2,0) circle (0.6cm);
		\node[fill=none] at (-2.7,0.5) {$T'$};
		\node[fill=none] at (2,0.5) {$S'$};
	\end{tikzpicture}
	\end{center}
	\caption{An illustration of the colour graph defined by $N$ in the proof of Proposition~\ref{prop:connected-components-colour-graph}.
	The formula $\exists y (G(y)\land N(x,y))$ defines in $\bC_\Phi$ a subset that does not intersect $G^{\bC_\Phi}$
	and intersects $R^{\bC_\Phi}$ and $B^{\bC_\Phi}$, a contradiction to the minimality of $(S,T,N)\in\mathcal S$.}
	\label{fig:case-1-N-relation}
\end{figure}

   Therefore, the colour graph defined by $N$ is bipartite.
    Since $(S,T,N)\in\mathcal{S}$, it must be that $N$ contains an edge between $S$ and $T$.
    If $N$ does not satisfy the conclusion of the proposition, there must be a path of even length $2k$ between $S$ and $T$.
    Let $G$ be the midpoint of this path. Let $Y\subseteq\bC_\Phi$ be the subset consisting of the elements of $\bC_\Phi$
    that are reachable by a path of length $k$ from an element in $G$. This set is pp-definable in $\bC_\Phi$,
    and intersects $S$ and $T$.
    Moreover, it is a proper subset of the support of $N$. Indeed, if $k$ is odd then no element of $G$ belongs to $Y$,
    otherwise the colour graph defined by $N$ would contain a cycle of length $k$.
    If $k$ is even, then no direct neighbour of an element in $G$ belongs to $Y$, for otherwise $N$ would contain a cycle of length $k+1$.
    By Proposition~\ref{prop:case-2} applied to $X:=Y$, we obtain a new binary symmetric relation $M\subseteq Y^2$ whose colour graph is loopless
    and contains an edge between $S'$ and $T'$. In particular, $(S',T',M)\in\mathcal{S}$, a contradiction to the minimality of $(S,T,N)$.
    Thus, it must be that $(S,T,N)$ satisfies the conclusion of the statement.
\end{proof}

The following lemma is the core argument that allows us to construct a uniformly continuous minor-preserving map $\Pol(\bC_\Phi)\to\Projs$ in Theorem~\ref{thm:conclusion-hardness}.
Given any $k$-ary map $f$ and any $\lambda\colon\{1,\dots,k\}\to\{1,\dots,\ell\}$, we write $f^{\lambda}$ for the map $(x_1,\dots,x_\ell)\mapsto f(x_{\lambda(1)},\dots,x_{\lambda(k)})$. 
\begin{lemma}\label{lem:correctness-extension}
	Let $\{S,T\}$ and $N$ be as in the conclusion of Proposition~\ref{prop:connected-components-colour-graph}.
	Let $(f_n)_{n\in\mathbb N}$ be a sequence of $k$-ary polymorphisms of $\bC_\Phi$ converging to some $f$.
	Let $\ell\geq 1$ and let $\lambda\colon\{1,\dots,k\}\to\{1,\dots,\ell\}$.
	Let $h\in \scrC$ be in ${\mathscr I}_{(f_n)^\lambda}$ for infinitely many $n$, and let $g\in \scrC\cap{\mathscr I}_{f}$.
	Then $(\xi^\typ_1(g))^\lambda$ and $\xi^\typ_1(h)$ induce the same function on $\{S,T\}$.
\end{lemma}
\begin{proof}
	For ease of notation, assume that $(\xi^\typ_1(g))^\lambda$ induces the first projection on $\{S,T\}$,
	that is, $(\xi^\typ_1(g))^\lambda(S,T,\dots,T) = S$.
	In order to prove that $\xi^\typ_1(h)$ is also the first projection on $\{S,T\}$, it suffices to prove that there exists $R\in S$ and $B\in T$
	such that $\xi^\typ_1(h)(R,B,\dots,B) \in S$.
	Let $R\in S$ and $B\in T$ be adjacent colours in the colour graph defined by $N$.
    Let $(a_1,\dots,a_k)$ be such that $(a_{\lambda(1)},\dots,a_{\lambda(k)})\in R^{\bC_\Phi}\times B^{\bC_\Phi}\times\cdots\times B^{\bC_\Phi}$.
    Let $\alpha,\beta_1,\dots,\beta_k\in\Aut(\bC_\Phi,<)$ be such that \[g(a_1,\dots,a_k)=\alpha f(\beta_1(a_1),\dots,\beta_k(a_k)).\]
    Since $(f_n)$ converges to $f$, for all large enough $n$ we have
    \begin{align*}
    	g(a_1,\dots,a_k) &= \alpha f(\beta_1(a_1),\dots,\beta_k(a_k))\\
				&= \alpha f_n(\beta_1(a_1),\dots,\beta_k(a_k)) & \tag{\mbox{$\dagger$}}.
	\end{align*}
    For some $n$ large enough, there exist $\gamma,\delta_1,\dots,\delta_k$  such that 
    \begin{align*}
    h(a_1,\dots,a_k)&=\gamma (f_n)^\lambda(\delta_1(a_1),\dots,\delta_k(a_k))\\
    &= \gamma f_n(\delta_{\lambda(1)}(a_{\lambda(1)}),\dots,\delta_{\lambda(k)}(a_{\lambda(k)})).
    \end{align*}
    
    \begin{figure}
    \begin{minipage}[c]{0.4\textwidth}
    \begin{tikzpicture}
    	\draw (0,0) -- (0,3) -- (2,3) -- (2,0) -- cycle;
	
		\draw (0.3,1) arc (-245:65:1.6cm and 0.9cm);
		\node[bv,label={below:{\scriptsize $x_2$}}] at (0.3,1) {};
		\node[bv,label={below:{\scriptsize $y_2$}}] at (1.7,1) {};
		
		\draw (1.7,2) arc (-65:245:1.6cm and 0.9cm);
		\node[rv,label={above:{\scriptsize $x_1$}}] at (0.3,2) {};
		\node[rv,label={above:{\scriptsize $y_1$}}] at (1.7,2) {};
		
		\node[draw=none] at (1,1.5) {$\theta^*$};
		
	\foreach\i in {90}
		\node[rv,scale=0.5,label={above:{\scriptsize $z^1_2$}}] at ($(1,2.8) + (\i:1.6cm and 0.9cm)$) {};
	
		\node[bv,scale=0.5,label={left:{\scriptsize $z^1_1$}}] at ($(1,2.8) + (150:1.6cm and 0.9cm)$) {};
		\node[bv,scale=0.5,label={right:{\scriptsize $z^1_3$}}] at ($(1,2.8) + (30:1.6cm and 0.9cm)$) {};
	
		\node[rv,scale=0.5,label={left:{\scriptsize $z^2_1$}}] at ($(1,0.2) + (-150:1.6cm and 0.9cm)$) {};
		\node[rv,scale=0.5,label={right:{\scriptsize $z^2_3$}}] at ($(1,0.2) + (-30:1.6cm and 0.9cm)$) {};
		
	\foreach\i in {-90}
		\node[bv,scale=0.5,label={below:{\scriptsize $z^2_2$}}] at ($(1,0.2) + (\i:1.6cm and 0.9cm)$) {};
	
    \end{tikzpicture}
    \end{minipage}
    \begin{minipage}[c]{0.4\textwidth}
    \begin{tikzpicture}[every  node/.style={},scale=0.5,
    vertex/.style={draw,circle,scale=0.3,fill}]
    	\draw (-5,0) -- (5,0);
	\draw (0,-5) -- (0,5);
	
	\node at (-0.8,-5) {$R^{\bC_\Phi}$};
	\node at (.8,-5) {$B^{\bC_\Phi}$};
	\node at (-6, -2.5) {$R^{\bC_\Phi}$};
	\node at (-6, 2.5) {$B^{\bC_\Phi}$};
	
	\node[inner sep=0,outer sep=0,label={below:{\scriptsize $\delta_{\lambda(1)}(a_\lambda(1))$}}] (d1a1) at (-4.7,-5.2) {};
	\node[inner sep=0,outer sep=0,label={below:{\scriptsize $\beta_1(a_{\lambda(1)})$}}] (b1a1) at (-2.3,-5.8) {};
	\node[inner sep=0,outer sep=0,label={left:{\scriptsize $\delta_{\lambda(2)}(a_{\lambda(2)})$}}] (d2a2) at (-5.2, 0.3) {};
	\node[inner sep=0,outer sep=0,label={left:{\scriptsize $\beta_2(a_{\lambda(2)})$}}] (b2a2) at (-5.2, 4.7) {};
	
	\draw[dashed] (d1a1) -- (-4.7,0.3);
	\draw[dashed] (b1a1) -- (-2.3,4.7);
	\draw[dashed] (d2a2) -- (-4.7,0.3);
	\draw[dashed] (b2a2) -- (-2.3,4.7);
	
	\node[vertex,label={below:{\scriptsize $(e(z^1_1),e(z^2_1))$}}] (i1) at (3,-1.3) {};
	\node[vertex] (i2) at (-1,1) {};
	\node[vertex] (i3) at (1,-2.7) {};
	\draw[red] (-2.3,4.7) -- (i1) -- (i2) -- (i3) -- (-4.7,0.3);
	\node[vertex,label={above:{\scriptsize $(e(x_1),e(x_2))$}}] at (-2.3,4.7) {}; 
	\node[vertex] at (-4.7,0.3) {}; 
    \end{tikzpicture}
    \end{minipage}
    \caption{Proof of Theorem~\ref{thm:conclusion-hardness}: A depiction of $\psi$ (left) in the case that $k=2$ and $2m=4$, and a view of $(R^{\bC_\Phi}\cup B^{\bC_\Phi})^2$ (right). The red edges on the right represent the relation $N$;
    these edges connect the images of the drawn points under $f_n$.}
    \label{fig:hard}
    \end{figure}

    {Since $(\bC_\Phi,\neq)$ is a model-complete core (Lemma~\ref{lem:C-phi-mc-core}), by Proposition~\ref{prop:mc-core}  
    the orbit of the tuple $(\beta_1a_{\lambda(1)},\dots,\beta_ka_{\lambda(k)},\delta_{\lambda(1)}a_{\lambda(1)},\dots,\delta_{\lambda(k)}a_{\lambda(k)})$ 
    has a primitive positive definition $\theta(x_1,\dots,x_k,y_1,\dots,y_k)$ 
    in $(\bC_\Phi,\neq)$.}
    Let $\theta^*$ be $\theta$ where the atomic conjuncts involving $\neq$ have been removed. 
    Let $\phi_N(x,y)$ be a primitive positive formula defining the relation $N\subseteq(\bC_\Phi)^2$
    in $\bC_\Phi$.
    Fix  an integer $m$ such that $2m > |\Phi|$.
    For every $i\in\{1,\dots,k\}$, let $z^i_1,\dots,z^i_{2m-1}$ be fresh variables. In the following, we also write $z^i_0$ for $x_i$ and $z^i_{2m}$ for $y_i$.
    Let $\psi(x_1,\dots,x_k,y_1,\dots,y_k)$ be the primitive positive formula whose conjuncts are {(see Figure~\ref{fig:hard}, left side)}: 
    \begin{itemize}
    \item $\theta^*(x_1,\dots,x_k,y_1,\dots,y_k)$,
    \item $\phi_N(z^i_j,z^i_{j+1})$, for every $i\in\{1,\dots,k\}$ and $j\in\{0,\dots,2m-1\}$,
    \item $R(z^1_j)$ for even $j\in\{1,\dots,2m-1\}$  and $B(z^1_j)$ for odd $j\in\{1,\dots,2m-1\}$,
    \item for $i\in\{2,\dots,k\}$, the conjunct $B(z^i_j)$ for even $j\in\{1,\dots,2m-1\}$ and $R(z^i_j)$ for odd $j\in\{1,\dots,2m-1\}$.
    \end{itemize}
    
    \new{We claim that $(\beta_1a_{\lambda(1)},\dots,\beta_ka_{\lambda(k)},\delta_{\lambda(1)}a_{\lambda(1)},\dots,\delta_{\lambda(k)}a_{\lambda(k)})$ satisfies $\psi$.}
    We first prove that $\psi$  is satisfiable in $\bB^{\ind}_{\mathcal F}$, where $\mathcal F$ is the
    coloured obstruction set of $\Phi$.
    Let $\mathfrak S'$ be the canonical database of $\psi$ (again see Figure~\ref{fig:hard}, left side),
    and let $\mathfrak S$ be the canonical database of $\theta^*$ (so that $\mathfrak S$ is a substructure of $\mathfrak S'$).
    By Corollary~\ref{cor:nf}, $\bS'$ is satisfiable is $\mathcal F$-free if and only if all the biconnected substructures of $\mathfrak S'$ are ${\mathcal F}$-free. 
    Suppose that there exists an obstruction $\mathfrak F\in\mathcal F$ and a homomorphism $e\colon\mathfrak F\to\mathfrak S'$.
    By the choice of $m$ we have that $|\mathfrak F|<2m$.
    Since $\Phi$ is in normal form, its obstructions are biconnected and we can suppose that the image of the homomorphism $e$ is a biconnected substructure of $\mathfrak S'$.
    It follows that either the image of $e$ is included in $\mathfrak S$, which would be a contradiction since $\mathfrak S$ has a homomorphism to $\bB^{\ind}_{\mathcal F}$,
    or it is included in the subset induced by the canonical database of some $N(z^i_j,z^i_{j+1})$ for some $i\in\{1,\dots,k\}$ and $j\in\{0,\dots,2m-1\}$.
    But the assumption on $N$ is that there is $(a,b)\in N$ such that $a\in R^{\bC_\Phi}$ and $b\in B^{\bC_\Phi}$.
    Therefore, the conjunct $\phi_N(z^i_j,z^i_{j+1}) $ is satisfiable by an assignment that maps $z^i_j$ and $z^i_{j+1}$
    to the appropriate colours.
    We conclude that there exists an embedding $e$ of $\mathfrak S'$ into $\bB^{\ind}_{\mathcal F}$.
    
    Let $d \colon \bB^{\ind}_{\mathcal F} \to \bB^{\hom}_{\mathcal F}$ be an injective homomorphism (whose existence follows from Theorem~\ref{thm:css-new}). Note that the image of the restriction of $d$ to the substructure
    $\bC_\Phi$ of $\bB^{\ind}_{\mathcal F}$ is in $\bC_\Phi$ since $d$ must preserve the colours. 
    Since $d \circ e$ is injective, the tuple $(e(x_1),\dots,e(x_k),e(y_1),\dots,e(y_k))$ satisfies 
    $\theta$.
    This means that $d \circ e \colon {\mathfrak S}' \to \bC_\Phi$ is a satisfying assignment that maps $(x_1,\dots,x_k,y_1,\dots,y_k)$ to a tuple in the same orbit as
    $(\beta_1a_{\lambda(1)},\dots,\beta_k a_{\lambda(k)}, \delta_{\lambda(1)}a_{\lambda(1)},\dots,\delta_{\lambda(k)} a_{\lambda(k)})$.
    By composing with an automorphism of $\bC_\Phi$, we can suppose that $(x_1,\dots,x_k,y_1,\dots,y_k)$ is exactly this tuple,
    \new{so that $\psi(\beta_1a_{\lambda(1)},\dots,\beta_k a_{\lambda(k)}, \delta_{\lambda(1)}a_{\lambda(1)},\dots,\delta_{\lambda(k)} a_{\lambda(k)})$ holds.}
    
    \new{We obtain from the previous claim that for every $i\in\{1,\dots,k\}$, there exists an $N$-path of even length that connects $\beta_i(a_{\lambda(i)})$ to $\delta_{\lambda(i)}(a_{\lambda(i)})$.}
    It must therefore be the case that  $f_n(\beta_1a_{\lambda(1)},\dots,\beta_ka_{\lambda(k)})$ and $f_n(\delta_{\lambda(1)}a_{\lambda(1)},\dots,\delta_{\lambda(k)}a_{\lambda(k)})$ 
    are connected by an $N$-path of even length, too. That is, there are $b_1,\dots,b_{2m-1}\in\bC_\Phi$ such that  
        \begin{align*}
    (b_j,b_{j+1}) & \in N  \quad \text{for all } j\in\{1,\dots,2m-2\} \\
    (f_n(\beta_1a_{\lambda(1)},\dots,\beta_ka_{\lambda(k)}),b_1) & \in N  
     \quad \text{and }  \\(b_{2m-1}, f_n(\delta_{\lambda(1)}a_{\lambda(1)},\dots,\delta_{\lambda(k)}a_{\lambda(k)})) & \in N  \quad {\text{(see Figure~\ref{fig:hard}, right side)}}. 
    \end{align*}
    By assumption, such a path of even length cannot connect $S$ and $T$.
    By ($\dagger$), the tuples $f_n(\beta_1a_{\lambda(1)},\dots,\beta_ka_{\lambda(k)})$ and $g(a_{\lambda(1)},\dots,a_{\lambda(k)})$ are in the same orbit,
    and since by assumption $(\xi^\typ_1(g))^\lambda(R,B,\dots,B)\in S$,
    we obtain that $f_n(\beta_1a_{\lambda(1)},\dots,\beta_ka_{\lambda(k)})$ belongs to a colour in $S$.
    Thus, $f_n(\delta_{\lambda(1)}a_{\lambda(1)},\dots,\delta_{\lambda(k)}a_{\lambda(k)})$ is in a colour that belongs to $S$.
    We obtain that $\xi^\typ_1(h)(R,B,\dots,B)\in S$, as desired.
\end{proof}

\begin{theorem}\label{thm:conclusion-hardness}
	Let $\Phi$ be a precoloured MMSNP sentence.
	Let $\scrC$ be the clone of polymorphisms of $\bC_\Phi$ that are canonical with respect to $(\bC_\Phi,<)$.
	If there is a clone homomorphism $\scrC^\typ_1\to\Projs$, then there exists a uniformly continuous minor-preserving map from
	$\Pol(\bC_\Phi)$ to $\Projs$ that preserves left-composition
    by $\overline{\Aut(\bC_\Phi)}$, and $\Csp(\bC^\tau_\Phi)$ is NP-hard.
\end{theorem}
\begin{proof}
As we have mentioned before, if the clone $\scrC^\typ_1$ of idempotent operations on a finite set has a homomorphism to $\Projs$, then $\scrC^\typ_1$
has a trivial subfactor $\{S,T\}$~(see \cite[Proposition 4.14]{BulatovJeavons}).

Let $\xi\colon \scrC^\typ_1\to\Projs$ be the clone homomorphism defined as follows.
	Let $R\in S$ and $B\in T$ be arbitrary. For a $k$-ary $f\in\scrC^\typ_1$, let $i\in\{1,\dots,k\}$ be the unique index such that $f(B,\dots,B,R,B,\dots,B)\in S$, where the argument $R$ is in
	the $i$th position. Such an $i$ exists because of the assumption that $\{S,T\}$ is a trivial subfactor of $\scrC^\typ_1$.
	Define $\xi(f)$ to be the $i$th projection. Note that the definition of $\xi$ does not depend on the choice of $R$ and $B$, by the fact that
	the equivalence relation on $S\cup T$ whose equivalence classes are $S$ and $T$ is assumed to be preserved by the operations in $\scrC^\typ_1$.
	It is straightforward to check that the map $\xi$ thus defined is a clone homomorphism.
	
    Let $\phi$ be the map   $\phi \colon \Pol(\bC_\Phi) \to \Pr$ 
    given by $\phi(f) := \xi(\xi^\typ_1(g))$ 
    for $g \in {\mathscr C} \cap {\mathscr I}_f$.
    By Lemma~\ref{lem:correctness-extension}, $\phi$ is well-defined.
    Indeed, let $f\in\Pol(\bC_\Phi)$ and let $g,h\in \scrC\cap{\mathscr I}_f$.
    Then $h\in \scrC\cap{\mathscr I}_{(f_n)^\lambda}$, by taking $\lambda$ to be the identity and $f_n:=f$ for all $n\in\mathbb N$.
    By Lemma~\ref{lem:correctness-extension}, $\xi(\xi^\typ_1(g))=\xi(\xi^\typ_1(h))$,
    so that $\phi(f)$ is well-defined. 
    Moreover, again by Lemma~\ref{lem:correctness-extension}  (using $f_n:=f$ for all $n$), $\phi$ is a minor-preserving map.
    
    We  verify that $\phi$ is uniformly continuous.
    For continuity, suppose that $(f_n)$ converges to $f$, where $f$ has arity $k$.
    For each $n \in {\mathbb N}$ there exists an $h_n \in {\mathscr C} \cap {\mathscr I}_{f_n}$, and since $\xi^{\typ}_\infty(h_n)$ can only take finitely many values, one of them appears for infinitely many $n \in {\mathbb N}$. 
    It follows that there exists an $h$ that lies in ${\scrC} \cap {\mathscr I}_{f_n}$ for infinitely many $n$. 
    By Lemma~\ref{lem:correctness-extension} with $\lambda$ being the identity, we have that $\xi^\typ_1(g)$ and $\xi^\typ_1(h)$ induce the same projection on $\{S,T\}$,
    that is, $\xi(\xi^\typ_1(g))=\xi(\xi^\typ_1(h))$.
    In particular, the sequence $(\phi(f_n))$ converges to $\phi(f)$ and $\phi$ is continuous.
    
     Next, we verify that $\phi$ preserves left-composition with $\overline{\Aut(\bC_\Phi)}$. Let $e$ be a unary operation in $\overline{\Aut(\bC_\Phi)}$,
	and let $f\in\scrC$.
	Let $g$ be canonical and interpolated by $e\circ f$ modulo $\overline{\Aut(\bC_\Phi)}$. Note that $g$ is also interpolated by $f$ modulo $\overline{\Aut(\bC_\Phi)}$, so that $\phi(f) = \xi(g) = \phi(e\circ f)$.
	Since the only unary operation in $\Projs$ is the identity operation, we finally have $\phi(e)\circ\phi(f)=\phi(f)$.
    It has been shown in~\cite{wonderland} (Proposition 6.4) that any continuous mapping from the polymorphism clone of a countable $\omega$-categorical structure
     to another function clone 
     is uniformly continuous if 
    it preserves left-composition
    with automorphisms of the structure. 
    Therefore, our map $\phi$ is uniformly continuous.
\end{proof}

\subsection{The dichotomy: conclusion}
Summing up the results of the previous two sections, we obtain the following dichotomy for precoloured MMSNP sentences. 

\begin{theorem}\label{thm:main-precoloured}
    Let $\Phi$ be a precoloured MMSNP sentence.
    Let $\scrC$ be the clone of polymorphisms of $\bC^\tau_\Phi$ that are canonical with respect to $(\bC_\Phi,<)$.
    Then the following equivalent statements hold:
    \begin{enumerate}[(1)]
        \item there is a clone homomorphism $\scrC^\typ_1\to\Projs$;
        \item there is a uniformly continuous minor-preserving map $\Pol(\bC_\Phi^\tau)\to\Projs$ that preserves left-composition by $\overline{\Aut(\bC_\Phi)}$;
    \end{enumerate}
    and $\Csp(\bC^\tau_\Phi)$ is NP-complete, or $\scrC^\typ_1$ contains a Siggers operation
    and $\Csp(\bC^\tau_\Phi)$ is in P.
\end{theorem}
\begin{proof}
     The implication $\neg (2) \Rightarrow \neg (1)$ is Theorem~\ref{thm:conclusion-hardness},
     and $\neg (1)$ is equivalent to the fact that $\scrC^\typ_1$ contains a Siggers operation by Theorem~\ref{thm:fin-ua}.      
\end{proof}
Note the existence of a Siggers operation in $\scrC^\typ_1$ is for given $\Phi$ clearly algorithmically decidable.

Via the facts about precolourings from Section~\ref{sect:precoloured}, Theorem~\ref{thm:main-precoloured} implies a more general result about  MMSNP sentences in normal form, Theorem~\ref{thm:main-connected} below.
We actually show a stronger formulation than the conjecture since we also provide a characterisation of the polynomial-time tractable cases
using pseudo-Siggers polymorphisms (which does not directly follow from~\cite{BartoPinskerDichotomy} since the structures under consideration need not be model-complete cores).

A $6$-ary operation $s$ is called a \emph{pseudo-Siggers modulo $e_1,e_2$} if the equality 
$$e_1(s(x,y,x,z,y,z))= e_2(s(y,x,z,x,z,y))$$ holds for all $x,y,z$.
In order to show that the two cases in Theorem~\ref{thm:main-connected} are disjoint, we need the following transfer for the existence of pseudo-Siggers polymorphisms of $\Pol(\bC^\tau_\Phi)$.

\begin{proposition}\label{prop:injectivisation-Siggers}
    The structure $\bC^\tau_\Phi$ has a pseudo-Siggers polymorphism modulo $\overline{\Aut(\bC_\Phi)}$ if, and only if,
    it has an injective polymorphism  that is pseudo-Siggers modulo $\overline{\Aut(\bC_\Phi,<)}$.
\end{proposition}
\begin{proof}
    Let $s\colon(\bC^\tau_\Phi)^6\to\bC^\tau_\Phi$ be the given pseudo-Siggers.
   Let $\bB$ be the $(\tau \cup \sigma)$-expansion
   of $(\bC^\tau_\Phi)^6$ where 
   $(a_1,\dots,a_6)$ has the same color as   $s(a_1,\dots,a_6)$ in $\bC_\Phi$. 
  We view $\bC_\Phi$ as a substructure of $\bB^{\ind}_{\mathcal F}$, and consequently
     $s$ as a homomorphism $\bB \to \bB^{\ind}_{\mathcal F}$.
    By Lemma~\ref{lem:construction-of-injective-maps}, we obtain an injective homomorphism $t\colon \bB \to\bB^{\ind}_{\mathcal F}$
    such that for all tuples $\tuple a,\tuple b$ with pairwise distinct entries in $\bB$, if $s(\tuple a)$ and $s(\tuple b)$ are in the same orbit in $\bB^{\ind}_{\mathcal F}$
    then so are $t(\tuple a)$ and $t(\tuple b)$ (call this property $(\dagger)$).

    We claim that for every finite substructure $\bA$ of $\bC^\tau_\Phi$, there exists an injective homomorphism 
    $t_A\colon \bA^6 \to \bC^\tau_\Phi$ that is pseudo-Siggers modulo $\Aut(\bC_\Phi,<)$.
    Let $\tuple a$ be the tuple whose entries are of the form $(x,y,x,z,y,z)$ for $x,y,z\in A$ (that is, $\tuple a$ is a tuple of $6$-tuples).
    Let $\tuple b$ be the tuple whose entries are of the form $(y,x,z,x,z,y)$ (using the same enumeration of the elements $(x,y,z)$ of $A^3$ as in $\tuple a$).
    Since $s$ is pseudo-Siggers modulo $\Aut(\bC_\Phi)$,
    the tuples $s(\tuple a)$ and $s(\tuple b)$ lie in the same orbit of $\Aut(\bC_\Phi)$, so they lie in the same orbit of 
     $\Aut(\bB^{\ind}_{\mathcal F})$ by Lemma~\ref{lem:C-phi-mc-core}.
    By $(\dagger)$, we obtain that $t(\tuple a)$ and $t(\tuple b)$ lie in the same orbit of $\Aut(\bB^{\ind}_{\mathcal F})$.
    Moreover,
    since $t$ is injective, there exists $\alpha\in\Aut(\bB^{\ind}_{\mathcal F})$
    such that the tuples $(\alpha t)(\tuple a)$ and $(\alpha t)(\tuple b)$ lie in the same orbit of
    $\Aut(\bB^{\ind}_{\mathcal F},<)$. 
    Let $h\colon (\bB^{\ind}_{\mathcal F},\neq)\to(\bB^{\hom}_{\mathcal F},\neq)$ be an injective homomorphism that is canonical from $(\bB^{\ind}_{\mathcal F},<)$ to $(\bB^{\hom}_{\mathcal F},<)$.
    We claim that $t_A:=h\circ\alpha\circ t$ is the desired injective homomorphism.

    We first prove that the range of $t_A$ is included in the domain of $\bC_\Phi$, that is, that all the elements that appear in the range are coloured.
    Let $a_1,\dots,a_6\in A$. Since the range of $s$ is included in the domain of $\bC_\Phi$, 
    there is an $M\in\sigma$ such that $s(a_1,\dots,a_6) \in M^{\bC_\Phi}$.
    By Lemma~\ref{lem:construction-of-injective-maps}, the element $t(a_1,\dots,a_6) \in M^{\bB^{\ind}_{\mathcal F}}$, so that $h(\alpha(t(a_1,\dots,a_6))) \in M^{\bC_\Phi}$ and hence lies in $\bC_\Phi$.
 
   We now show that $t_A\colon\bA^6\to\bC^\tau_\Phi$ is pseudo-Siggers modulo $\Aut(\bC_\Phi,<)$.
    Note that since $(\alpha t)(\tuple a)$ and $(\alpha t)(\tuple b)$ lie in the same orbit in $\Aut(\bB^{\ind}_{\mathcal F},<)$,
    the tuples $t_A(\tuple a)$ and $t_A(\tuple b)$ lie in the same orbit in $\Aut(\bB^{\hom}_{\mathcal F},<)$ by the canonicity of $h$.
    Therefore, there exists $\beta\in\Aut(\bB^{\hom}_{\mathcal F},<)$ such that $\beta t_A(\tuple a) = t_A(\tuple b)$.
    Since the domain of $\bC_\Phi$ is preserved by automorphisms of $(\bB^{\hom}_{\mathcal F},<)$ 
	the restriction of $\beta$ to the domain of
	$\bC_\Phi$ is an automorphism of $(\bC_\Phi,<)$.
    In conclusion, $t_A$ is pseudo-Siggers modulo $\Aut(\bC_\Phi,<)$.

A standard compactness argument now shows that there exists $t'\colon(\bC_\Phi)^6\to\bC_\Phi$ that is on every finite subset pseudo-Siggers modulo $\Aut(\bC_\Phi,<)$.  
    Another compactness argument
    (the lift lemma; see, e.g., Lemma 4.2 in \cite{BartoPinskerDichotomy}) shows that 
    $t'$ is pseudo-Siggers modulo $\overline{\Aut(\bC_\Phi,<)}$.
\end{proof}

\begin{theorem}[\cite{BartoPinskerDichotomy} and \cite{BKOPP}]
\label{thm:inf-dichotomy}
Let $\bB$ be an $\omega$-categorical model-complete core. Then the following are equivalent:
\begin{itemize}
\item $\bB$ has an expansion $\bC$ by finitely many unary singleton relations such that $\Pol(\bC)$ has a continuous clone homomorphism to $\Pr$, or 
\item $\bB$ has no pseudo-Siggers polymorphism.
\end{itemize}
If $\bB$ is moreover a first-order reduct of a homogeneous structure with finite relational signature, the two items
are equivalent to:
\begin{itemize}
\item $\Pol(\bB)$ has a uniformly continuous minor-preserving map to $\Pr$. 
\end{itemize}
\end{theorem}

\begin{theorem}\label{thm:main-connected}
    Let $\Phi$ be an MMSNP sentence in strong normal form.
    Then either 
    \begin{itemize}
    \item 
    there is a uniformly continuous minor-preserving map $\Pol(\bC_\Phi^\tau)\to\Projs$ 
    and $\Csp(\bC^\tau_\Phi)$ is NP-complete, or 
    \item  $\Pol(\bC^\tau_\Phi)$ contains a pseudo-Siggers operation modulo $\overline{\Aut(\bC_\Phi)}$ 
    and $\Csp(\bC^\tau_\Phi)$ is in P.
    \end{itemize}
    In particular, Conjecture~\ref{conj:inf-dichotomy} holds for all CSPs in MMSNP.
\end{theorem}
\begin{proof}
If there is a uniformly continuous minor-preserving map $\Pol(\bC_\Phi^\tau)\to\Projs$, then the NP-hardness of $\Csp(\bC_\Phi^\tau)$  follows from Theorem~\ref{thm:wonderland}.
Otherwise, let $\Psi$ be the standard precolouration of $\Phi$ with input signature $\rho \supseteq \tau$. 
By Theorem~\ref{thm:precol} there is no uniformly continuous minor-preserving map $\Pol(\bC^\rho_\Psi) \to \Projs$,
so that Theorem~\ref{thm:main-precoloured} above states that $\Csp(\bC^\rho_\Psi)$ is in P, which implies that $\Csp(\bC^\tau_\Phi)$ is also in P.
Moreover, Proposition~\ref{prop:neq-is-algebraic} implies that there is no uniformly continuous minor-preserving map from $\Pol(\bC^\tau_\Phi,\neq)$ to $\Projs$.
Since $(\bC^\tau_\Psi,\neq)$ is a model-complete core by Theorem~\ref{thm:mc-core},
it follows from Theorem~\ref{thm:inf-dichotomy} that $\Pol(\bC^\tau_\Phi,\neq)$ contains a pseudo-Siggers operation modulo $\overline{\Aut(\bC^\tau_\Phi)}$. 
Therefore, also $\Pol(\bC^\tau_\Phi)$ contains a pseudo-Siggers operation modulo $\overline{\Aut(\bC^\tau_\Phi)}$.

To show that the two cases are mutually exclusive, suppose that $\Pol(\bC^\tau_\Phi)$ contains 
a pseudo-Siggers operation $g$. 
Then $\Pol(\bC^\tau_\Phi,\neq)$ has a pseudo-Siggers by Proposition~\ref{prop:injectivisation-Siggers}.
Since $(\bB,\neq)$ is a model-complete core, Theorem~\ref{thm:inf-dichotomy} implies that there is no uniformly continuous minor-preserving map from $\Pol(\bB,\neq)\to\Projs$.
By Proposition~\ref{prop:neq-is-algebraic}, there is no uniformly continuous minor-preserving map $\Pol(\bB) \to \Projs$.

Finally, we show that the above implies
Conjecture~\ref{conj:inf-dichotomy} for CSPs in MMSNP. Suppose that $\bB$ is an $\omega$-categorical structure such that $\Phi$ describes 
$\Csp(\bB)$. Since $\bB$ and $\bC^{\tau}_\Phi$ are $\omega$-categorical and have the same CSP, they are homomorphically equivalent. 
Proposition~\ref{prop:uch1} then implies 
that there are uniformly continuous minor-preserving maps $\Pol(\bB) \to \Pol(\bC^{\tau}_\Phi)$ and $\Pol(\bC^{\tau}_\Phi) \to \Pol(\bB)$.
\end{proof}

\medskip 
Finally, we obtain the following theorem for MMSNP in general.

\begin{theorem}
\label{thm:main}
Let $\Phi$ be an MMSNP $\tau$-sentence. 
Then $\Phi$ is logically equivalent to a 
finite disjunction $\Phi_1 \vee \cdots \vee \Phi_k$ of connected MMSNP sentences;
for each $i \leq k$ there exists an $\omega$-categorical structure $\bB_i$ such that 
$\Phi_i$ describes $\Csp(\bB_i)$, and 
either
\begin{itemize}
\item $\Pol(\bB_i)$ has a 
\blue{uniformly continuous} 
minor-preserving map to $\Pr$, for some $i \in \{1,\dots,k\}$, and $\Phi$ is NP-complete.
\item $\Pol(\bB_i)$ contains a
pseudo-Siggers polymorphism, for each $i \in \{1,\dots,k\}$, and 
$\Phi$ is in P. 
\end{itemize}
\end{theorem}
In particular, every problem in MMSNP is in P or NP-complete.
\begin{proof}
By Proposition~\ref{prop:connected}, 
the sentence $\Phi$ is logically equivalent to a finite disjunction $\Phi_1 \vee \cdots \vee \Phi_k$ of connected MMSNP sentences. By Theorem~\ref{thm:snf-existence},
we may assume that each of the $\Phi_i$ is in strong normal form. The sentence $\Phi_i$ describes $\Csp(\bC^\tau_{\Phi_i})$. Theorem~\ref{thm:main-connected}
above states that either $\Pol(\bC^\tau_{\Phi_i})$ 
has a uniformly continuous minor-preserving map to $\Projs$, and $\Phi_i$ is NP-complete, or $\Pol(\bC^\tau_{\Phi_i})$ contains 
a pseudo-Siggers polymorphism. 
Then Proposition~\ref{prop:connected-compl}
states that $\Phi$ is in P if the second case applies for all $i \leq k$, and is NP-hard otherwise.
\end{proof}

Again, it is clear from the proof that given
an MMSNP sentence $\Phi$, the two cases in Theorem~\ref{thm:main} can be distinguished algorithmically. The reason is that the connected
MMSNP sentences $\Phi_1, \dots, \Phi_k$ can be computed from $\Phi$ (Proposition~\ref{prop:connected}), 
and also each of the $\Phi_i$ can be effectively rewritten into strong normal form (Theorem~\ref{thm:snf-existence}), and so 
the claim follows from our observations above.

\section{Conclusion and Open Problems}
Every problem in MMSNP is in P or NP-complete.
Our new proof of this complexity dichotomy does not rely on the expander constructions of Kun~\cite{Kun} and allows to explain the dichotomy by algebraic properties.
On the other hand, we use a number of other results from the literature, most notably the powerful recent Ramsey result by \Nesetril\ and Hubi\v{c}ka;
the usage of this result via canonical functions~\cite{BodPin-CanonicalFunctions} is deeply used in our approach at numerous places.
We consider it as an interesting question whether the result of \Nesetril\ and Hubi\v{c}ka can be used to give another proof of Kun's result on expander structures. 

We want to point out again that our approach of the dichotomy for MMSNP leads to a stronger result:
we prove the universal-algebraic tractability conjecture for CSPs in MMSNP.
This result implies the dichotomy for MMSNP since every problem in MMSNP is a finite union of CSPs.
We close with two interesting problems related to MMSNP that remain open. 

\begin{enumerate} 
\item Can we drop uniform continuity in the statement of our main result? In all other existing classifications, this was possible (see~\cite{BKOPP}), but it has recently been shown in~\cite{TopologyIsRelevant} that this is not possible for general $\omega$-categorical structures (which might have an infinite relational signature). 
\item Is the following computational problem decidable:
Given an MMSNP sentence, decide whether there exists an equivalent Datalog program?
This problem has been called \emph{Datalog rewritability} and has been asked in~\cite{FeierKuusistoLutz} 
(also see~\cite{Qualitative-Survey} for a discussion of the important challenge of algebraically characterising the power of Datalog for CSPs with $\omega$-categorical templates). 
We conjecture that an MMSNP sentence is Datalog-rewritable if, and only if, the clone $\scrC^\typ_1$ does not have a clone homomorphism to a clone of linear maps on a finite prime field
(where $\scrC$ is the clone of polymorphisms of $\bC_\Phi$ that are canonical with respect to $(\bC_\Phi,<)$.
If this is true, then the Datalog rewritability problem is indeed decidable.
\end{enumerate} 
\clearpage
\bibliographystyle{alpha} 
\bibliography{local.bib}

\clearpage
\appendix

\section{The Hubi\v{c}ka-Ne\v{s}et\v{r}il theorem}
\label{sect:hn}
We need additional terminology to properly explain how to deduce the Ramsey statement that we need (Theorem~\ref{thm:HN-Ramsey}) from 
the results of 
Hubi\v{c}ka and Ne\v{s}et\v{r}il~\cite{Hubicka-Nesetril-All-Those}. 
Throughout this section, let $\tau$ be a relational signature. 
A class ${\mathcal K}$ 
of finite $\tau$-structures 
\begin{itemize}
\item has the \emph{Ramsey property} if for every $r \in \mN$
and all finite substructures
$\bA,\bB \in {\mathcal K}$ there exists $\bC  \in {\mathcal K}$ such that 
$\bC \to (\bB)^\bA_r$ holds; 
\item is called \emph{hereditary} if it is closed under substructures.
\end{itemize}

A primitive positive formula $\phi(x_1,\dots,x_n)$ is called \emph{irreducible} if it cannot be equivalently written as $\phi_1(x_1,\dots,x_n) \wedge \phi_2(x_1,\dots,x_n)$ such that $\phi_1$
and $\phi_2$ have \blue{strictly less existentially quantified variables than $\phi$. 
A $\tau$-structure $\bA$ is called  \emph{irreducible} if for all $a_1,\dots,a_k \in A$ 
the formula $\exists a_1,\dots,a_k. \phi$,
for $\phi$ the canonical query of $\bA$, 
is irreducible.}
Note that $\bA$ is irreducible if and only if 
for all $a,b \in A$ there is an $R \in \tau$ 
and a $t \in R^{\bA}$ such that both $a$ and $b$
appear in entries of $t$ (this is the original definition in the literature of structural Ramsey theory).  

A homomorphism $f$ from $\bA$ to $\bB$ is called
a \emph{homomorphism-embedding} if $f$ restricted to any irreducible substructure of $\bA$ is an embedding into $\bB$.
Let $\bC$ be a $\tau$-structure and ${\mathcal K}$ a class of irreducible $\tau$-structures. An irreducible 
$\tau$-structure $\bC' \in {\mathcal K}$ is a \emph{(strong) ${\mathcal K}$-completion}
of $\bC$ if there exists an (injective) homomorphism-embedding from $\bC$ into $\bC'$. 
\begin{definition}
Let $\mathcal R$ be a class of finite irreducible $\tau$-structures and $\mathcal K$ a subclass of ${\mathcal R}$. We say that ${\mathcal K}$ is a \emph{locally finite subclass of ${\mathcal R}$} if for every $\bC_0 \in {\mathcal R}$ there exists an $n \in {\mathbb N}$ such that every 
$\tau$-structure $\bC$ with a homomorphism-embedding into $\bC_0$ also has a strong ${\mathcal K}$-completion, provided that 
every substructure of $\bC$ with at most $n$ vertices has a strong ${\mathcal K}$-completion. 
\end{definition}

\begin{theorem}
[Theorem 2.1 from~\cite{Hubicka-Nesetril-All-Those}]
\label{thm:locally-finite-Ramsey}
Let $\tau$ be a relational signature, let
$\mathcal R$ be a Ramsey class of irreducible finite $\tau$-structures, and let ${\mathcal K}$ be hereditary locally finite subclass of ${\mathcal R}$ with strong amalgamation. Then ${\mathcal K}$ is Ramsey. 
\end{theorem}

Let $\tau$ be a relational signature, and let ${\mathfrak R}$ be the homogeneous 
$\tau$-structure whose age is the class of all finite $\tau$-structures. 
The following theorem is known as the \Nesetril-R\"odl theorem for relational structures. 
\begin{theorem}[Theorem~3.6 in~\cite{Hubicka-Nesetril-All-Those}]
\label{thm:nr}
The structure $\mathfrak R * ({\mathbb Q};<)$ is Ramsey. 
\end{theorem}

\begin{theorem}[Consequence of Theorem 2.1 from~\cite{Hubicka-Nesetril-All-Those}\footnote{The authors thank Jan Hubi\v{c}ka for helpful discussions.}]
Let ${\mathcal F}$ be a finite set of finite $\tau$-structures. 
Then the structure $(\HN,<)$ is Ramsey. 
\end{theorem}
\begin{proof}
Let $m$ be the size of the maximal structure in
${\mathcal F}$, and let $\rho$ be the signature
of $\HN$. 
Recall that $\rho$ contains a relation symbol $R_\phi \in \rho$ for 
every primitive positive $\tau$-formula $\phi$ 
with at most $m$ variables. 
Let ${\mathfrak R}^{*}$ be the $\rho$-expansion of 
${\mathfrak R}$ defined by setting $R_\phi^{{\mathfrak R}^*}$ to be the relation defined by $\phi$ over ${\mathfrak R}$. 
Clearly, ${\mathcal K} := \Age(\HN,<)$ is hereditary and has strong amalgamation, and ${\mathcal R} := \Age({\mathfrak R^*} * ({\mathbb Q};<))$
is a Ramsey class of irreducible structures by Theorem~\ref{thm:nr} (note that ${\mathfrak R}^* * ({\mathbb Q};<)$ is Ramsey if and only if 
$\mathfrak R * ({\mathbb Q};<)$ is Ramsey since 
the two structures have the same automorphism group). 
So by Theorem~\ref{thm:locally-finite-Ramsey}
it suffices to verify that ${\mathcal K}$ is a locally finite subclass of
${\mathcal R}$. Let $\bC_0 \in {\mathcal R}$ be arbitrary. We choose $n := m$. 
Let $\bC$ be a $\rho$-structure with a homomorphism-embedding $e$ into $\bC_0$ such that every substructure of $\bC$ with at most $n$ vertices has a strong ${\mathcal K}$-completion. 
Let $\psi$ be the canonical query of
$\bC$, and replace every formula $R_\phi(x_1,\dots,x_k)$ in $\psi$ by $\phi(x_1,\dots,x_k)$. 
Let $\bC'$ be the canonical database for the resulting $(\tau \cup \{<\})$-formula; replace
$<^{\bC'}$ by any linear extension. 
Let 
$\bC''$ be the $\rho$-expansion of $\bC'$
where $R_\phi \in \rho$ denotes 
the relation defined by $\phi$ in $\bC'$,
and let $\bC'''$ be the substructure of $\bC''$
induced by $C$. 

{\bf Claim.} $\bC''' \in \Age(\HN,<)$. To show the claim it
suffices to show that no structure in 
${\mathcal F}$ homomorphically maps into 
the $\tau$-reduct of $\bC'''$. Suppose that
there exists a homomorphism $h$ from 
$\mathfrak F \in {\mathcal F}$ to $(\bC''')^{\tau}$.
By assumption, the substructure $\mathfrak F'$ induced by
the image of $h$ in $\bC'''$ has a strong ${\mathcal K}$-completion, i.e., there exists
an injective homomorphism-embedding $g$
from $\mathfrak F'$ to a structure in $\mathcal K$.
But then $g \circ h$ is a map from ${\mathfrak F}$ to a structure in ${\mathcal K}$, a contradiction. 

So to show that $\bC$ has a strong ${\mathcal K}$-completion it therefore suffices to show
that the natural inclusion map $\id_C$ from $\bC$ to $\bC'''$ is
an (injective) homomorphism-embedding. 
It is clear from the construction that 
$\id_C$ is a homomorphism, and that
$\bC'''$ and $\bC$ have the same $\tau$-reducts. 
So suppose that there exists a tuple $t = (t_1,\dots,t_k) \in C^k$
such that $t \in R^{\bC'''}_\phi$ for some
$R_\phi \in \rho$. We claim that 
then $t \in R^{\bC}_\phi$. 
Clearly, it suffices
to show the claim for irreducible $\phi$. 
The structure $\bC'$
must contain vertices that witness that 
the primitive positive formula $\phi$ holds in
$\bC'$ on the tuple $t$. 
Each of those vertices
is either a vertex of $\bC$ or has been introduced
for the existentially quantified variables of 
some conjunct $R_{\psi}(s_1,\dots,s_\ell)$
of the canonical query of $\bC$. Since
$\phi$ is irreducible, $\psi$ can be chosen
so that $\{s_1,\dots,s_{\ell}\}$ contains 
$t_1,\dots,t_k$ and $\psi(s_1,\dots,s_\ell)$
implies $\phi(t_1,\dots,t_k)$. 
Let $\bD$ be the substructure of $\bC$ induced by 
$s_1,\dots,s_\ell$. 
Note that because of the tuple 
$(s_1,\dots,s_\ell) \in R^{\bD}_{\psi}$ the
structure $\bD$ is irreducible. 
Since $\ell \leq n$ the structure $\bD$ 
has a strong ${\mathcal K}$-completion,
and since it is irreducible we must have that
$\bD \in {\mathcal K}$. In particular, we
must have that $t \in R_\phi^{\bD}$
and hence $t \in R_\phi^{\bC}$, which
is what we wanted to show. 
\end{proof}

\end{document}